\documentclass[12pt]{article}
\usepackage{amsmath,amssymb,exscale,cancel}
\usepackage{slashed}
\allowdisplaybreaks
\usepackage{graphicx}
\usepackage{epsfig}
\usepackage{multicol}
\usepackage{color}
\usepackage{mathrsfs}
\usepackage{blindtext}
 \usepackage{fancyhdr}
\usepackage{hyperref}
\usepackage{cite}
\usepackage{mathtools}

\usepackage{floatrow}

\usepackage{graphicx}
\usepackage{sidecap}

\usepackage[latin1]{inputenc} 

\usepackage{multicol}  
 
 \usepackage{titlesec}
 
 \usepackage{rotating,slashed,xcolor,amsfonts,expdlist,charter}

\numberwithin{equation}{section}

\usepackage{xcolor}
\usepackage{sectsty}

\usepackage{mdframed}
\usepackage{titletoc}

\numberwithin{equation}{section}

\usepackage{xcolor}
\usepackage{sectsty}

\usepackage{mdframed}
\usepackage{titletoc}

\definecolor{secnum}{RGB}{13,151,225}
\definecolor{ptcbackground}{RGB}{212,237,252}
\definecolor{ptctitle}{RGB}{0,177,235}

\titlecontents{lsection}
  [5.8em]{\sffamily}
  {\color{secnum}\contentslabel{2.3em}\normalcolor}{}
  {\titlerule*[1000pc]{.}\contentspage\\\hspace*{-5.8em}\vspace*{5pt}%
    \color{white}\rule{\dimexpr\textwidth-15.5pt\relax}{1pt}}

\usepackage{hyperref}
\hypersetup{colorlinks,bookmarksopen,bookmarksnumbered,citecolor=rossos,
linkcolor=redy,pdfstartview=FitH,urlcolor=green-go}
\usepackage{slashed}

\definecolor{blus}{cmyk}{1,0.9,0,0.1}
\definecolor{verdes}{cmyk}{0.99,0,0.59,0.65}
\definecolor{rossos}{cmyk}{0,1,1,0.55}
\definecolor{redy}{cmyk}{0,1,1,0.7}
\definecolor{greeny}{cmyk}{0.99,0,0.59,0.98}
\definecolor{green-go}{cmyk}{0.79,0,0.59,0.5}

\usepackage{titlesec}

\def\Lag{\mathscr{L}}

\newcommand{\beq}{\begin{equation}}
\newcommand{\eeq}{\end{equation}}

\def\hhref#1{\href{http://arxiv.org/abs/#1}{arXiv:#1}} 

 \def\Lag{\mathscr{L}}
 
\newcommand{\tmtextbf}[1]{{\bfseries{#1}}}
\newcommand{\tmtextrm}[1]{{\rmfamily{#1}}}

\def\be{\begin{equation}}
\def\ee{\end{equation}}
\def\ba{\begin{array} }
\newcommand{\Tr}{\,{\rm Tr}}
\def\bac{\begin{array} {c}}
\def\bacc{\begin{array} {cc}}
\def\baccc{\begin{array} {ccc}}
\def\bacccc{\begin{array} {cccc}}
\def\ea{\end{array}}
\def\bea{\begin{eqnarray}}
\def\eea{\end{eqnarray}}

\definecolor{red}{rgb}{1,0,0}

\def\psl{\hbox{\hbox{${p}$}}\kern-1.9mm{\hbox{${/}$}}}
\def\dsl{\hbox{\hbox{${\partial}$}}\kern-2.2mm{\hbox{${/}$}}}
\def\Dsl{\hbox{\hbox{${D}$}}\kern-2.6mm{\hbox{${/}$}}}

\def\Lag{\mathscr{L}}

\newcommand{\gappeq}{{\rlap{{\raise}.5ex\text{\ensuremath{>}}}{{\lower}.5ex\text{\ensuremath{\sim}}}}}
\newcommand{\lappeq}{{\rlap{{\raise}.5ex\text{\ensuremath{<}}}{{\lower}.5ex\text{\ensuremath{\sim}}}}}
\newcommand{\I}{\tmtextrm{1{\kern}-.24em l}}

\begin{document}
\topmargin -1.0cm
\oddsidemargin 0.9cm
\evensidemargin -0.5cm

{\vspace{-1cm}}
\begin{center}

\vspace{-1cm}


 {\Huge \tmtextbf{ 
\color{redy} Introduction to Thermal Field Theory:\\ From First Principles to Applications}} {\vspace{.5cm}}\\

\vspace{1.9cm}

{\large  {\bf Alberto Salvio }

\vspace{.4cm}
{\em  

Physics Department, University of Rome Tor Vergata, \\ 
via della Ricerca Scientifica, I-00133 Rome, Italy\\

\vspace{0.2cm}

I. N. F. N. -  Rome Tor Vergata,\\
via della Ricerca Scientifica, I-00133 Rome, Italy

\vspace{.4cm}

Email: alberto.salvio@roma2.infn.it 

\vspace{.7cm}

\today

\vspace{0.4cm}


 \vspace{0.5cm}
}

\vspace{.3cm}

\vspace{0.5cm}

}
\vspace{0.cm}

%
%
%
%

%
 \vspace{1.5cm}

\end{center}

%
%
 \noindent---------------------------------------------------------------------------------------------------------------------------------
 \begin{center}
{\bf \large Abstract}
\end{center}

\noindent This review article provides the basics and discusses some important applications of thermal field theory, namely the combination of statistical mechanics and relativistic quantum field theory. In a first part the fundamentals are covered: the density matrix, the corresponding averages and the treatment of fields of various spin in a medium. A second part is dedicated to the computation of thermal Green's function for scalars, vectors and fermions with path-integral methods. These functions play a crucial role in thermal field theory, as explained here. 
A more applicative part of the review is dedicated to the production of particles in a medium and to phase transitions in field theory, including the process of vacuum decay in a general theory featuring a first-order phase transition. To understand this review, the reader should only  have a good knowledge of non-statistical quantum field theory.

  \vspace{0.5cm}
  
\noindent---------------------------------------------------------------------------------------------------------------------------------

%

  \vspace{-0.9cm}
  
  \newpage
  \noindent --------------------------------------------------------------------------------------------------------------------------------
\tableofcontents

\noindent --------------------------------------------------------------------------------------------------------------------------------

\vspace{0.5cm}


%




\newpage

\subsection*{Notation}
\begin{itemize}
\item Dirac's notation is used to denote states.
\item Repeated indices understand a summation (unless otherwise stated).
\item Latin letters denote space indices, while greek letters denote spacetime indices. A label $0$ indicates the time component of a four-vector.
\item A spatial three-vector is written with an arrow on top, e.g.~$\vec{x}$ for the spatial coordinates.
\item $\epsilon_{ijk}$ is the totally antisymmetric Levi-Civita symbol (with $\epsilon_{123}=1$).
\item The convention for the spacetime metric is $\eta=\,$diag$(1,-1,-1,-1)$.  Its components $\eta_{\mu\nu}$ are used as usual to lower and raise the spacetime indices.
\item We use natural units where the speed of light $c$, the reduced Planck constant $\hbar$ and the Boltzmann constant $k_B$ are all equal to one, unless otherwise stated.
\item $O^\dagger$: Hermitian conjugate of the  generic linear operator $O$.
\item $A^*$: complex conjugate of the  generic  matrix $A$.
\item $A^T$: transpose of $A$.
\item $A^{\dag}$: Hermitian conjugate of $A$.
\item $[O_1,O_2]$: commutator of two generic operators $O_1$ e $O_2$.
\item $\{O_1,O_2\}$: anticommutator of two generic operators $O_1$ e $O_2$.
\end{itemize}

\newpage

%
%

\section{Introduction}\label{Basic}

This review provides an introduction to the fundamentals and some important applications of thermal field theory (TFT), namely the combination of statistical mechanics and relativistic quantum field theory. 

The combination of statistical mechanics and quantum field theory was first developed in the non-relativistic setting in the late 1950s to theoretically describe condensed and nuclear matter in standard laboratory experiments. This was important to bridge the gap between the theoretical description of microscopic and macroscopic systems. The formalism of second quantization, which leads to quantum field theory, can be applied within both Galilean and Einsteinian relativity and is the most effective language to describe many-body systems, where the number of particles is allowed to change. In 1955 Matsubara~\cite{Matsubara:1955ws} showed how a statistical equilibrium description of quantum field theory can be obtained by formally substituting time with an imaginary quantity. This paved the way for the imaginary-time formalism in equilibrium TFT, where the theory is described as an Euclidean quantum field theory with periodic (for bosonic fields) or antiperiodic (for fermionic fields) boundary conditions with respect to the imaginary time.

In 1965 Fradkin~\cite{Fradkin:1960lgt} pioneered the statistical study of relativistic quantum field theory (i.e.~TFT). Several years later it was realized that TFT is an essential tool to describe processes in the cosmological plasma. Indeed, in that context  the presence of a huge number of particles prevents the exact (non-statistical) description and relativistic effects are non negligible. During the subsequent decades much work was done to understand the nature of the electroweak phase transition using TFT, which is of great interest both in particle physics and cosmology.

Moreover, at some point during the history of the universe, when the temperature was sufficiently high, quarks and gluons behaved as weakly interacting particles, almost free to move in their plasma, the quark-gluon plasma. TFT was and still is an essential tool to describe such physical situation. Experimental activities conducted in laboratories, such as CERN, 
aim to reproduce the quark-gluon plasma through  collisions of heavy ions, such as gold or lead nuclei. The application of TFT to heavy-ion collisions further boosted the interest in this theory. The study of the  frontier (the QCD transition) between the regime where quarks and gluons form such a plasma and the regime where they are confined in colorless bound particles needs non-perturbative methods. It was then natural to combine lattice gauge theory and TFT. The first studies of this type were conducted during the beginning  of 1980s. In the recent years, several lattice studies have agreed on the evidence that the QCD transition temperature is about 156.5 MeV at negligible densities~\cite{Aarts:2023vsf}.

Today TFT remains the standard theoretical tool to study particle physics processes (decays, scattering processes, particle production, phase transitions, etc.) in a thermal medium.  In cosmology, it is believed that such a  medium was originally formed shortly after a primordial exponential expansion of the spacetime volume (a.k.a.~inflation), through a mechanism called reheating. 

The aim of this review is not to give an exhaustive overview of the TFT literature, but is rather to introduce in a constructive way the fundamentals of TFT starting from the basic principles of quantum mechanics and relativity. Once this is done, it appears natural to provide some of the most important physical applications of TFT that can already been understood. Here two main applications are discussed: particle production in a thermal bath and the theory of phase transitions in field theory. To achieve these goals, a detailed treatment of the thermal Green's functions is provided.  These functions, whose importance in TFT goes even beyond the applications considered here, are defined as the statistical average of the expectation values of time-ordered product of fields taken on a complete set of states. Both the real-time and imaginary-time formalisms are explained, as they are, respectively, the most suitable approaches for describing particle production in a thermal bath and phase transitions.

 As any other review\footnote{For other reviews and textbooks related to the subjects discussed here see~\cite{Quiros:1994dr,Bellac:2011kqa,Nair:2005iw,Landsman:1986uw,Das,Laine:2016hma,Ghiglieri:2020dpq} (see also~Ref.~\cite{Caron-Huot:2007zhp} for a related discussion).}, 
 most of the results presented here are not original, but actually some derivations and explanations are  first provided here. These include, for example,  the derivation and interpretation of the full equilibrium density matrix given in Sec.~\ref{Density matrix and ensemble averages}, the explanation, given in the same section, of the general relevance in TFT of the thermal Green's functions, which are then studied in  Sec~\ref{Thermal Green's functions}, the full derivation of the path-integral representation of these  thermal functions in the fermionic case, and the description of vacuum decay for a generic first-order phase transition provided in Sec.~\ref{Thermal vacuum decay in first-order phase transitions}.
 
The writing of this review originally began as part of Ph.D. lectures on TFT given by the author in the spring of 2023 at the University of Rome Tor Vergata. No previous knowledge of TFT is required to understand this monograph, but a good knowledge of quantum field theory without statistical mechanics is assumed.

\newpage

\section{Density matrix and ensemble averages}\label{Density matrix and ensemble averages}

In quantum field theory (QFT) one usually focuses on processes with a small number of  particles.
However, there are many physical situations where one encounters processes in a medium such as, for example, the propagation of particles and scattering processes in a gas at finite temperature and density. Of course, microscopically, such situations involve a very large number of particles.

 If the system is initially in a pure state $|\alpha\rangle$, we can still write the transition amplitude for a scattering process as
\be \mathscr{A}_{\alpha\to \beta} =\langle \beta|\hat S|\alpha\rangle,  \label{Aif}\ee
where $|\beta\rangle$ is the final state in question\footnote{We take as usually $|\alpha\rangle$ and $|\beta\rangle$ with unit norm without loss of generality.} and $\hat S$ is the scattering operator, which transforms $|\alpha\rangle$ into the state (represented in the interaction picture) at time $+\infty$. The evaluation of $\mathscr{A}_{\alpha\to \beta}$ may be computationally difficult, but Eq.~(\ref{Aif}) gives the right formula for the transition amplitude in general. 

However, if the initial state is only statistically specified we are interested in averaging the rates of the processes over initial states with appropriate probabilities $p_\alpha$ (or over final states with probabilities $p_\beta$). It is important to note that these are the probabilities of choosing  particular states in a statistical ensemble and are not the quantum probabilities of the collapse of the wave function onto a specific eigenstate of the observable being measured. 

If we want to compute the rates of inclusive processes (such as decay or scattering rates summed over final states or production rates summed over initial states) we can use the optical theorem, and express them
 in terms of the imaginary part of the $\alpha\to\alpha$ amplitude,
\be  \sum_\beta \langle\alpha|\hat T^\dagger|\beta\rangle\langle\beta|\hat T|\alpha\rangle=2\, {\rm Im}\left(\langle\alpha|\hat T|\alpha\rangle\right), \label{ratealpha}\ee
where $\hat T\equiv -i(\hat S-1)$ and the unitarity condition $1=\hat S^\dagger \hat S = 1+i(\hat T-\hat T^\dagger)+\hat T^\dagger \hat T$ has been used, or in terms of the $\beta\to\beta$ amplitude \be  \sum_\alpha \langle\beta|\hat T|\alpha\rangle\langle\alpha|\hat T^\dagger|\beta\rangle=2\, {\rm Im}\left(\langle\beta|\hat T|\beta\rangle\right), \label{ratebeta}\ee
where the unitarity condition in the form $1=\hat S \hat S^\dagger = 1+i(\hat T-\hat T^\dagger)+\hat T \hat T^\dagger$ has been used.
If the system is only statistically specified we should sum over $\alpha$ with weights $p_\alpha$ (using~(\ref{ratealpha}))
\be \sum_{\alpha,\beta}p_\alpha \langle\alpha|\hat T^\dagger|\beta\rangle\langle\beta|\hat T|\alpha\rangle=2\, {\rm Im}\left(\sum_\alpha p_\alpha\langle\alpha|\hat T|\alpha\rangle\right)\label{Toptical1}\ee
or equivalently sum over $\beta$ with weights $p_\beta$ (using~(\ref{ratebeta}))
\be  \sum_{\beta,\alpha} p_{\beta} \langle\beta|\hat T|\alpha\rangle\langle\alpha|\hat T^\dagger|\beta\rangle=2\, {\rm Im}\left(\sum_\beta p_\beta\langle\beta|\hat T|\beta\rangle\right). \label{Toptical2}\ee

From these results it  is clear that in TFT important quantities are the so called thermal Green's functions for the field operators $\Phi(x)$,
\be\langle {\cal T}\Phi(x_1)...\Phi(x_n) \rangle \equiv\sum_\alpha p_\alpha\langle\alpha|{\cal T}\Phi(x_1)...\Phi(x_n)|\alpha\rangle, \label{TGF}\ee
which depend on $n$ spacetime variables $x_1, ... , x_n$. Here ${\cal T}$ is the time-ordered product. The functions in~(\ref{TGF}) are also known as $n$-point functions and the one with $n=2$ is called the thermal propagator. 

In quantum statistical mechanics the above-mentioned probabilities are given by the density matrix (which is actually an operator)
\be \rho = \sum_\alpha p_\alpha |\alpha\rangle\langle\alpha|, \label{rhointro}\ee
where the unit-norm states $|\alpha\rangle$  form an orthonormal basis and  $p_\alpha$ is the probability of finding $|\alpha\rangle$  in the statistical ensemble. The expectation value of an operator $A$ with this statistical distribution is 
\be \langle A\rangle = \sum_\alpha p_\alpha\langle\alpha|A|\alpha\rangle, \ee
which, using the orthonormality of the basis $\left\{|\alpha\rangle\right\}$, can be rewritten as
\be \langle A\rangle = \Tr(\rho A). \label{averageA} \ee
Note that the thermal Green's function, Eq.~(\ref{TGF}), can be written as 
\be \langle {\cal T}\Phi(x_1)...\Phi(x_n) \rangle =\Tr\left(\rho\, T\Phi(x_1)...\Phi(x_n)\right). \label{TGFtrace} \ee
Since the $p_\alpha$ represent a probability distribution, we have that all $p_\alpha$ are real and
\be 0\leq p_\alpha\leq 1, \quad \sum_\alpha p_\alpha =1 \implies \Tr\rho =1. \label{Trr1}
\ee
The reality of $p_\alpha$ implies the hermiticity of $\rho$.
Note that choosing a statistical probability distribution $p_\alpha$ for the states $|\alpha\rangle$ is equivalent to choosing a density matrix $\rho$. Moreover, note that $p_\alpha$ can be written as $\exp(-h_\alpha)$ where $h_\alpha$ is a non-negative real number. So we can write 
\be\rho=\exp(-{\cal H}),\label{rhoToH}\ee 
where ${\cal H}$ is a Hermitian operator with real eigenvalues (which are the $h_\alpha$).
The trace is independent of the basis, so Eq.~(\ref{averageA}) and $\Tr\rho =1$ hold in any basis. The case of a pure state $|\lambda\rangle$ corresponds to $p_\lambda=1$ and $p_\alpha=0$  for all $\alpha\neq\lambda$, and in this case $\rho^2=\rho$. The equation $\rho^2=\rho$ is not only a necessary but also a sufficient condition for the system to be in a pure state. Indeed, for $\rho^2=\rho$
\be \sum_\alpha p_\alpha |\alpha\rangle\langle\alpha| =
 \rho = \rho^2= \sum_\alpha p_\alpha^2  |\alpha\rangle\langle\alpha|, \label{rr2}
\ee
where in the last step the orthonormality of the basis $\left\{|\alpha\rangle\right\}$ has been used. Since the $p_\alpha$ form a probability distribution, Eq.~(\ref{rr2}) tells us that one of the  $p_\alpha$ is one and all the others are zero, namely the system is in a pure state. Therefore, for $\rho\neq\rho^2$ we necessarily have a non-trivial mixing of states.

From Eq.~(\ref{rhointro}) it follows that under a generic unitary symmetry operator $U$ the density matrix transforms as
\be \rho\to \sum_\alpha p_\alpha U|\alpha\rangle\langle\alpha| U^\dagger=U\rho U^\dagger, \label{rhoTr}\ee
which resembles but is different from the usual transformation of a generic operator ${\cal O}$ introduced in basic courses on quantum mechanics, i.e.~${\cal O}\to U^\dagger {\cal O} U$ because $U^\dagger$ in~(\ref{rhoTr}) is on the right rather than on the left.
Suppose now that at time $t_0$ the density matrix $\rho(t_0)$ is given by~(\ref{rhointro}). Then at time $t$ the density matrix is
\be \rho(t) 
= U(t)\rho(t_0)U(t)^\dagger,\ee
where $U(t)=\exp(-iH(t-t_0))$ is the time-evolution operator. So $\rho(t)$ satisfies the time-evolution equation
\be i \dot\rho = [H,\rho], \ee
where a dot represents the derivative with respect to $t$.

Let us suppose now that the system is in thermal equilibrium. In this case $\rho$ must be independent of time and so $[H,\rho]=0$.
This equation tells us that the density matrix is a function of the form $\rho=\rho({\cal O}_1, {\cal O}_2 ...)$, where the ${\cal O}_i$ are operators that commute with the Hamiltonian, namely they are conserved quantities. An important information on this function can be obtained by considering additively conserved quantities ${\cal O}_i$, such as the energy itself, the (linear) momentum, the angular momentum, the charges, etc. Now, if one divides the system under study into independent subsystems I, II, ... the probability distribution is the product of the probability distribution of I, II, ... and so the density matrix is the direct product of the density matrices of I, II, ... : $\rho=\rho_{\rm I} \times \rho_{\rm II} \times ... $ . On the other hand, since ${\cal O}_i$ are additively conserved quantities, they can be expressed as the direct sum of conserved quantities for the subsystems, ${\cal O}_i={\cal O}_{i\rm I}+{\cal O}_{i\rm II}+...$. The only\footnote{To show that there no other solutions consider the density matrix, $\rho_0({\cal O})$,  with another normalization. i.e.~$\rho_0(0)=1$ (here the index $i$ that labels the conserved quantities is understood for simplicity). This is related to the density matrix $\rho({\cal O})$ with normalization Tr$\rho$=1 through $\rho=\rho_0/$Tr$\rho_0\equiv\rho_0/Z$. Now the conditions above give $\rho_0({\cal O}_I+{\cal O}_{II}+...)=\rho_0({\cal O}_I)\times\rho_0({\cal O}_{II})\times...$, which implies $\rho_0=\exp(-\beta {\cal O})$ and so $\rho=\exp(-\beta {\cal O})/Z$. Indeed, for generic commuting variables $x$ and $h$ the condition $\rho_0(x+h)=\rho_0(x)\rho_0(h)$ tells us 
\be \rho'_0(x) \equiv \lim_{h\to 0} \frac{\rho_0(x+h)-\rho_0(x)}{h} = \rho_0(x) \lim_{h\to 0}\frac{\rho_0(h)-1}{h} \equiv \rho_0(x)\rho'(0),\ee
which gives $\rho_0(x)=\exp(-\beta x)$, where $\beta\equiv - \rho_0'(0)$.} function of the ${\cal O}_i$ with this property is the exponential\footnote{In these notes repeated indices understand a summation (unless otherwise stated).}, 
 \be \rho= \frac1{Z}\exp\left(-\beta_i{\cal O}_i\right), \ee 
 where $Z$ and the $\beta_i$ are constants. If the ${\cal O}_i$ are Hermitian the $\beta_i$ and thus $Z$ are real because, as we have seen, we can write $\rho$ as in~(\ref{rhoToH}). 
 From $\Tr\rho =1$ it follows 
 \be Z= \Tr\exp\left(-\beta_i{\cal O}_i\right),\ee
 which is known as the partition function.

  In the relativistic\footnote{In this review we will only consider special relativity, leaving aside the (nevertheless interesting) study of gravity and non-inertial frames.} case, the conserved quantities are $H$, the (linear) momentum $\vec P$, the angular momentum $\vec J$ and, possibly, a set of charges $Q_a$, which generate possible continuous internal symmetry groups. Thus the equilibrium density matrix can be written as
  \be \rho= \frac1{Z}\exp\left(-\beta_\mu P^\mu-\beta_{ij}J_{ij} - \beta_a Q_a\right), \label{EqDen}\ee 
  where $P^\mu$ is the four momentum, $J_{ij}=\epsilon_{ijk}J_k$, the $J_k$ are the three components of the angular momentum $\vec J$ and $\epsilon_{ijk}$ is the totally antisymmetric Levi-Civita symbol (we choose $\epsilon_{123}=1$). We take $\beta_{ij}=-\beta_{ji}$ without loss of generality because $J_{ij}=-J_{ji}$ (a symmetric part of $\beta_{ij}$ would not contribute to~(\ref{EqDen})).
  
  Let us consider now a covariant generalization of (\ref{EqDen})
    \be \rho= \frac1{Z}\exp\left(-\beta_\mu P^\mu-\beta_{\mu\nu}J^{\mu\nu} - \beta_a Q_a\right), \label{EqCov}\ee 
    where the $J^{\mu\nu}$ are the generators of the full Lorentz group. Recalling $J_{\mu\nu}=-J_{\nu\mu}$ we take $\beta_{\mu\nu}=-\beta_{\nu\mu}$ without loss of generality. The density matrix in~(\ref{EqCov}) does not generically describe a system at equilibrium because some generators of the Lorentz group, the boosts $J_{0i}$, do not commute with $H$. However, this density matrix allows us to easily obtain useful information on $\langle P^\mu\rangle$ and $\langle J^{\mu\nu}\rangle$ and so, as particular cases, on $\langle \vec P\rangle$ and $\langle \vec J\rangle$. Let us consider a proper\footnote{In these notes a proper Lorentz transformation has by definition $\Lambda^0_{~0}\geq 1$ and det$\Lambda=1$, where $\Lambda$ is the Lorentz matrix (with elements in the $\mu$th line and $\nu$th column given by $\Lambda^{\mu}_{~\nu}$).} Lorentz transformation $x^\mu\to\Lambda^{\mu}_{~\nu} x^{\nu}$ (the $x^{\mu}$ are the spacetime coordinates) and let $U(\Lambda)$ be the corresponding unitary operator on the Hilbert space. According to the general rule~(\ref{rhoTr}) the density matrix in~(\ref{EqCov}) transforms as
    \be\rho\to U(\Lambda)\rho U(\Lambda)^\dagger = \frac1{Z} \exp\left(-\beta_\mu U(\Lambda)P^\mu U(\Lambda)^\dagger-\beta_{\mu\nu}U(\Lambda)J^{\mu\nu}U(\Lambda)^\dagger - \beta_a Q_a\right) \label{rhoLorentz}
    \ee 
    where the Lorentz invariance of charges, namely
    \be Q_a= U(\Lambda)Q_aU(\Lambda)^\dagger, \ee
     has been used. 
     %
   Now $P^\mu$ and $J^{\mu\nu}$ transform respectively as a vector and a rank two tensor under a proper Lorentz transformations, i.e.
   \be U(\Lambda)^\dagger P^\mu U(\Lambda) = \Lambda^\mu_{~\rho}P^\rho, \quad U(\Lambda)^\dagger J^{\mu\nu}U(\Lambda) =\Lambda^\mu_{~\rho}\Lambda^\nu_{~\sigma}J^{\rho\sigma}. \ee 
   But in~(\ref{rhoLorentz}) $U^\dagger$ appears on the right rather than on the left as a consequence of the general rule in~(\ref{rhoTr}), so we need the inverse transformation rules 
     \be U(\Lambda)P^\mu U(\Lambda)^\dagger = \Lambda_{\rho}^{~\mu}P^\rho, \quad U(\Lambda)J^{\mu\nu}U(\Lambda)^\dagger =  \Lambda_{\rho}^{~\mu}\Lambda_{\sigma}^{~\nu} J^{\rho\sigma},\label{LorPJ}\ee 
     where we used that the elements  in the $\mu$th line and $\nu$th column of the inverse Lorentz matrix $\Lambda^{-1}$ is  
     \be \Lambda^{-1 \mu}_{\quad~\nu} = \eta_{\nu\sigma}\Lambda^\sigma_{~\rho} \eta^{\rho \mu}  \equiv\Lambda_\nu^{~\mu} \label{InvLor} \ee 
     and $\eta_{\mu\nu}$ is the flat metric, which is used as usual to lower and raise the spacetime indices. From~(\ref{LorPJ}) it follows that the Lorentz transformation of the density matrix in~(\ref{rhoLorentz}) can be written as 
        \be\rho\to  \frac1{Z} \exp\left(-\Lambda_\mu^{~\rho}\beta_\rho P^\mu -\Lambda_\mu^{~\rho}\Lambda_\nu^{~\sigma}\beta_{\rho\sigma}J^{\mu\nu} - \beta_a Q_a\right). \label{rhoLorentz2}
    \ee 
    Therefore, performing this  Lorentz transformation of the density matrix is equivalent to replacing $\beta_\mu\to \Lambda_\mu^{~\rho}\beta_\rho$ and $\beta_{\mu\nu}\to \Lambda_\mu^{~\rho}\Lambda_\mu^{~\sigma}\beta_{\rho\sigma}$.   
    Lorentz covariance then tells us that 
    \be \langle P^\mu \rangle = f_1\beta^\mu + f_2 \beta^{\mu\rho}\beta_{\rho}, \qquad  \langle J^{\mu\nu} \rangle  = f_J \beta^{\mu\nu},   \ee 
    where $f_1$, $f_2$ and $f_J$ are functions of the Lorentz invariant quantities $\beta_a$, $\beta_\mu\beta^\mu$, $\beta_{\mu\nu}\beta^{\mu\nu}$, $\beta_{\mu\nu} \beta^\nu \beta_\rho\beta^{\mu\rho}$ only.
    
    Taking now the generic equilibrium density matrix in~(\ref{EqDen}) we have to set $\beta_{0i}=0$ that gives 
    \be\langle P^\mu \rangle = f_1\beta^\mu + f_2 \beta^{\mu j}\beta_j, \qquad  \langle J_{ij} \rangle  = f_J \beta_{ij}. \label{VEVPJ}  \ee 
    In particular, for the components of the momentum
    \be\langle P^i \rangle = f_1\beta^i + f_2 \beta^{i j}\beta_j \label{VEVP}\ee 
    and we see that setting $\beta^i =0$ corresponds to having a system with zero (average) momentum. Moreover, the second equation in~(\ref{VEVPJ}) indicates 
 that zero (average) angular momentum corresponds to $\beta_{ij}=0$. 
        
     From now on we set $\beta_{ij}=0$ and consider a system at equilibrium without angular momentum. In this case we simply have 
    \be \rho =\frac1{Z}\exp(-\beta_\mu P^\mu-\beta_aQ_a). \ee
    The convergence of the traces in~(\ref{Trr1}) and~(\ref{averageA}) requires $\beta_0>0$ and $\beta_i\beta_i<\beta_0^2$. This can be easily seen by choosing a basis of momentum, energy and charge eigenstates
     $|n\rangle$ to compute the traces, such that
   \be Z = \sum_n \exp(-\beta_\mu p^\mu_n-C_{n}), \label{Zn}\ee
   where $p_n^\mu$ (and $C_n$) is the eigenvalue of $P^\mu$ (and, respectively, $\beta_aQ_a$) corresponding to the eigenstate $|n\rangle$. Focusing on the terms in~(\ref{Zn}) with zero spatial momentum and charge one can easily see that the convergence of the sum requires $\beta_0>0$. Moreover, if we had $\beta_i\beta_i\geq\beta_0^2$ there would be directions in the space of states where the sum in~(\ref{Zn}) does not converge (for large momenta one can neglect the masses and the absolute value of the momentum can approach the total energy 
  and also can feature a negative $p_n^i\beta_i$ of the form $-|\vec{p}_n||\vec \beta|$ even for vanishing charges). Therefore, there exist a proper Lorentz transformation that transforms $\beta_i\to 0$ and $\beta_0\to \beta$, where $\beta$ is some positive number.  According to~(\ref{VEVP}) this transformation corresponds to moving to the rest frame of the system. Therefore, in the rest frame an equilibrium density matrix reads
  \be \rho = \frac1{Z}\exp(-\beta H-\beta_a Q_a). \ee 
  The inverse of $\beta$ is the temperature $T$, while $\mu_a\equiv -\beta_a/\beta$ is the chemical potential associated with $Q_a$.

 \newpage
 
 
 \section{Thermal free fields}\label{Thermal free fields}

 Let us start studying what happens in a field theory by considering the simple case of free fields. The generalization to the interacting case will be provided in Sec.~\ref{Thermal Green's functions}.
 
 \subsection{Real scalar field}\label{Real scalar field}
 
 The simplest field we can consider is a real scalar $\varphi$ with  Lagrangian given by 
 \be \mathscr{L}= \frac12\partial_\mu \varphi \partial^\mu\varphi - \frac12m^2\varphi^2, \ee
  $m$ is the mass of the scalar.
In this case
the only conserved quantities are $P^\mu$ and $\vec J$. Going to the rest frame of the system, i.e.~$\langle P^i\rangle = 0$, the equilibrium density matrix is 
\be \rho = \frac1{Z} \exp(-\beta H). \ee 

In this simple case the Hamiltonian reads 
\be H = \sum_k \omega_k a^\dagger_k a_k, \ee
 where $a_k$ is the annihilation operator of a particle (a quantum of the field $\varphi$) with four-momentum 
 \be k = \left ( \bac\omega_k \\ \vec k \ea\right)= \left ( \bac\sqrt{m^2+\vec k^2} \\ \vec k \ea\right).\ee
 The operator $a_k^\dagger$ is the corresponding  creation operator. Note that the vacuum $|0\rangle$ (the state that is annihilated by all $a_k$) has zero energy, i.e.~$H|0\rangle=0$. Here we have put the system in a space with finite volume $V$; it will be soon clear that it is computationally convenient to do so and then let $V\to\infty$ at the end of the calculation. As well known, the operators $a_k$ and $a_k^\dagger$ satisfy $[a_k,a_l^\dagger]=\delta_{kl}, [a_k,a_l]=[a_k^\dagger,a_l^\dagger]=0$. 
 
 The partition function in this case reads
\be Z=\Tr\exp(-\beta H) = \Tr\exp(-\beta \sum_k\omega_kN_k), \ee
where the $N_k\equiv a_k^\dagger a_k$  are the number operators. 
We can then compute the trace
by choosing the orthonormal basis of the $N_k$ eigenstates: denoting here these eigenstates $|n\rangle$ and  the corresponding eigenvalues $n_k (= 0,1,2,...)$  we have
\bea Z &=& \sum_n\langle n |\exp(-\beta \sum_k\omega_kN_k)|n\rangle =\sum_n\langle n |\exp(-\beta \sum_k\omega_kn_k)|n\rangle =\sum_n \exp(-\beta \sum_k\omega_kn_k)\nonumber\\
&=& \sum_n \prod_k\exp(-\beta \omega_kn_k) = \prod_k\sum_{n_k}\exp(-\beta \omega_kn_k)= \prod_k \frac1{1-e^{-\beta\omega_k}}. \label{ZB}\eea

The average number of particles, $\langle N_k\rangle$, with a given momentum $k$ and the average energy of the full system, $\langle H\rangle$, can now be easily computed:
\bea \langle N_k\rangle &=& \Tr (\rho N_k) 
=\frac1Z\sum_n \exp(-\beta \sum_{k'}\omega_{k'}n_{k'})n_k= -\frac1\beta\frac{\partial}{\partial\omega_k}\log Z=\frac1{e^{\beta \omega_k}-1}= f_B(\omega_k), \nonumber \\
\langle H\rangle &=&  \sum_k\omega_k \langle N_k\rangle=
\sum_k \omega_k f_B(\omega_k),\nonumber \eea	where the function $f_B$ is defined by
\be f_B(\omega) \equiv \frac1{e^{\beta\omega}-1}.\ee
These are the Bose-Einstein formul\ae~for $\langle N_k\rangle$ and $\langle H\rangle$ because a scalar field describes bosons. The function $f_B$ satisfies 
\be f_B(-\omega) = -e^{\beta \omega} f_B(\omega) = -1-f_B(\omega),\label{fBprop} \ee
which will be useful in Sec.~\ref{Thermal Green's functions}.

As we have seen the thermal Green's functions in~(\ref{TGF}) are useful tools in thermal field theory.
Let us compute now the thermal propagator (the thermal Green's function for $n=2$) for a real free scalar
\bea \langle {\cal T}\varphi(x)\varphi(y)\rangle&=&\theta(x_0-y_0)\langle \varphi(x)\varphi(y)\rangle+\theta(y_0-x_0)\langle \varphi(y)\varphi(x)\rangle, 
 \eea
 where \be \varphi(x) = \sum_k \frac{a_ke^{-ikx}+{\rm h.c.}}{\sqrt{2V\omega_k}} \label{varphiDec}\ee
and $V$ is the  volume of the full space. The factor $\sqrt{2V\omega_k}$ has been inserted in the denominator to ensure the equal-time canonical commutator between $\varphi$ and its conjugate momentum, $\dot \varphi$, where a dot represents a time derivative. In our case the (average) angular momentum vanishes and $\rho$ commutes with $P^\mu$.  
So $\langle {\cal T}\varphi(x)\varphi(y)\rangle$ is a function of $x-y$ and we can thus set $y=0$.  When $T\to 0$ only the vacuum $|0\rangle$ gives a non-vanishing contribution in the trace and $\rho |0\rangle \to |0\rangle$. So in this limit $\langle {\cal T}\varphi(x)\varphi(0)\rangle$ should go to the zero-temperature propagator $\langle 0 | {\cal T}\varphi(x)\varphi(0)|0\rangle$.  Using 
\be \langle a_k^\dagger a_l\rangle = f_B(\omega_k)\delta_{kl}, \quad \langle a_k a_l^\dagger\rangle = (1+f_B(\omega_k))\delta_{kl}, \quad \langle a_k a_l\rangle = \langle a_k^\dagger a_l^\dagger\rangle = 0 \ee
one obtains for the  ``non time-ordered" 2-point function
\bea &&G^>(x)\equiv \langle \varphi(x)\varphi(0)\rangle = \sum_{kl}\frac{\langle a_k a_l^\dagger\rangle e^{-ikx}+\langle a_k^\dagger a_l\rangle e^{ikx}}{2V\sqrt{\omega_k\omega_l}}=\sum_{k}\frac{(1+f_B(\omega_k)) e^{-ikx}+f_B(\omega_k)e^{ikx}}{2V\omega_k} \nonumber  \\
&=&\int \frac{d^3k}{2(2\pi)^3\omega_k}\left[(1+f_B(\omega_k)) e^{-ikx}+f_B(\omega_k)e^{ikx}\right],  \label{Wf} \eea
 where the limit $V\to \infty$ has been taken\footnote{We recall that in order to take the limit $V\to \infty$ one can first use the finite $V$ formula $1/V=d^3k/(2\pi)^3$ and then let $V\to \infty$.}. By using
 \be \delta(k^2-m^2)= \frac{\delta(k_0-\omega_k)+\delta(k_0+\omega_k)}{2\omega_k}\label{delta3}\ee
 we can rewrite $G^>$ as follows
 \be G^>(x) =  \int \frac{d^4 k}{(2\pi)^4} \Delta_B^>(k) e^{-ikx}, \quad \mbox{where} \quad  \Delta_B^>(k) \equiv(\theta(k_0)+n_B(k_0))2\pi  \delta(k^2-m^2), \ee
and the Bose-Einstein distribution, 
\be n_B(x)\equiv f_B(|x|),\ee
has been introduced. Analogously, for $G^<(x)\equiv \langle \varphi(0)\varphi(x)\rangle = G^>(-x)$ we have
\be G^<(x) = \int \frac{d^4 k}{(2\pi)^4} \Delta_B^<(k) e^{-ikx}, \quad \mbox{where} \quad  \Delta_B^<(k) \equiv \Delta_B^>(-k)=(\theta(-k_0)+n_B(k_0))2\pi  \delta(k^2-m^2).  \nonumber\ee
 From these results, recalling the expression of the propagator at zero temperature, one easily finds
 \be \hspace{-0.25cm}\langle {\cal T}\varphi(x)\varphi(0)\rangle = \int \frac{d^4 k}{(2\pi)^4} \Delta_B(k) e^{-ikx}, \label{FreePropSc} \ee
where
 \be \quad \Delta_B(k)\equiv \Delta_0(k) +2\pi\, n_B(k_0)\delta(k^2-m^2) \quad \mbox{and} \quad  \Delta_0(k)\equiv\frac{i}{k^2-m^2+i\epsilon}. \label{DeltaB}\ee
 The second term in $\Delta_B$  is the finite-temperature contribution to the propagator in the momentum space. We see that term is not invariant under the full Lorentz group, this happens because we are working in a specific frame, the rest frame of the system, $\langle \vec P\rangle = 0$ (see the end of Sec.~\ref{Density matrix and ensemble averages}). 
  
 The simplest case of a scalar field is not sufficient to describe realistic models, such as the Standard Model (SM). To do so we extend the discussion to gauge and fermion fields in the next sections.
 
 \subsection{Gauge field}

 Introducing a gauge field is a method to describe in a Lorentz covariant way spin-1 massless particles. Such particles have two helicity states, $\pm 1$, so the creation and annihilation operators $a_{kr}$ and $a_{kr}^\dagger$, are now labeled by an extra helicity index, $r=1,2$. Here $k$ is the light-like four-momentum of the massless spin-1 particle 
 \be k = \left ( \bac\omega_k \\ \vec k \ea\right)= \left ( \bac|\vec k| \\ \vec k \ea\right).\ee
 Starting again with a finite space volume $V$, the $a_{kr}$ and $a_{kr}^\dagger$ satisfy  
 \be [a_{kr},a_{ls}^\dagger]=\delta_{rs}\delta_{kl},  \quad [a_{kr},a_{ls}]=[a_{kr}^\dagger,a_{ls}^\dagger]=0. \label{CommRulA} \ee
 
 Therefore, the Hamiltonian for a free system of massless spin-1 particles is
 \be H = \sum_{kr} \omega_k a^\dagger_{kr} a_{kr}. \ee
Also in this case the only conserved quantities are $P^\mu$ and $\vec J$. A simple generalization of the calculation performed in~(\ref{ZB}) to particles with two helicity states gives
 \be Z = \prod_{kr} \frac1{1-e^{-\beta\omega_k}}. \label{ZB2}\ee
 and for the averages of the number operators $N_{kr}\equiv a_{kr}^\dagger a_{kr}$ and the Hamiltonian
 \bea \langle N_{kr}\rangle &=& f_B(\omega_k),\\
\langle H\rangle &=&\sum_{kr} \omega_k f_B(\omega_k) = 2\sum_{k} \omega_k f_B(\omega_k) .\eea
The factor of $2$ above is the number of helicity states.

 Let us compute now the gauge-field thermal propagator 
\bea \langle {\cal T}A_\mu(x)A_\nu(0)\rangle&=&\theta(x_0)G^>_{\mu\nu}(x)+\theta(-x_0)G^<_{\mu\nu}(x), 
 \eea
 where 
 \be G^>_{\mu\nu}(x)\equiv  \langle A_\mu(x)A_\nu(0)\rangle, \qquad G^<_{\mu\nu}(x)\equiv  \langle A_\nu(0)A_\mu(x)\rangle =G^>_{\mu\nu}(x)^*= G^>_{\nu\mu}(-x) \ee
 and we have set one of the two spacetime points to zero using translation invariance, which is not broken as the angular momentum of the system vanishes. The gauge field $A_\mu$ can be expressed in terms of the creation and annihilation operators as follows: 
 \be A_\mu(x) = \sum_{kr} \frac{\epsilon_{r\mu}(k)a_{kr}e^{-ikx}+{\rm h.c.}}{\sqrt{2V\omega_k}}. \label{AmuDec} \ee
We take the polarization vectors $\epsilon_{r\mu}(k)$ real without loss of generality: any phase of $\epsilon_{r\mu}(k)$ can be included in $a_{kr}$ without changing the commutation rules in~(\ref{CommRulA}).
Using 
\be \langle a_{kr}^\dagger a_{ls}\rangle = f_B(\omega_k)\delta_{rs}\delta_{kl}, \quad \langle a_{kr} a_{ls}^\dagger\rangle = (1+f_B(\omega_k))\delta_{rs}\delta_{kl}, \quad \langle a_{kr} a_{ls}\rangle = \langle a_{kr}^\dagger a_{ls}^\dagger\rangle = 0 \ee
one obtains (taking $V\to \infty$)
\be G^>_{\mu\nu}(x)=\int \frac{d^3k}{2(2\pi)^3\omega_k}P_{\mu\nu}(k)\left[(1+f_B(\omega_k)) e^{-ikx}+f_B(\omega_k)e^{ikx}\right],\ee
where 
\be P_{\mu\nu}(k) = \sum_r \epsilon_{r\mu}(k)\epsilon_{r\nu}(k).  \ee
From $G^<_{\mu\nu}(x)=G^>_{\nu\mu}(x)^*$ it also follows 
\be G^<_{\mu\nu}(x) = \int \frac{d^3k}{2(2\pi)^3\omega_k}P_{\mu\nu}(k)\left[(1+f_B(\omega_k)) e^{ikx}+f_B(\omega_k)e^{-ikx}\right].\ee
 Note that the polarization vectors $\epsilon_{r\mu}(k)$ are dimensionless and the only dimensionful quantity they depend on is $\vec k$. So for a given $\vec k$ and $r$ the polarization vector can be taken to be either even or odd under $\vec k\to -\vec k$.  
 As a result $P_{\mu\nu}(k)=P_{\mu\nu}(-k)$.
 By using now~(\ref{delta3}) and setting $m=0$, one finds 
 \be G^>_{\mu\nu}(x) =  \int \frac{d^4 k}{(2\pi)^4} P_{\mu\nu}(k)\Delta_B^>(k) e^{-ikx}, \qquad G^<_{\mu\nu}(x) =  \int \frac{d^4 k}{(2\pi)^4} P_{\mu\nu}(k)\Delta_B^<(k) e^{-ikx}. \ee
 For the propagator one then obtains
\be \langle {\cal T}A_\mu(x)A_\nu(0)\rangle =   \int \frac{d^4 k}{(2\pi)^4} P_{\mu\nu}(k)\Delta_B(k) e^{-ikx}.\ee

As known from zero-temperature QFT, $P_{\mu\nu}$ is gauge dependent. Indeed, changing the gauge\footnote{At the non-interacting level, which we are discussing here, that gauge transformation also applies to non-Abelian gauge fields.}, $A_\mu\to A_\mu+\partial_\mu \alpha$ introduces another scalar field $\alpha$, which can also be expanded in terms of its creation and annihilation operators, $a_{k0}^\dagger$ and $a_{k0}$ (satisfying $[a_{k0},a_{l0}^\dagger] = \delta_{kl}$, $[a_{k0},a_{l0}]=[a_{k0}^\dagger,a_{l0}^\dagger]=0$ and commuting with $a_{kr}$ for $r=1,2$): 
\be \alpha(x) = c_\alpha \sum_k \frac{ia_{k0}e^{-ikx}+{\rm h.c.}}{\sqrt{2V\omega_k}}, \ee
where $c_\alpha$ is some real constant with the dimension of a length.
As a result
\be A_\mu(x) \to  \sum_{k}  \frac{\sum_{r=0}^2 \epsilon_{r\mu}(k)a_{kr}e^{-ikx}+{\rm h.c.}}{\sqrt{2V\omega_k}},\ee 
where $\epsilon_{0\mu}(k) = c_\alpha k_\mu$. Including the extra degrees of freedom carried by $\alpha$ in the Hilbert space and repeating the steps above, one finds that the gauge transformation changes $P_{\mu\nu}$ as follows
\be P_{\mu\nu}(k)\to \sum_{r=0}^2 \epsilon_{r\mu}(k)\epsilon_{r\nu}(k),\label{Pmunugauge}\ee
which again satisfies $P_{\mu\nu}(k)=P_{\mu\nu}(-k)$.
This example\footnote{A more general treatment will be provided in Sec.~\ref{Gauge theory} using path-integral methods.} illustrates that changing gauge generically changes $P_{\mu\nu}$, but in a temperature independent way. From zero-temperature QFT we know that we can effectively bring $P_{\mu\nu}$ into several different forms (several gauges). For example,
\be P_{\mu\nu} = -\eta_{\mu\nu} \qquad \mbox{(Feynman gauge).} \ee

 \subsection{Fermion field}\label{Fermion field}

 Let us now consider a fermion field, which we assume here to be free. In this case we have annihilation and creation operators  for the particle, $c_{kr}, c_{kr}^\dagger$, and for the antiparticle, $d_{kr}, d_{kr}^\dagger$, where the index $r=1,2$ corresponds to the  spin state, $k$ is the fermion four-momentum 
 \be k = \left ( \bac E_k \\ \vec k \ea\right)= \left ( \bac\sqrt{m^2+\vec k^2} \\ \vec k \ea\right)\ee
 and $m$ here is the fermion mass. Putting  the system again in a space with finite volume $V$, these operators satisfy the anticommutation rules 
 \bea &&\{c_{kr},c_{ls}^\dagger\} = \{d_{kr},d_{ls}^\dagger\} =\delta_{kl}\delta_{rs}, \label{antNot}\\ 
&&\{c_{kr},c_{ls}\}=\{d_{kr},d_{ls}\}=\{c_{kr},d_{ls}\}=\{c_{kr},d_{ls}^\dagger\}=0.\label{ant} \eea 
 The Hamiltonian is 
 \be H = \sum_{kr} E_k(N_{kr}+\bar N_{kr}), \label{HF}\ee
 where the number operators $N_{kr}$ and $\bar N_{kr}$ are given by
 \be N_{kr} = c_{kr}^\dagger c_{kr}, \qquad \bar N_{kr}=d_{kr}^\dagger d_{kr}. \label{Nferm} \ee

  In the fermion case we generally have the conservation of fermion number $N$. This conserved quantity is expressed in terms of the annihilation and creation operators as follows:
 \be N = \sum_{kr}(N_{kr}-\bar N_{kr}). \ee
 Indeed, one can easily verify that the anticommutator rules in~(\ref{antNot}) and~(\ref{ant}) imply $[N,H]=0$.
 Given the presence of a conserved quantity besides $P^\mu$ and $J_{ij}$ we can introduce here a corresponding chemical potential $\mu$ and  the density matrix reads
\be \rho =\frac1Z \exp(-\beta (H-\mu N)).  \ee
We will keep $\mu$ generic in this section to provide an example of thermal field theory at finite chemical potential.

 In this case the partition function is 
 \be Z=\Tr\exp(-\beta (H-\mu N)) = \Tr\exp\left(-\beta \sum_{kr}[(E_k-\mu)N_{kr}+(E_k+\mu)\bar N_{kr}]\right). \ee
 In order to compute $Z$ it is again convenient to use the orthonormal basis of eigenstates $|n,\bar n\rangle$ of the number operators, i.e.~$N_{kr} |n,\bar n\rangle = n_{kr}|n,\bar n\rangle$, $\bar N_{kr} |n,\bar n\rangle = \bar n_{kr}|n,\bar n\rangle$. In this case, however, the corresponding eigenvalues can only be $n_{kr}=0,1$ and $\bar n_{kr}=0,1$ because we are dealing with a fermion and so need to take into account Pauli's exclusion principle. So
 {\allowdisplaybreaks\bea Z &=& \sum_{n \bar n} \langle n,\bar n|\exp\left(-\beta \sum_{kr}[(E_k-\mu)N_{kr}+(E_k+\mu)\bar N_{kr}]\right)|n,\bar n\rangle \nonumber  \\
 &=& \sum_{n \bar n} \exp\left(-\beta \sum_{kr}[(E_k-\mu)n_{kr}+(E_k+\mu)\bar n_{kr}]\right) \nonumber \\
 &=& \left[\sum_n \prod_{kr} \exp\left(-\beta(E_k-\mu)n_{kr}\right)\right]\left[\sum_{\bar n} \prod_{kr} \exp\left(-\beta(E_k+\mu)\bar n_{kr}\right)\right]\nonumber\\
 &=& \left[\prod_{kr} \sum_{n_{kr}}  \exp\left(-\beta(E_k-\mu)n_{kr}\right)\right]\left[\prod_{kr} \sum_{{\bar n}_{kr}} \exp\left(-\beta(E_k+\mu)\bar n_{kr}\right)\right]\nonumber\\
 &=& \left[\prod_{kr} \left(1+e^{-\beta(E_k-\mu)}\right)\right] \left[\prod_{kr} \left(1+e^{-\beta(E_k+\mu)}\right)\right].
 \eea}
 
 By using again the basis of the $|n,\bar n\rangle$ it is easy to show that 
 \be \langle c_{kr}^\dagger c_{ls}\rangle =\delta_{kl}\delta_{rs} \frac1{e^{\beta(E_k-\mu)}+1}, \quad  \langle d_{kr}^\dagger d_{ls}\rangle =\delta_{kl}\delta_{rs} \frac1{e^{\beta(E_k+\mu)}+1} \label{dddccd}\ee
 and all other averages of pairs of annihilation and/or creation operators vanish.
So in particular 
 \be \langle N_{kr}\rangle =\frac1{e^{\beta(E_k-\mu)}+1} \equiv  f_F(E_k-\mu), \quad  \langle \bar N_{kr}\rangle = \frac1{e^{\beta(E_k+\mu)}+1} \equiv  f_F(E_k+\mu). \label{dddccd2} \ee
  The Fermi-Dirac distribution is recovered because we are dealing with a fermion. Note that the function $f_F$ satisfies 
  \be f_F(-x)=1-f_F(x).\ee 
A  physical interpretation of $\mu$ can be obtained by noting  that for a positive (negative) $\mu$ we have a net excess of fermions (antifermions) over antifermions (fermions) for each $k$ and $r$. For $\mu=0$ we have $\langle N_{kr}\rangle=\langle \bar N_{kr}\rangle$ for all $k$ and $r$. It is now easy to compute the (average) energy  and conserved fermion number
 \bea \langle H\rangle &=& \sum_{kr} E_k(\langle N_{kr}\rangle +\langle\bar N_{kr}\rangle), = 2\sum_kE_k(f_F(E_k-\mu)+f_F(E_k+\mu))\nonumber \\
 \langle N\rangle &=& \sum_{kr} (\langle N_{kr}\rangle -\langle\bar N_{kr}\rangle) = 2\sum_k(f_F(E_k-\mu)-f_F(E_k+\mu)). \eea
 We see that, as expected, $\langle N\rangle$ vanishes when $\mu=0$. When $\mu$ is positive (negative) $\langle N\rangle$ is positive (negative).
 
 
 We can now compute the  thermal fermion propagator
\be  \langle {\cal T} \psi_\alpha(x)\bar \psi_\beta(0)\rangle = \theta(x_0)S_{\alpha\beta}^>(x)  +\theta(-x_0)  S_{\alpha\beta}^<(x) \ee
where the ``non time-ordered" 2-point functions are defined by
\be S_{\alpha\beta}^>(x) \equiv \langle \psi_\alpha(x)\bar \psi_\beta(0)\rangle, \quad S_{\alpha\beta}^<(x) \equiv -\langle \bar \psi_\beta(0)\psi_\alpha(x)\rangle.\ee
As well-known, the expression of the Dirac field $\psi$ in terms of the annihilation and creation operators is
\be \psi(x) =\sum_{kr}\frac{c_{kr}u_{kr} e^{-ikx}+d^\dagger_{kr}v_{kr} e^{ikx}}{\sqrt{VE_k}}, \label{DiracFieldDec}\ee
where $u_{kr}$ and $v_{kr}$ are constant spinors, whose form is constrained by requiring $\psi$ to be a solution of the Dirac equation. We choose conventions such that 
\be \sum_r u_{kr}\bar u_{kr} =\frac{\slashed{k}+m}{2}, \quad \sum_r v_{kr}\bar v_{kr} =\frac{\slashed{k}-m}{2},   \label{sumuv}\ee
where $\slashed{k}\equiv \gamma^\mu k_\mu$ and $\gamma^\mu$ are the Dirac matrices, which satisfy $\{\gamma^\mu,\gamma^\nu\}=2\eta^{\mu\nu}$, and we defined as usual $\bar\psi\equiv \psi^\dagger \gamma^0$. The factors $\sqrt{VE_k}$ in the denominators of~(\ref{DiracFieldDec}) have been inserted  to ensure that the Hamiltonian derived from the Dirac field in normal ordering reproduces~(\ref{HF}). 
Using now~(\ref{dddccd}),~(\ref{dddccd2}) and~(\ref{sumuv})  one finds (in the limit $V\to \infty$) 
\bea  S^>(x) &=& \int \frac{d^3k}{2(2\pi)^3 E_k} \left[(1-f_F(E_k-\mu)) (\slashed{k}+m)e^{-ikx} + f_F(E_k+\mu)(\slashed{k}-m)e^{ikx}\right] \nonumber \\
&=& \int \frac{d^4k}{(2\pi)^4}(\slashed{k}+m) \Delta^>_F(k) e^{-ikx}, \label{S>x} \\S^<(x) &=&- \int \frac{d^3k}{2(2\pi)^3 E_k} \left[f_F(E_k-\mu) (\slashed{k}+m)e^{-ikx} +(1- f_F(E_k+\mu))(\slashed{k}-m)e^{ikx}\right]\nonumber\\
&=& \int \frac{d^4k}{(2\pi)^4}(\slashed{k}+m) \Delta^<_F(k) e^{-ikx}, \label{S<x}
 \eea
 where
 \be \Delta^>_F(k) \equiv (\theta(k_0) - n_F(k_0))2\pi \delta(k^2-m^2), \quad \Delta^<_F(k) \equiv (\theta(-k_0) - n_F(k_0))2\pi \delta(k^2-m^2),\ee
 $n_F(x)\equiv f_F(|x|-\mu\, s(x))$ and $s(x)$ is the sign of $x$. We have found that the non time-ordered 2-point functions are the sum of the known zero-temperature and zero-chemical-potential expressions plus a correction. Recalling the expression of the fermion propagator at $T=0$ and $\mu=0$, the thermal fermion propagator can then be written as 
 \bea \hspace{-1cm} \langle {\cal T} \psi(x)\bar \psi(0)\rangle &=& i\int \frac{d^4k}{(2\pi)^4} \frac{\slashed{k}+m}{k^2-m^2+i\epsilon} e^{-ikx} \nonumber \\ &&+\int \frac{d^3k}{2(2\pi)^3 E_k} \left[-f_F(E_k-\mu) (\slashed{k}+m) e^{-ikx} +f_F(E_k+\mu) (\slashed{k} - m) e^{ikx}\right]\nonumber \\
&=&  \int \frac{d^4k}{(2\pi)^4}(\slashed{k}+m)\Delta_F(k) e^{-ikx},\eea
where
\be  \quad \Delta_F(k) = \Delta_0(k) -2\pi n_F(k_0) \delta(k^2-m^2). \ee

An interesting aspect to note is that at zero temperature, when $\beta\to\infty$, the propagator still differs from the result in non-statistical QFT if $\mu\neq 0$. This is  because  $n_F(k_0)\to 0$ if $|k_0|>\mu \, s(k_0)$, but $n_F(k_0)\to 1$ if $|k_0|<\mu \, s(k_0)$. In this limit the particle densities vanish for $E_k>|\mu|$ (as $\langle N_{kr}\rangle \to 0$ and $\langle \bar N_{kr}\rangle \to 0$ in this case), but for all $k$ such that $E_k<|\mu|$ the average number of fermions or antifermions goes to one for $\mu$ positive or negative, respectively, as clear from~(\ref{dddccd2}). 

\newpage

\section{Thermal Green's functions}\label{Thermal Green's functions}

As we have seen in Sec.~\ref{Density matrix and ensemble averages} in TFT we are interested in computing thermal Green's functions. In this section we provide some methods to do so, mainly using the path-integral approach. For simplicity, we assume that the system is in thermal equilibrium and the chemical potentials are negligible, as is typically the case in a standard cosmological setup. Furthermore, we work in the rest frame, such that the density matrix is simply $\rho=\exp(-\beta H)/Z$.

\subsection{Scalar theory}\label{Scalar theory}

Let us start from a theory with only scalar fields, which we collectively represent by an array\footnote{Here, unlike in Sec.~\ref{Real scalar field}, we added a hat on top of the field operators because later we will derive a path-integral formula and we will need to distinguish between the field operators and the field integration variables in the path integral.} $\hat \varphi$ of Hermitian field operators. 

The simplest thermal Green's function is the thermal propagator,
\bea \langle {\cal T}\hat\varphi(x)\hat\varphi(0)\rangle&=&\theta(x_0)G^>(x)+\theta(-x_0)G^<(x). 
 \eea
  So we start by making some general considerations about it. Here $\langle {\cal T}\hat\varphi(x)\hat\varphi(0)\rangle$ is the exact thermal propagator, and $G^>(x)\equiv \langle\hat\varphi(x)\hat\varphi(0)\rangle$ and $G^<(x)\equiv \langle\hat\varphi(0)\hat\varphi(x)\rangle$ are the exact non time-ordered 2-point functions. 
  
  We can formally see $e^{-\beta H}$ as an imaginary-time evolution operator. So, $e^{\beta H}\hat\varphi(t,\vec x)e^{-\beta H} = \hat\varphi(t-i\beta,\vec x)$. Using then the ciclicity of the trace in the thermal Green's function leads to 
\be G^>(t-i\beta,\vec x) = G^<(t,\vec x),\label{KMS}\ee
which is known as the Kubo-Martin-Schwinger (KMS) condition~\cite{KMS}. 
The KMS condition implies a relation between the Fourier transforms
\be \tilde G^>(k) = \int d^4x\, e^{ikx} G^>(x), \quad \tilde G^<(k) = \int d^4x\, e^{ikx} G^<(x).
\ee
To find this relation first note that 
\bea \tilde G^<(k) &=& \int_{-\infty}^{+\infty} dt \int d^3x\,e^{ik_0t} e^{i\vec k \cdot \vec x} G^<(t,\vec x) =   \int_{-\infty}^{+\infty} dt\,e^{ik_0t} \int d^3x \, e^{i\vec k \cdot \vec x} G^>(t-i\beta,\vec x)\nonumber \\ &=& e^{-\beta k_0}\int_{-\infty-i\beta}^{+\infty-i\beta} dt\,e^{ik_0t} \int d^3x \, e^{i\vec k \cdot \vec x} G^>(t,\vec x),\label{Gfbtilde}\eea
where in the second step we used the KMS condition.
The latter integral on the straight line Im$(t)=-\beta$ in the complex time plane can be rewritten as the integral of the same integrand but on the real axis using the analyticity of $G^>$ in the strip $-\beta<{\rm Im}(t)<0$. To understand why this strip is in the analyticity domain of $G^>$ let us consider the (orthonormal) eigenstates of $H$, 
 which we call $|n\rangle$, and denote the corresponding eigenvalues $E_n$, 
   so
\be G^>(x) = \frac1{Z}\sum_{n m} \langle m |\hat\varphi(0,\vec x) |n\rangle\langle n|\hat\varphi(0)|m\rangle  e^{iE_m(t+i\beta)} e^{-i E_n t}. \label{Gfexp}\ee
To write this expression for $t$ real and an arbitrary value of $\beta > 0$ one is implicitly assuming that it converges on the real time axis and for all values of $\beta> 0$. 
Then the same expression for $-\beta<{\rm Im}(t)<0$ also converges as in this strip the exponentials appearing in~(\ref{Gfexp}) help the convergence of the series. As a result,~(\ref{Gfbtilde}) then implies
\be \tilde G^<(k) =e^{-\beta k_0} \int_{-\infty}^{+\infty} dt\,e^{ik_0t} \int d^3x \, e^{i\vec k \cdot \vec x} G^>(t,\vec x) =e^{-\beta k_0}\tilde G^>(k),\label{Gtfb} \ee
where in the first step we used the residue theorem. This relation will be useful to study particle production in Sec.~\ref{Weakly-coupled particle production}.

\subsubsection{Path integral}\label{Path integral}

Let us now look for a path-integral formula to compute arbitrary thermal Green's functions.
 Since $\hat\varphi$ is Hermitian there exists an orthogonal basis of eigenstates of $\hat \varphi$, namely $\hat\varphi(0,\vec{x})|\varphi\rangle = \varphi(0,\vec{x}) |\varphi\rangle$ and, representing the fields in the Heisenberg picture $\hat\varphi(t,\vec{x})|\varphi,t\rangle = \varphi(0,\vec{x}) |\varphi,t\rangle$, where $\hat\varphi(t,\vec{x})\equiv \exp(iHt)\hat\varphi(0,\vec{x})\exp(-iHt)$ and $|\varphi, t\rangle \equiv \exp(iHt)|\varphi\rangle$. As usual, we choose a normalization of these states such that 
\be \int \delta\varphi(t) |\varphi,t\rangle\langle\varphi,t| = 1, \qquad \mbox{where} \qquad \delta\varphi(t) \equiv \prod_{\vec{x}} d\varphi(t,\vec{x}).\ee
The product on the right of the expression above is over all space points $\vec{x}$, so one is implicitly assuming some sort of lattice regularization in deriving a path-integral formula.
Choosing the basis $\{|\varphi,t_0\rangle\}$ to represent the trace in~(\ref{TGFtrace}), the thermal Green's functions can, therefore, be written as follows:
\bea \langle {\cal T}\hat\varphi(x_1)...\hat\varphi(x_n) \rangle &=& \int \delta \varphi(t_0) \langle\varphi,t_0|\rho\,  {\cal T}\hat\varphi(x_1)...\hat\varphi(x_n)|\varphi,t_0\rangle \nonumber \\
&=&\frac1{Z}\int \delta \varphi(t_0) \langle\varphi,t_0|e^{-\beta H}  {\cal T}\hat\varphi(x_1)...\hat\varphi(x_n)|\varphi,t_0\rangle \nonumber\\&=&\frac1{Z}\int \delta \varphi(t_0) \langle\varphi,t_0-i\beta|  {\cal T}\hat\varphi(x_1)...\hat\varphi(x_n)|\varphi,t_0\rangle. \label{ScalarTGF}\eea 
 
  
  
 By adapting the derivation (see e.g.~Sec.~9.1 of~\cite{Weinberg1}) of the general path-integral formula to this case, one finds that this object can be represented by a path integral 
 \be \hspace{-0.3cm}\langle {\cal T}\hat\varphi(x_1)...\hat\varphi(x_n) \rangle =\frac1{Z}\int \delta\varphi \, \delta p_\varphi  \, \varphi(x_1)...\varphi(x_n) \exp\left(i \int_C d^4x \left(\dot\varphi(x)p_\varphi(x) - \mathcal{H}_c(\varphi(x), p_\varphi(x))\right)\right), \label{PIscalars}\ee 
 where a dot represents the time derivative, $p_\varphi$ is the momentum conjugate to $\varphi$ and the path integration is only on fields satisfying  the periodicity condition
  \be \varphi(t_0,\vec{x}) = \varphi(t_0-i\beta,\vec{x}).\label{PerCon}\ee
  Note that the partition function that appears in the denominator in~(\ref{PIscalars}) is
  \be Z =\int \delta\varphi \, \delta p_\varphi  \,  \exp\left(i \int_C d^4x \left(\dot\varphi(x)p_\varphi(x) - \mathcal{H}_c(\varphi(x), p_\varphi(x))\right)\right),  \ee
  which is nothing but the path integral in~(\ref{PIscalars}) without $\varphi(x_1)...\varphi(x_n)$. So the different normalization in the integration measures on $\varphi$ and $p_\varphi$
  \be \delta\varphi = \prod_x d\varphi(x), \quad \delta p_\varphi= \prod_x d(p_\varphi(x)/2\pi)\ee is irrelevant here.
The function $\mathcal{H}_c$ is, in the classical limit, the classical Hamiltonian density of the theory under study, which is defined in terms of the Hamiltonian density operator $\mathcal{H}$ as
\be \mathcal{H}_c(\varphi(x), p_\varphi(x))\equiv 
\frac{\langle \varphi|\mathcal{H}(\hat\varphi(x), \hat p_\varphi(x))|p_\varphi\rangle}{\langle \varphi|p_\varphi\rangle}, 
\ee 
where $\hat p_\varphi$ is the conjugate momentum operator  and $|p_\varphi\rangle$ the corresponding eigenstates.  

Moreover,  in the path integral in~(\ref{PIscalars}), while the space integral has  no restriction,
 the integral over $t$ is performed on a contour $C$ in the complex $t$ plane that connects the two points $t_0$ and $t_0-i\beta$ and includes the times $x_1^0, ... , x_n^0$. This is because we have to include $t_0$, $x_1^0, ... , x_n^0$ and $t_0-i\beta$ in the set of discrete times (that we introduce in deriving the path integral) and $t_0$ and $t_0-i\beta$ play the role of initial and final time, respectively. The time-ordered product $\mathcal{T}$ should then be understood as the product of fields ordered according to the orientation of $C$: one introduces a parametrization $t(v)$ of $C$ where $v$ is a real parameter such that $t(v)$ proceeds along $C$ following its orientation as $v$ increases. The ordering along the path is the ordering in $v$. For example, given two points $t$ and $t'$ on $C$, which correspond to $v$ and $v'$, respectively, the Heaviside step function is here defined as
\be \theta(t'-t)\equiv \theta(v'-v), \label{contourT}\ee
where $\theta$ in the right-hand side is the ordinary Heaviside step function for real argument. The Dirac's delta function $\delta$ here is defined for the contour $C$ as follows: given two points $t$ and $t'$ on $C$, which correspond to the values $v$ and $v'$  of the real variable that parameterizes $C$, namely $t=t(v)$ and $t'=t(v')$, 
\be \delta(t-t') \equiv \left(\frac{d t}{d v}\right)^{-1} \delta(v-v'), \label{contourDF}\ee
where $\delta$ on the right-hand side is the usual Dirac's delta function for real argument.
With this definition we have, for a generic function $f$ on $C$,
\be \int_C dt' \delta(t-t') f(t') = f(t) \ee
and 
\be \frac{d}{dt'}\theta(t'-t) = \delta(t'-t). \ee

There are several ways to choose the contour $C$ with the properties above and, also, the starting time $t_0$ is arbitrary. 

If we set $t_0=0$ and we go from $0$ to $-i\beta$ along the imaginary axis (as a possible choice for $C$) we have a formula for the thermal Green's functions in Euclidean spacetime. This choice is known as the {\it imaginary-time formalism}~\cite{Matsubara:1955ws}.

\begin{figure}[t]
\begin{center}
  \includegraphics[scale=0.7]{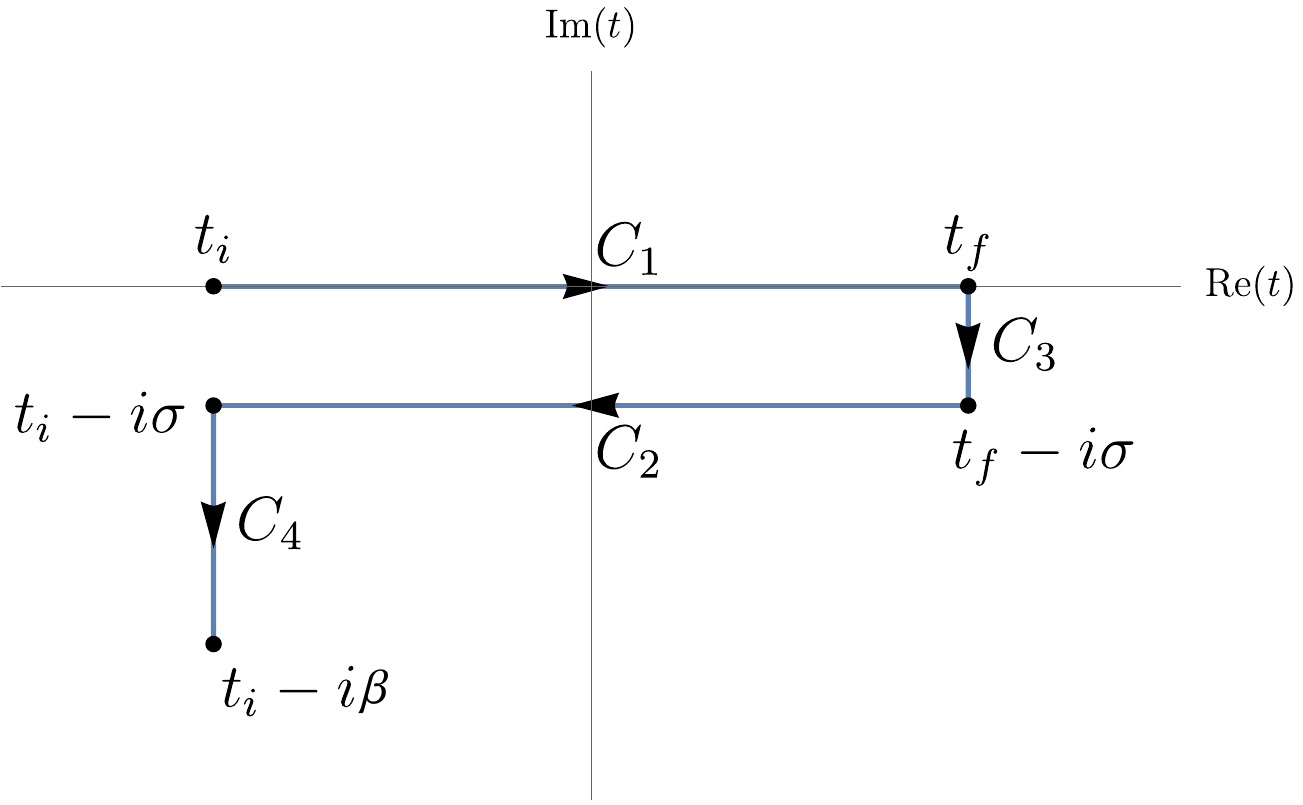}  
    \caption{\em  A contour $C$ which allows us to compute the thermal Green's function directly at real times. Here $t_0 = t_i\to-\infty$.}
\label{contour}
  \end{center}
\end{figure}

 However, it is often useful to have a formula for the thermal Green's functions directly computed at real times ({\it real-time formalism}). To do so $C$ must include the real axis. A way to do so is represented in Fig.~\ref{contour}~\cite{Matsumoto:1982ry}: take $t_0 =t_i$, where $t_i$ is a large and negative time (we will take $t_i\to-\infty$ at the end), then go from $t_i$ to $t_f$ along the real axis (segment $C_1$), where $t_f$ is a large positive time (we will take $t_f\to\infty$ at the end). Next go from $t_f$ to  $t_f-i\sigma$ parallel to the imaginary axis (segment $C_3$), where $\sigma\in (0,\beta)$, next go from $t_f-i\sigma$ to $t_i -i\sigma$ parallel to the real axis (segment $C_2$) and finally reach the point $t_i-i\beta$ going parallel to the imaginary axis (segment $C_4$). Note that the choice of $\sigma\in (0,\beta)$ is arbitrary, so contours with different $\sigma$ in that interval lead to equivalent theories.
 
 Whatever choice we make for $C$ we can generate the thermal Green's function by performing functional derivatives of the generating functional 
 \be \mathcal{Z} (J) =\frac1{``J\to 0"} \int \delta\varphi \, \delta p_\varphi  \, \exp\left(i \int_C d^4x \left(\dot\varphi(x)p_\varphi(x) - \mathcal{H}_c(\varphi(x), p_\varphi(x))+J(x)\varphi(x)\right)\right), \ee  
 where the denominator $``J\to 0"$  is the numerator for $J\to 0$.
 The explicit formula is
 \be \langle {\cal T}\hat\varphi(x_1)...\hat\varphi(x_n) \rangle  = \left.\frac1{ i^n}\frac{\delta^n}{\delta J(x_1) ... \delta J(x_n)}  \mathcal{Z}(J) \right|_{J=0}, \label{ZJ0}\ee
 where here we use the Dirac's delta function  on the contour $C$, Eq.~(\ref{contourDF}), to define functional derivatives.

 As in the zero-temperature case, we can analytically compute the integral over $p_\varphi$ in~(\ref{PIscalars}) when $\mathcal{H}_c$ has a term quadratic in $p_\varphi$, a (possibly vanishing) term linear in $p_\varphi$ and a (possibly vanishing) term constant with respect to $p_\varphi$. Since this  is typically the case in the theories of interest in physics we will assume it to be true from now on. Performing the integral over $p_\varphi$ one obtains the Lagrangian path integral 
 \be \langle {\cal T}\hat\varphi(x_1)...\hat\varphi(x_n) \rangle = \frac{\int \delta\varphi  \, \varphi(x_1)...\varphi(x_n) \exp\left(i S_C(\varphi)\right)}{``\varphi(x_1)...\varphi(x_n) \to 1"},\ee
 where we have also assumed that the term quadratic in $p_\varphi$ inside $\mathcal{H}_c$ is independent of $\varphi$, because, again, that is typically the case in theories of interest.
%
 The denominator in the expression above reminds us that we have to divide by the path integral without the field insertion $\varphi(x_1)...\varphi(x_n)$. Again the path integration is only on fields satisfying  the periodicity condition in~(\ref{PerCon}) and the time integration is on the contour $C$. The action $S_C$ is given in terms of  the Lagrangian (density) $\mathscr{L}$ by
 \be S_C(\varphi)=\int_C d^4x\,\mathscr{L}(\varphi,\partial\varphi).\ee
 The corresponding expression for the generating functional is given by
 \be\mathcal{Z}(J) = \frac1{``J\to 0"} \int \delta\varphi \exp\left(iS_C(\varphi)+i\int_C d^4x J(x)\varphi(x)\right).  \ee

 
 There are two possible ways to compute the path integral. One way is to employ the lattice approximation and use a computer. In this case one has to choose the imaginary-time formalism to obtain a well-defined integral. 
 
 The second way is to use perturbation theory in its range of validity. One assumes that the action is a small deformation of an action $S_0$ for which we can evaluate the path integral exactly. This is the case when the functional $S_0$  contains only a piece quadratic in $\varphi$, plus a (possibly vanishing) term linear in $\varphi$ and a (possibly vanishing) term constant with respect to $\varphi$. So we write
 \be S_C(\varphi) = S_0(\varphi) + \Delta(\varphi), \ee
 where $\Delta$ is a local functional of $\varphi$ proportional to a small  coupling constant. By local functional we mean
\be \Delta(\varphi) = \int_C d^4x \mathscr{L}_I(\varphi), \label{DeltaAss} \ee
where $\mathscr{L}_I$ 
is a functional of $\varphi$ which is constructed only from $\varphi$  and its derivatives at the spacetime point $x$.   This quantity represents the interaction Lagrangian.
  In both $S_0$ and $\Delta$ the time integration is on the contour $C$. The full perturbative series for the generating functional can then be written
 \be \mathcal{Z}(J) = \frac1{``J\to 0"} \exp\left(i\Delta\left(\frac1{i}\frac{\delta}{\delta J}\right)\right)\mathcal{Z}_0(J), \label{ZZ0} \ee 
 where $\mathcal{Z}_0(J)$ is the generating functional for the unperturbed theory:
 \be \mathcal{Z}_0(J) =  \frac1{``J\to 0"}  \int \delta\varphi \exp\left(iS_0(\varphi)+i\int_C d^4x J(x)\varphi(x)\right). \ee
 For a scalar theory one typically has 
 \be S_0(\varphi) = \frac12\int_C d^4x\,  \varphi \mathcal{O} \varphi, \qquad \mathscr{L}_I(\varphi)= -  \mathcal{V}(\varphi),\ee
 where 
 \be\mathcal{O} = -\Box -m^2\label{Ochoice},\ee the operator $\Box\equiv\partial_\mu\partial^\mu$ is the d'Alembertian, $m^2$ is the scalar squared mass (matrix) and $\mathcal{V}$ is the potential (density) minus the mass term. We take the matrix $m^2$ diagonal without loss of generality (we can always diagonalize it through an orthogonal redefinition of $\varphi$, which leaves the measure $\delta \varphi$ invariant).
 
 
 The generating functional $\mathcal{Z}_0$ can be computed analytically with the usual stationary point method, to obtain
  \be \mathcal{Z}_0(J) = \exp\left(-\frac{i}2\int_Cd^4xd^4y\, J(x) D(x-y) J(y)\right), \label{Z0f}\ee
 where $D$ is the Green's function of the operator $\mathcal{O}$ 
\be  \mathcal{O}D(x-y) = \delta(x-y). \label{GreenDef}\ee
By taking the second functional derivative of $\mathcal{Z}_0(J)$ and then setting $J=0$ one finds that $iD(x-y)$ equals the thermal propagator in the unperturbed theory, $\Delta=0$. This is the reason why $D$ only depends on $x-y$ at least for real values of $x^0$ and $y^0$: the density matrix $\rho$ commutes with $P^\mu$ and so the 2-point thermal Green's function only depends on $x-y$. Below we will check, by direct calculation of the Green's function $D$, that the same is true for any $x^0$ and $y^0$ on $C$.


When $\mathcal{O}$ is chosen as in~(\ref{Ochoice})  the only solution of~(\ref{GreenDef}) satisfying the KMS condition in~(\ref{KMS})  is given by
\be iD(x) = \int \frac{d^3k}{2(2\pi)^3\omega_k}\left[\left(e^{-ikx}+2f_B(\omega_k)\cos(kx)\right)\theta(t)+\left(e^{ikx}+2f_B(\omega_k)\cos(kx)\right)\theta(-t)\right],\label{PropCom}\ee
which reproduces the propagator found in Sec.~\ref{Real scalar field} when $t$ is taken on the real axis. However, this formula holds  for any $t$ on $C$. 

In order to show~(\ref{PropCom}) it is convenient to work with the spacial Fourier transform $D_s(t,\vec k)$, which is related to $D(x)$ by
\be D(x) = \int \frac{d^3k}{(2\pi)^3} e^{-i\vec k\cdot \vec x}D_s(t,\vec k). \ee
By inserting this representation in~(\ref{GreenDef}) one finds (for any $\vec k$)
\be -\left(\frac{\partial^2}{\partial t^2} + \omega_k^2\right)D_s(t) = \delta(t) \ee
and using the ansatz
\be D_s(t) = D^>_s(t)\theta(t)+D^<_s(t)\theta(-t)\ee
one finds
\bea -\dot\delta(t)\left(D^>_s(t)-D^<_s(t)\right)-2\delta(t)\left(\dot D^>_s(t)-\dot D^<_s(t)\right)\nonumber\\-\theta(t)\left(\ddot D^>_s(t)+\omega_k^2D^>_s(t)\right)-\theta(-t)\left(\ddot D^<_s(t)+\omega_k^2D^<_s(t)\right)=\delta(t), \label{StepProp}\eea
which leads to the conditions\footnote{Note that applying the distribution in~(\ref{StepProp})  proportional to $\dot\delta(t)$ to a test function $f$ gives
\be -\int dt\,  \dot\delta(t)\left(D^>_s(t)-D^<_s(t)\right) f(t)  = \int dt\,\delta(t)\left[\left(\dot D^>_s(t)-\dot D^<_s(t)\right)f(t)+ \left(D^>_s(t)-D^<_s(t)\right) \dot f(t)\right]. \ee 
The second term in the square bracket gives the first condition in~(\ref{CondPropCom}), while the first term together with the terms proportional to $\delta(t)$ in~(\ref{StepProp}) leads to the second condition in~(\ref{CondPropCom}).} (for any $\vec k$)
\be D^>_s(0)=D^<_s(0), \qquad \dot D^>_s(0)=\dot D^<_s(0)-1\label{CondPropCom}\ee
as well as the equations
\be \ddot D_s^{>(<)}(t)=-\omega_k^2 D_s^{>(<)}(t).\ee
The solutions of these equations are
\be D_s^{>(<)}(t) = D_p^{>(<)}e^{-i\omega_k t}+D_n^{>(<)}e^{i\omega_k t}\ee
and the conditions in~(\ref{CondPropCom}) lead respectively to
\be D_p^>+D_n^>=D_p^<+D_n^<, \qquad D_p^>-D_n^>=D_p^<-D_n^<-\frac{i}{\omega_k}. \ee
Summing and subtracting these equations gives us, respectively
\be D_p^>=D_p^<-\frac{i}{2\omega_k}, \qquad D_n^>=D_n^<+\frac{i}{2\omega_k}.\label{PropStep2}\ee 
On the other hand, the KMS condition in~(\ref{KMS}) implies
\be D_p^>e^{-\omega_k\beta}=D_p^<, \qquad D_n^> e^{\omega_k\beta} = D_n^<.\ee
Using this result in~(\ref{PropStep2}) leads to
\be D_p^> =\frac{-ie^{\omega_k\beta}f_B(\omega_k)}{2\omega_k}, \qquad D_n^> =\frac{-if_B(\omega_k)}{2\omega_k }, \ee
which, using~(\ref{fBprop}), gives~(\ref{PropCom}).

Let us now consider the contour in Fig.~\ref{contour}, which allows us to directly compute the thermal Green's functions at real times, and send $t_i\to -\infty$ and $t_f\to +\infty$ to recover the full real axis and be able to compute thermal Green's functions for arbitrary times. In this limit both the segments $C_3$ and $C_4$ go to infinity and it is possible to show that~\cite{Niemi:1983nf}
\be D(t_{12} - t_{34}, \vec{x}) \to 0, \ee 
where $t_{12}\in C_1\cup C_2$ and $t_{34}\in C_3\cup C_4$. To understand this result let us represent $D$ through its Fourier transform $\tilde D$, 
\be D(x) = \int \frac{d^4k}{(2\pi)^4} \tilde D(k) e^{-ikx}, \ee
and compute it at a time $x^0=t_{12} - t_{34}$. In the limit $t_i\to -\infty$ and $t_f\to +\infty$ we have that $u\equiv$ Re$(x^0)\to\pm\infty$, while $v\equiv$ Im$(x^0)$ remains finite, and so
\be D(x) = \int \frac{dk_0}{2\pi} e^{-ik_0u}  \int \frac{d^3k}{(2\pi)^3}e^{k_0 v} \tilde D(k_0,\vec k) e^{i\vec{k}\cdot\vec{x}}\to 0\ee
because infinite oscillations due to $e^{-ik_0u}$ kill the above integral over $k_0$. As a result the generating functional becomes the product of a functional computed with the contour $C_1\cup C_2$ and a functional computed  with the contour $C_3\cup C_4$ (see Eqs.~(\ref{DeltaAss}),~(\ref{ZZ0}) and~(\ref{Z0f})). But we are here interested in generating the thermal Green's functions computed at real times so only the first factor is relevant for us. We can, therefore, restrict the time integration in~(\ref{Z0f}) to  $C_1\cup C_2$.


It is now convenient to introduce the notation (for real $x^0$)
\be J_1(x) = J(x), \qquad J_2(x) = J(x^0-i\sigma, \vec x), \label{J12Def} \ee
such that the integral in~(\ref{Z0f}) can be substituted as follows
\be \int_Cd^4xd^4y\, J(x) D(x-y) J(y) \to \int d^4xd^4y J_a(x) D_{ab}(x-y)J_b(y), \ee
where $a, b= 1, 2$ and we have introduced (for real $x^0$ and $y^0$)
\bea   D_{11}(x-y)= D(x-y), \quad D_{22}(x-y) =-D^*(x-y), \label{D1122} \\
 D_{12}(x-y)=  D^<(x^0-y^0+i\sigma, \vec x - \vec y),  \quad D_{21}(x-y)=  D^>(x^0-y^0-i\sigma, \vec x - \vec y).\label{D1212}\eea
The second equation in~(\ref{D1122}) comes from the fact that when we invert the orientation of $C_2$ to rewrite all time integrations as integrations on the real axis from $-\infty$ to $+\infty$ the step functions $\theta(t)$ and $\theta(-t)$ are exchanged and this, according to Eq.~(\ref{PropCom}), corresponds to taking the complex conjugate. One would perhaps have expected a minus sign in the right-hand sides of~(\ref{D1212}): the signs there take into account that the definition of functional derivatives on $C_2$ changes by an overall sign if one uses the Dirac's delta function on the contour defined in Eq.~(\ref{contourDF}) (as we do when using the source $J$) or the ordinary Dirac's delta function (as we do when using the source $J_1$ and $J_2$). This sign change on $C_2$ is due to the factor $\left(\frac{d t}{d v}\right)^{-1}$ in~(\ref{contourDF}). We take this into account by switching the sign of $D_{12}$ and $D_{21}$.  Going now to four-dimensional momentum space we find 
\bea i\tilde D_{11}(k)= \Delta_B(k), \quad  i\tilde D_{22}(k)= \Delta_B(k)^*, \label{Prop1122}\\
i\tilde D_{12}(k) = e^{\sigma k_0}\Delta^<_B(k), \quad i\tilde D_{21}(k) = e^{-\sigma k_0}\Delta^>_B(k), \label{Prop1221}\eea
where $\Delta_B$, $\Delta^<_B$ and $\Delta^>_B$ have been introduced in Sec.~\ref{Real scalar field}. 

Finally, the full generating functional in~(\ref{ZZ0}) can be written as follows
 \bea \mathcal{Z}(J) &=& \frac1{``J_{1,2}\to 0"} \exp\left(i\int d^4x\left(\mathscr{L}_I\left(\frac1{i}\frac{\delta}{\delta J_1}\right)-\mathscr{L}_I\left(\frac1{i}\frac{\delta}{\delta J_2}\right)\right)\right) \nonumber \\
 &&\exp\left(-\frac{i}2\int d^4xd^4y J_a(x) D_{ab}(x-y)J_b(y)\right) \\
 &=&
 \frac1{``J_{1,2}\to 0"} \int \delta\varphi_1 \delta\varphi_2 \exp\left(\frac{i}2 \int d^4xd^4y \varphi_a D^{(-1)}_{ab}(x-y)\varphi_b(y) \right. \nonumber \\
 && \left. i \int d^4x \left(\mathscr{L}_I(\varphi_1)-\mathscr{L}_I(\varphi_2)\right) + i \int d^4 x J_a(x)\varphi_a(x)\right), \label{ZZ02} \eea 
 where $D^{(-1)}_{ab}(x-y)$ satisfies
  \be \int d^4z\, D^{(-1)}_{ac}(x-z)D_{cb}(z-y) =\delta_{ab}\delta(x-y).\ee
 
 Therefore, working in the real-time formalism requires doubling the degrees of freedom (recall $a=1,2$). Looking at the first integral in the last line of~(\ref{ZZ02}), we also notice that in the Feynman diagram representation of perturbation theory only vertices of type 1 or type 2 exist (no mixed vertices arises) and the type 2 vertices have sign opposite to the type 1 ones. However, there are propagator components $\{a, b\}=\{1, 2\}$
  and $\{a, b\}= \{2, 1\}$, which mix type 1 and type 2 particles (see Eq.~(\ref{Prop1221})). One should keep in mind, however, that external lines should only be of type 1, because we are interested here in the thermal Green's function at real times.
  
  \subsubsection{Cutting rules at finite temperature}\label{Cutting rules at finite temperature}
As we have explained in Sec.~\ref{Density matrix and ensemble averages}, one reason why thermal Green's functions are important is because they allow us to compute inclusive rates through Eqs.~(\ref{Toptical1}) and~(\ref{Toptical2}). Therefore, we are interested in computing imaginary parts of amplitudes (multiplied by\footnote{The $-i$ comes from the $-i$ in the definition $\hat T\equiv -i(\hat S-1)$, which we have used in deriving the optical theorem in Sec.~\ref{Density matrix and ensemble averages}.} $-i$). The cutting rules are methods to extract these imaginary parts. We now derive them.

First, let us briefly discuss the zero temperature case. 
At a given order in perturbation theory we consider a Feynman diagram with $p$ vertices in coordinate space and amputated of its external legs, we call this quantity $G(x_1, ..., x_p)$. The $p$ vertices $x_1, ..., x_p$ are linked by the free propagator $G_0(x_i-x_j)$ (with $G_0(x)$ given by Eq.~(\ref{FreePropSc})). As usual we can decompose the propagator in the sum of two terms,
\be G_0(x)= \theta(x_0)G_0^>(x)+
\theta(-x_0)G_0^<(x), \ee
where $G_0^>(x)$ is the corresponding non time-ordered 2-point function and $G_0^<(x) = G_0^>(-x)$. From $G(x_1, ..., x_p)$ we build new quantities $F(x_1, ..., \underline{x}_k,..., x_p)$ having the same topology as $G(x_1, ..., x_p)$, but where some of the spacetime coordinates are underlined. In the graphical representation (the Feynman diagram) of $F$, we represent underlined spacetime coordinates by circled vertices. The expression for $F$ is built according to the following rules
\begin{itemize}
\item Reverse the sign  of the   vertex, if it is underlined.
\item If in $G(x_1, ..., x_p)$ two vertices $x_i$ and $x_j$ are linked by $G_0(x_i-x_j)$ then in $F$
\begin{enumerate}
\item use $G_0(x_i-x_j)$ if neither $x_i$ nor $x_j$ are underlined,
\item use $G_0^>(x_i-x_j)$ if $x_i$ is underlined, but $x_j$ is not,
\item use $G_0^<(x_i-x_j)$ if $x_j$ is underlined, but $x_i$ is not,
\item use the complex conjugate $G_0(x_i-x_j)^*$ if both $x_i$ and $x_j$ are underlined.
\end{enumerate}
\end{itemize}
The rules $1$-$4$ above can be equivalently stated by saying that in going from $G$ to $F$ the propagator $G_0$ is substituted by a new propagator $\bar G_0$ defined by
\bea \bar G_0(x_i-x_j) &\equiv& G_0(x_i-x_j), \quad \bar G_0(\underline{x}_i-\underline{x}_j) \equiv G_0(x_i-x_j)^*, \\ \quad\bar G_0(\underline{x}_i-x_j) &\equiv& G_0^>(x_i-x_j),\quad \bar G_0(x_i-\underline{x}_j) \equiv G_0^<(x_i-x_j). \eea
Note that, using $G_0^<(x) = G_0^>(-x)$, one finds
\be \bar G_0(\underline{x}_i-x_j) = G_0^<(x_j-x_i) \equiv \bar G_0(x_j-\underline{x}_i). \ee
Now, let us call $x_m$ the vertex with the largest time component, that is $x_m^0 > x_j^0$ for all $j\neq m$. Then  for all $j\neq m$
\bea \bar G_0(\underline{x}_m-x_j)&\equiv& G_0^>(x_m-x_j)=G_0(x_m-x_j)\equiv \bar G_0(x_m-x_j), \\
\bar G_0(\underline{x}_m-\underline{x}_j) &\equiv& G_0(x_m-x_j)^* = G_0^>(x_m-x_j)^* =G_0^<(x_m-x_j) \equiv \bar G_0(x_m-\underline{x}_j),\eea
where in the second line we used the property $G_0^<(x)=G_0^>(x)^*$. So it does not matter if we underline the vertex with the largest time component in the propagators or we do not.
From this result it follows
\be F(x_1, ... , \underline{x}_k, ... , \underline{x}_m, ... , x_p) = -  F(x_1, ... , \underline{x}_k, ... , x_m, ... , x_p) \label{largesttime}\ee
that is known as the largest time equation. In the two $Fs$ above the only difference is that $x_m$ is underlined on the left and not underlined on the right. The minus sign comes from the fact that one should reverse the sign of  the coupling associated with a vertex, when it is underlined. The largest time equation~(\ref{largesttime}) implies
\be \sum_{\rm O} 
F(x_1, ... , x_p) = 0, \label{LTEimp}\ee
where the sum is over all possible ways of underlining all vertices. This result allows us to compute the imaginary part of $-iG$: 
\bea {\rm Im}(-i G(x_1, ... , x_p)) &=& -{\rm Re}(G(x_1, ... , x_p)) = -\frac12\left(G(x_1, ... , x_p)+G(x_1, ... , x_p)^*\right) \nonumber\\ &=&-\frac12\left(F(x_1, ... , x_p)+F(\underline{x}_1, ... , \underline{x}_p)\right) = \frac12 \sum_{({\rm O})} F(x_1, ..., x_p), \label{cuttingT0}\eea
where the sum is over all possible ways of underlining vertices, except for the case with no vertex or all vertices underlined. In the third step we have used the fact that by changing all circled vertices with uncircled vertices and viceversa we take the complex conjugate of all vertices and all propagators. Eq.~(\ref{cuttingT0}) summarizes the zero-temperature cutting rules.


Let us now consider the momentum space. At $T=0$ the Fourier transforms of $G_0^>$ and $G_0^<$ are, respectively,
\be 
\theta(k_0) 2\pi\delta(k^2-m^2), \quad 
\theta(-k_0) 2\pi\delta(k^2-m^2) \ee
and, in the convention where $\exp(-i k(x_i-x_j))$ corresponds to a momentum $k$ flowing from the  $j$th vertex into the $i$th vertex, the momentum always flows from uncircled to circled vertices. This in particular implies that at zero temperature circled and uncircled vertices must form connected sets. 

At finite temperature, the Fourier transforms of $G_0^>$ and $G_0^<$ are, respectively,
\be 
\Delta_B^>(k)\equiv (\theta(k_0)+n_B(k_0)) 2\pi\delta(k^2-m^2), \quad 
\Delta_B^<(k)\equiv (\theta(-k_0)+n_B(k_0)) 2\pi\delta(k^2-m^2) \ee
and momentum can flow in both directions because of $n_B$. At this point it is useful to recall Eqs.~(\ref{Prop1122}),~(\ref{Prop1221}) and~(\ref{ZZ02}) and choose $\sigma = 0$~\cite{Keldish}. With this choice, uncircled vertices may be identified with type 1 vertices and circled ones with type 2 vertices. Now, let us call $y_1, ... , y_l$ the vertices attached to external lines and $z_1, ... , z_n$ the internal ones. While $y_1, ... , y_l$ are necessarily of type 1,  $z_1, ... , z_n$ are either of type 1 or type 2 and the relevant amplitude  is 
\be \mathcal{G}\equiv\sum_{z_j \in \{1,2\}} G( y_1, ... , y_l; z_1, ... , z_n), \ee
where the sum is over all possible ways of underlining the vertices $z_1, ... , z_n$. Next, one introduces the quantities
\be F(y_1, ... ,\underline{y}_k, ... , y_l; z_1, ... , \underline{z}_p, ... ,z_n) \ee
defined like in the $T=0$ case, but now using the finite-temperature propagators. Then $\mathcal{G}$ can be rewritten as follows
\be \mathcal{G}= \sum_{{\rm O}\,  z} F(y_1, ... , y_l; z_1, ... , z_n), \ee
where the sum is over all possible ways of circling the $z$ vertices. Note that 
\be \mathcal{G}^*= \sum_{{\rm O}\,  z} F(\underline{y}_1, ... , \underline{y}_l; z_1, ... , z_n). \ee 
This is because by changing all circled vertices with uncircled vertices and viceversa we take the complex conjugate of all vertices and all propagators.
If we now use the largest time equation~(\ref{largesttime}) and~(\ref{LTEimp}), which still holds at $T\neq 0$, we obtain
\be {\rm Im}(-i \mathcal{G}) = \frac12 \sum_{{\rm O} (y) z} F(y_1, ... , y_l; z_1, ... , z_n),\label{KobesSemenoff}\ee
where the sum is over all possible ways of circling the $y$ and $z$ vertices, except for the cases with no $y$ vertex or all $y$ vertices circled. 

Eqs.~(\ref{KobesSemenoff}) gives us the cutting rules at finite temperature. They were first derived by Kobes and Semenoff~\cite{KS}. For this reason they are also known as Kobes-Semenoff rules. They can be used, among other things, to compute production or, more generally, interaction rates of particles. Some physical examples will be provided in Sec.~\ref{Weakly-coupled particle production}.

 \subsection{Fermions}\label{Fermions}
  
Let us now move to the Green's functions for fermion field operators\footnote{Just like in the  scalar case of Sec.~\ref{Scalar theory}, here, unlike in Sec.~\ref{Fermion field}, we added hats on top of the fermion operators because later we will derive a path-integral formula and we will need to distinguish between the field operators and the field integration variables in the path integral.} $\hat\psi$ and $\hat{\bar\psi}$ (where we understand an index running over all species of fermion particles).

As we did for scalar fields, we start by considering the simplest Green's function, the thermal propagator
\bea S_{\alpha\beta}(x)\equiv\langle {\cal T}\hat\psi_\alpha(x)\hat{\bar\psi}_\beta(0)\rangle&=&\theta(x_0)S_{\alpha\beta}^>(x)  +\theta(-x_0)  S_{\alpha\beta}^<(x) . 
 \eea
where the ``non time-ordered" 2-point functions are defined by
\be S_{\alpha\beta}^>(x) \equiv \langle \hat\psi_\alpha(x)\hat{\bar\psi}_\beta(0)\rangle, \quad S_{\alpha\beta}^<(x) \equiv -\langle \hat{\bar \psi}_\beta(0)\hat\psi_\alpha(x)\rangle. \label{Sforback}\ee
  Here $\langle {\cal T}\hat\psi_\alpha(x)\hat{\bar\psi}_\beta(0)\rangle$ is the exact thermal propagator, and $S_{\alpha\beta}^>(x)$ and $S_{\alpha\beta}^<(x)$ are the exact non time-ordered 2-point functions. The minus sign in~(\ref{Sforback}) implies that the fermionic thermal propagator is antisymmetric in the exchange of the two fermion fields. Just like in the scalar case, we can formally see $e^{-\beta H}$ as an imaginary-time evolution operator. So, $e^{\beta H}\hat\psi(t,\vec x)e^{-\beta H} = \hat\psi(t-i\beta,\vec x)$. Using now the ciclicity of the trace in the thermal Green's function leads to 
\be S^>(t-i\beta,\vec x) = -S^<(t,\vec x).\label{KMSf}\ee
This relation is known as the Kubo-Martin-Schwinger (KMS) condition for fermions and it differs from the KMS condition for bosons in~(\ref{KMS}) because of the minus sign. 

The KMS condition implies, like in the bosonic case, a relation between the Fourier transforms, defined as usual by
\be \tilde S^>(k) = \int d^4x\, e^{ikx} S^>(x), \quad \tilde S^<(k) = \int d^4x\, e^{ikx}S^<(x).
\ee
To find this relation first note that 
\bea \tilde S^<(k) &=& \int_{-\infty}^{+\infty} dt \int d^3x\,e^{ik_0t} e^{i\vec k \cdot \vec x} S^<(t,\vec x) =  -  \int_{-\infty}^{+\infty} dt\,e^{ik_0t} \int d^3x \, e^{i\vec k \cdot \vec x} S^>(t-i\beta,\vec x)\nonumber \\ &=& - e^{-\beta k_0}\int_{-\infty-i\beta}^{+\infty-i\beta} dt\,e^{ik_0t} \int d^3x \, e^{i\vec k \cdot \vec x} S^>(t,\vec x) =-e^{-\beta k_0}\tilde S^>(k),\label{Sfbtilde}\eea
where in the second step we used the KMS condition for fermions and in the last step we used the same analiticity argument presented below Eq.~(\ref{Gfbtilde}). Eq.~(\ref{Sfbtilde}) differs from the scalar case in~(\ref{Gtfb})  because of the minus sign.

\subsubsection{Fermionic path integral}\label{Fermionic path integral}

In deriving the path integral in a scalar theory we started from the eigenstates of the field operators $\hat\varphi|\varphi\rangle = \varphi|\varphi\rangle$ and used the completeness of these states, which is guaranteed by the Hermiticity of $\hat\varphi$. The fermionic field operator $\hat\psi$, however, is not Hermitian and so nothing guarantees the existence of a complete set of eigenstates of $\hat\psi$. This problem can be overcome by introducing anticommuting c-numbers (a.k.a. Grassmann variables) as we now show.

 The creation and annihilation operators for fermions, introduced in Sec.~\ref{Fermion field}, satisfy anticommutation rather commutation rules (Eqs.~(\ref{antNot}) and~(\ref{ant})). Let us simply denote here the annihilation operators $c_{kr}$ and $d_{kr}$ with $a_i$, where $i$ is a collective index running over both particles and antiparticles as well as all values of $k$ and $r$ (and also over all species of fermions). The anticommutation rules in~(\ref{antNot}) and~(\ref{ant})  then simply read
 \be \{a_i,a_j^\dagger\} =\delta_{ij}, \quad \{a_i,a_j\} = 0. \label{AnticommF} \ee 
 At this point we introduce the Grassmann variables $\eta_i$,
 \be \{\eta_i,\eta_j\} = 0, \ee
 which are assumed to satisfy
 \be \{\eta_i,a_j\} = 0, \quad \{\eta_i,a^\dagger_j\} = 0. \ee
 Note that for all $i$ we have $\eta_i^2=0$.
 
 Let us now define the state
 \be |\eta\rangle \equiv \exp\left(-\sum_i \eta_i a^\dagger_i\right)|0\rangle, \ee
 where $|0\rangle$ here is a state without fermions, i.e.~$a_i|0\rangle=0$. The state $|\eta\rangle$ is an ``eigenstate" of $a_l$ with ``eigenvalue" $\eta_l$:
 \bea  a_l |\eta\rangle &=&  a_l\left(\prod_{i\neq l} e^{-\eta_ia_i^\dagger}\right) e^{-\eta_la_l^\dagger} |0\rangle = \left(\prod_{i\neq l} e^{-\eta_ia_i^\dagger}\right) a_le^{-\eta_la_l^\dagger} |0\rangle  = \left(\prod_{i\neq l} e^{-\eta_ia_i^\dagger}\right) a_l(1-\eta_la_l^\dagger) |0\rangle   \nonumber \\
  &=&\left(\prod_{i\neq l} e^{-\eta_ia_i^\dagger}\right)\eta_l  |0\rangle = \left(\prod_{i\neq l} e^{-\eta_ia_i^\dagger}\right)\eta_l e^{-\eta_l a_l^\dagger} |0\rangle  =\eta_l |\eta\rangle. \label{etaeig} \eea  
 
 We also introduce independent Grassmann variables $\eta^*_i$, 
 \be \{\eta^*_i,\eta^*_j\} = 0, \ee
 which are also assumed to satisfy
 \be \{\eta^*_i,a_j\} = 0, \quad \{\eta^*_i,a^\dagger_j\} = 0. \ee
 These additional Grassmann variables allow us to define a ``bra" state
 \be\langle\eta| \equiv \langle 0|\exp\left(-\sum_i a_i\eta_i^*\right), \ee
which is a ``left eigenstate" of $a^\dagger_l$ with ``eigenvalue" $\eta_l$
 \be\langle\eta|a^\dagger_l = \langle\eta|\eta^*_l. \label{etaeigs} \ee
 The proof of this statement is very similar to the one in~(\ref{etaeig}).
 

 For $|0\rangle$ we assume the simple normalization 
 \be \langle 0|0\rangle =1.\ee
 Since $\langle 0|a^\dagger_i =0$, this also implies 
 \be \langle 0|\eta\rangle = \langle 0|0\rangle =1.\ee
 Analogously, since $a_i|0\rangle =0$, 
 \be \langle\eta|0\rangle = \langle 0|0 \rangle = 1 \ee
 Note also that 
 \bea \langle\eta|\eta'\rangle 
 &=&  \langle 0|\left(\prod_i e^{-a_i\eta^*_i} e^{-\eta'_i a_i^\dagger}\right)|0\rangle = 
  \langle 0|\left(\prod_i (1-a_i\eta^*_i)(1-\eta'_i a_i^\dagger)\right)|0\rangle  \nonumber \\
  &=& \langle 0|\left(\prod_i (1-a_i\eta^*_i -\eta'_i a_i^\dagger+a_i\eta^*_i \eta'_i a_i^\dagger)\right)|0\rangle = \langle 0|\left(\prod_i (1+\eta^*_i \eta'_i )\right)|0\rangle \nonumber \\ &=& e^{\eta^*\eta'},
 \label{etapeta} \eea
  where we defined $\eta^*\eta'\equiv \sum_i \eta^*_i\eta'_i$.
  
  One can also introduce an integration over the variables $\eta_i$ and $\eta^*_i$, which satisfy the rules 
\be   \int d\eta_i  =\int d\eta_i^* 
= 0, \quad  \int d\eta_i \eta_j  =\int d\eta_i^*\eta_j^* = \delta_{ij}.\ee
With these rules the completeness relation reads 
\be \int d\eta^*d\eta \, e^{-\eta^*\eta} | \eta\rangle \langle\eta| = 1, \label{CompletEta} \ee
where we defined $d\eta^* d\eta \equiv \prod_i d\eta_i^*d\eta_i$. To show~(\ref{CompletEta}) first write
 \be \int d\eta^*d\eta \, e^{-\eta^*\eta} | \eta\rangle \langle\eta| = \int \left(\prod_i d\eta_i^*d\eta_i\right)\left(\prod_i e^{-\eta_i^*\eta_i}\right)\left(\prod_i e^{-\eta_ia^\dagger_i}\right)|0\rangle\langle 0|\left(\prod_i e^{-a_i\eta_i^*}\right) \ee
 and then start integrating over the various $\eta_i^*$ and $\eta_i$ one by one. Let us start from say $i=1$, the integral over $\eta_1^*$ and $\eta_1$ is 
 \be \int d\eta_1^*d\eta_1\, e^{-\eta_1^*\eta_1}e^{-\eta_1a^\dagger_1}|0\rangle\langle 0|e^{-a_1\eta_1^*} = \int d\eta_1^*d\eta_1(-\eta_1^*\eta_1+1)e^{-\eta_1a^\dagger_1}|0\rangle\langle 0|e^{-a_1\eta_1^*} =  |0\rangle \langle 0| + |1_1\rangle\langle 1_1|, \nonumber  \ee
 where $|1_i\rangle$ is the state with one fermion of type $i$. Next, perform the integration over $\eta_2^*$ and $\eta_2$: 
 \be \int d\eta_2^*d\eta_2 \, e^{-\eta_2^*\eta_2}e^{-\eta_2a^\dagger_2}(|0\rangle \langle 0| + |1_1\rangle\langle 1_1|)e^{-a_2\eta_2^*} =  |0\rangle \langle 0| + |1_1\rangle\langle 1_1|+ |1_2\rangle\langle 1_2|+ |1_1,1_2\rangle\langle 1_1,1_2|,   \ee
 where $|1_i,1_j\rangle$ is the state with one fermion of type $i$ and one of type $j$. Reiterating this procedure one finds~(\ref{CompletEta}).
 
 Similarly, one can also show that for any operator $A$ commuting\footnote{If $A$ instead anticommutes with the $\eta_i$ and $\eta_i^*$
one has 
 \be \Tr (A)=\int d\eta^*d\eta \, e^{-\eta^*\eta} \langle \eta|A|\eta\rangle . \label{TrFera} \ee
 But, as we will see in this section, for our physical purposes it is not restrictive to assume that $A$ commutes with the $\eta_i$ and $\eta_i^*$.} with the $\eta_i$ and $\eta_i^*$
 \be  \Tr (A)=\int d\eta^*d\eta \, e^{-\eta^*\eta} \langle -\eta|A|\eta\rangle . \label{TrFer} \ee
 Indeed, like for the completeness relation one can decompose
\be \int d\eta^*d\eta \, e^{-\eta^*\eta} \langle -\eta|A|\eta\rangle = \int \left(\prod_i d\eta_i^*d\eta_i\right)\left(\prod_i e^{-\eta_i^*\eta_i}\right)\langle 0|\left(\prod_i e^{a_i\eta_i^*}\right)A\left(\prod_i e^{-\eta_ia^\dagger_i}\right)|0\rangle. \ee
and  start  performing the integral over $\eta_1^*$ and $\eta_1$: 
\be \int d\eta_1^*d\eta_1 e^{-\eta_1^*\eta_1} e^{a_1\eta_1^*}Ae^{-\eta_1a^\dagger_1} =  A +a_1 A a_1^\dagger. \ee
Next, the integral over $\eta_2^*$ and $\eta_2$ reads 
\be \int d\eta_2^*d\eta_2 e^{-\eta_2^*\eta_2} e^{a_2\eta_2^*}(A+a_1 A a_1^\dagger)e^{-\eta_2a^\dagger_2} =  A +a_1 A a_1^\dagger+a_2 A a_2^\dagger+a_2a_1 A a_1^\dagger a_2^\dagger. \ee
Reiterating this procedure and taking the expectation value with the state $|0\rangle$ one finds~(\ref{TrFer}).

 Now, let us define the time-dependent ket and bra
 \be |\eta,t\rangle \equiv e^{iHt}|\eta\rangle, \quad \langle \eta,t|\equiv\langle\eta|e^{-iHt}. \ee
 Using~(\ref{etaeig}) and~(\ref{etaeigs}) one finds
 \be a_l(t)|\eta,t\rangle=\eta_l|\eta,t\rangle, \quad \langle \eta,t| a_l^\dagger(t) = \langle \eta,t| \eta_l^*,  \label{etaeigts}\ee
 where $a_l(t)\equiv e^{iHt}a_le^{-iHt}$ and $a^\dagger_l(t)\equiv e^{iHt}a_l^\dagger e^{-iHt}$.
 Then, applying  $e^{iHt}$ from the left and $e^{-iHt}$ from the right to~(\ref{CompletEta}), one obtains
 \be \int d\eta^*d\eta \, e^{-\eta^*\eta} | \eta,t\rangle \langle\eta,t| = 1. \label{CompletEtat} \ee
 On the other hand, using the ciclicity of the trace in~(\ref{TrFer}),
  \be  \Tr (A)=\int d\eta^*d\eta \, e^{-\eta^*\eta} \langle -\eta,t|A|\eta,t\rangle. \label{TrFert} \ee
  Here, we are interested in computing the thermal Green's functions
   \be  \Tr (\rho \mathcal{T}\hat O_1(x_1) ... \hat O_n(x_n)), \label{GreenO} \ee
   where the $\hat O_i$ are constructed with $\hat \psi$ and $ \hat{\bar\psi}$. The time-ordered product $\mathcal{T}$ is antisymmetric in the exchange of an odd number of fermion field operators $\hat \psi$ and $ \hat{\bar\psi}$. In particular we are interested in the case where the operators $\hat O_i(x_i)$ emerge from the transition operator $\hat T$ defined in Sec.~\ref{Density matrix and ensemble averages}. For our purposes it is, therefore, not restrictive to assume that $\hat O_1(x_1)\hat O_2(x_2) ... $ commutes with the $\eta_i$ and $\eta_i^*$.

The rest of the derivation of the fermionic path integral parallels the one of the scalar path integral. One sets $t$ equal to some reference value $t_0$ and introduces a partition (which includes the times $x_i^0$) of a contour $C$ connecting $t_0$ with $t_0-i\beta$: that is one introduces $\mathcal{N}$ times $t_k$ on $C$ with $k=1, ..., \mathcal{N}$, which include the $x_i^0$, ordered as $t_\mathcal{N}>t_{\mathcal{N}-1}> ... >t_2>t_1 $, where $t_k>t_{k-1}$ means that $t_k$ is after $t_{k-1}$ in $C$. 
The $t_k-t_{k-1}$ become infinitesimal as $\mathcal{N}\to\infty$. 
It is also useful to define $t_{\mathcal{N}+1}\equiv t_0-i\beta.$ One then inserts the identity in the form~(\ref{CompletEtat}) $\mathcal{N}$ times obtaining (for $k=1, ... \,  , \mathcal{N}+1$)
\be e^{-\eta_k^*\eta_k}\langle\eta_k,t_k|\eta_{k-1},t_{k-1}\rangle = e^{-\eta_k^*\eta_k}\langle\eta_k|e^{-i(t_k-t_{k-1})H}|\eta_{k-1}\rangle. \ee

Now, the Hamiltonian, like any linear operator, can always be written in terms of the creation and annihilation operators, $H=H(a^\dagger,a)$. Taking $\mathcal{N}\to\infty$ so that $t_k-t_{k-1}$ becomes infinitesimal one can, therefore, write
\be \langle\eta_k|e^{-i(t_k-t_{k-1})H}|\eta_{k-1}\rangle =  \langle\eta_k|1-i(t_k-t_{k-1})H|\eta_{k-1}\rangle =  e^{-i(t_k-t_{k-1})H_c(\eta_k^*,\eta_{k-1})}\langle\eta_k|\eta_{k-1}\rangle, \ee
where 
\be H_c(\eta^*,\eta') \equiv \frac{\langle\eta|H(a^\dagger,a) |\eta' \rangle}{\langle\eta|\eta' \rangle}, \ee
plays the role of a classical ($c$-number) Hamiltonian.
It is always possible to compute $H_c(\eta^*,\eta')$ explicitly because, by repeatedly applying~(\ref{AnticommF}), we can always put all annihilation operators on the right of all creation operators and then use~(\ref{etaeig}),~(\ref{etaeigs}) and~(\ref{etapeta}). 

When $t_k=x_i^0$ we also have an operator $\hat O_i(x_i)$ inside the inner product 
\be e^{-\eta_k^*\eta_k}\langle\eta_k,t_k|\hat O_i(x_i)|\eta_{k-1},t_{k-1}\rangle. \ee
The $\hat O_i(x_i)$ are operator in the Heisenberg picture 
\be  \hat O_i(x_i) =  e^{iHx_i^0}  \hat O_i(0, \vec x_i)e^{ - iHx_i^0}\ee
and the $\hat O_i(0, \vec x_i)$, like any linear operator, can be written in terms of the creation and annihilation operators. Using~(\ref{etaeigts}) one then finds 
\be \langle\eta_k,t_k|\hat O_i(x_i)|\eta_{k-1},t_{k-1}\rangle = O_i(x_i)\langle\eta_k,t_k|\eta_{k-1},t_{k-1}\rangle,\ee
where $O_i(x_i)$ is the $c$-number field obtained by substituting $a(t)$ with $\eta_{k-1}$ and $a^\dagger(t)$ with $\eta_{k}^*$ after putting all annihilation operators on the right of all creation operators. Note also that $\eta_{k-1}$ and $\eta_k$ tend to a common value as $\mathcal{N}\to\infty$. Putting everything together one obtains the following path-integral representation of the thermal Green's functions
\bea  &&\hspace{-1cm} \langle \mathcal{T}\hat O_1(x_1) ... \hat O_n(x_n)\rangle \nonumber \\
&&\hspace{-1cm}= \frac1{Z} \int \delta\eta^*\delta\eta \exp\left(\int_C dt\left(-\eta^*(t)\dot \eta(t)-iH_c(\eta^*(t),\eta(t)\right) \right)O_1(x_1) ... O_n(x_n), \label{PIGFF} \eea
where 
\be \delta\eta^*\delta\eta \equiv \prod_{k=0}^\mathcal{N} d\eta^*_kd\eta_k. \ee 
The main difference with respect to the bosonic case is the presence of antiperiodic boundary conditions, 
\be \eta(t_0-i\beta)=-\eta(t_0), \quad  \eta^*(t_0-i\beta)=-\eta^*(t_0) \label{ABC}\ee
 rather than periodic ones. The antiperiodicity comes from the minus sign in the formula~(\ref{TrFert}) for the trace.


Let us now rewrite the argument of the exponential in~(\ref{PIGFF}) in terms of fields. First, note that from the boundary conditions in~(\ref{ABC}) and the fact that the $\eta$ and $\eta^*$ are Grassmann variables it follows, among other things,
\be \int_Cdt \, \eta^*(t)\dot \eta(t) = \int_Cdt  \,\frac{d}{dt}\left(\eta^*(t)\eta(t)\right) -\int_Cdt \, \dot\eta^*(t) \eta(t) = -\int_Cdt \, \dot\eta^*(t) \eta(t)= \int_Cdt \, \eta(t) \dot\eta^*(t).\nonumber\ee
One can  introduce a $c$-number Grassmann field 
\be \psi(x) \equiv \sum_{kr} \frac{\gamma_{kr}(t)u_{kr} e^{-i\vec k\cdot \vec x}+\delta^*_{kr}(t)v_{kr} e^{i\vec k\cdot \vec x}}{\sqrt{VE_k}}, \label{psigd}\ee
where the $\gamma_{kr}(t)$ and $\delta_{kr}(t)$ are the Grassmann variables $\eta_i(t)$ in our case and the $u_{kr}$ and $v_{kr}$ are the constant spinors introduced in Sec.~\ref{Fermion field}.
Using $u^\dagger_{kr}u_{ks} = E_k \delta_{rs}$, $v^\dagger_{kr}v_{ks} = E_k \delta_{rs}$ and $u^\dagger_{kr}v_{-ks} = 0$, which hold in our convention for the constant spinors $u_{kr}$ and $v_{kr}$, one can thus substitute
\be \eta^*(t)\dot \eta(t) \to \int d^3x \, \psi^\dagger(x)\dot \psi(x). \ee
Moreover, we introduce a classical ($c$-number) Hamiltonian density $\mathcal{H}_c$ that satisfies
\be H_c(\eta^*,\eta) = \int d^3x \, \mathcal{H}_c(\bar\psi,\psi).  \label{etapsitr} \ee 
 The path integral in~(\ref{PIGFF}) can then be written 
\bea  &&\hspace{-1cm} \langle \mathcal{T}\hat O_1(x_1) ... \hat O_n(x_n)\rangle \nonumber \\
&&\hspace{-1cm}= \frac1{``O_i\to 1"} \int \delta\bar\psi\delta\psi\exp\left(\int_C d^4x\left[-\psi^\dagger(x)\dot \psi(x)-i\mathcal{H}_c(\bar\psi(x),\psi(x))\right] \right)O_1(x_1) ... O_n(x_n), \label{PIGFFF} \eea
subject to the antiperiodic boundary conditions,
\be \psi(t_0-i\beta,\vec x)=-\psi(t_0,\vec x), \quad \bar \psi(t_0-i\beta,\vec x)=-\bar\psi(t_0,\vec x). \label{AntiPpsi}\ee
In deriving~(\ref{PIGFFF}) we have used the fact that the relation between the $\gamma_{kr}$ and $\delta_{kr}$ and the fields $\psi$ and $\bar\psi$ is linear (see Eq.~(\ref{psigd})) and thus the Jacobian cancels with the one coming from the denominator, $Z$.

Now, in Dirac's theory of fermions the momentum $P_\psi$ conjugate to $\psi$ is given by $P_\psi = i\psi^\dagger$  and so we can write 
\be   \langle \mathcal{T}\hat O_1(x_1) ... \hat O_n(x_n)\rangle  = \frac1{Z}
 \int \delta\bar\psi\delta\psi
 \exp\left(i\int_C d^4x\, \mathscr{L}(\bar\psi(x),\psi(x))\right)O_1(x_1) ... O_n(x_n), \label{PIGFFFL} 
\ee
where $\mathscr{L}$ represents here the fermionic Lagrangian,
\be  \mathscr{L}(\bar\psi,\psi) = P_\psi \dot\psi -  \mathcal{H}_c(\bar\psi,\psi). \label{LagHamPsi}\ee

Note that these Green's functions can be obtained by taking functional derivatives of
\be \mathcal{Z}(\bar\kappa,\kappa) = \frac1{``\{\bar\kappa,\kappa\}\to0"} \int \delta\bar\psi\delta\psi
 \exp\left(i\int_C d^4x\, \mathscr{L}(\bar\psi(x),\psi(x))+i\int_Cd^4x(\bar\kappa\psi+\bar\psi\kappa)\right)\ee 
 with respect to the Grassmann sources $\kappa$ and $\bar\kappa$. 
 
 In the case of a free Dirac fermion, recalling that the Hamiltonian operator $H$ has the form given in Eqs.~(\ref{HF}) and~(\ref{Nferm}), one obtains
 \be H_c(\eta^*,\eta) = \sum_i E_i \eta_i^*\eta_i = \sum_{kr}  E_k\left(\gamma_{kr}^*\gamma_{kr} + \delta_{kr}^*\delta_{kr} \right),\ee
 which reproduces~(\ref{etapsitr}) for 
 \be\mathcal{H}_c(\bar\psi,\psi) = \bar\psi\left(-i\gamma^j\partial_j+m\right)\psi . \ee
 Note that this is precisely the Hamiltonian density that one would have obtained from~(\ref{LagHamPsi}) using the Dirac expressions for the conjugate momentum,  $P_\psi = i\psi^\dagger$,  and the Lagrangian, 
  $\mathscr{L} =\bar \psi (i\slashed{\partial}-m)\psi$. 
 
  Just like in the scalar theory previously discusses, If we set $t_0=0$ and we go from $0$ to $-i\beta$ along the imaginary axis (as a possible choice for $C$) we have a formula for the thermal Green's functions in the imaginary-time formalism. On the other hand, choosing $C$ as in Fig.~\ref{contour} (with $t_0=t_i \to -\infty$ and $t_f\to\infty$) we can directly obtain the thermal Green's functions at real time.

Let us now compute the thermal propagator for a free Dirac fermion with $\mathscr{L} =\bar \psi (i\slashed{\partial}-m)\psi$ in the real-time formalism. We choose $C$ as in Fig.~\ref{contour} taking $\sigma\to 0$, which leads to a simpler propagator. The discussion parallels the one given for scalars in Sec.~\ref{Path integral}. Also for fermions the contour $C$ leads to a doubling of the degrees of freedom and so a four-component propagator: working in momentum space we have
\bea \tilde S_{11}(k)= (\slashed{k}+m) \Delta_F(k), \quad  \tilde S_{22}(k)= (\slashed{k}+m) \Delta_F(k)^*, \label{Propf1122}\\
\tilde S_{12}(k) = (\slashed{k}+m) \Delta^<_F(k), \quad \tilde S_{21}(k) =  (\slashed{k}+m) \Delta^>_F(k), \label{Propf1221}\eea 
where $\Delta_F$, $\Delta^>_F$ and $\Delta^<_F$ have been defined in Sec.~\ref{Fermion field}.
It is easy to understand where these formul\ae~come from. The $11$ component comes from setting both times of the propagator on the real axis and thus one simply recovers the propagator computed in Sec.~\ref{Fermion field}. The $12$ component comes from setting the first time on the real axis and the second one on the $C_2$ segment in Fig.~\ref{contour}; as  $\sigma\to0$ this segment goes to the real axis, but the contour $\theta$ function in~(\ref{contourT}) selects $\Delta^<_F$. On the other hand, the $21$ component comes from setting the first time on $C_2$ and the second one on the real axis and then the contour $\theta$ function selects $\Delta^>_F$. Finally, the $22$ component comes from setting both times on $C_2$; when one sends $\sigma\to0$ the segment $C_2$ goes to the real axis, but the contour $\theta$ function exchanges $\Delta_F^>$ and $\Delta_F^<$ (the orientation of $C_2$ is opposite to the one of the real axis) and this corresponds to taking the complex conjugate of $\Delta_F(k)$ in the propagator: indeed, setting $T=0$ (as the finite temperature parts of $\Delta_F^>$ and $\Delta_F^<$ are the same and give a real contribution to $\Delta_F(k)$) one finds
\bea \int \frac{d^4k}{(2\pi)^4} \Delta_0^*(k)(\slashed{k}+m) e^{-ikx}=\int \frac{d^4k}{(2\pi)^4} \Delta_0^*(k)(-\slashed{k}+m) e^{ikx} \nonumber\\= \theta(x_0) \int \frac{d^3k}{2(2\pi)^3E_k}(-\slashed{k}+m) e^{ikx}+\theta(-x_0) \int \frac{d^3k}{2(2\pi)^3E_k}(\slashed{k}+m) e^{-ikx}.\eea
Comparing this expression with Eqs.~(\ref{S>x})-(\ref{S<x}) for $T=0$ we see that, indeed, taking the complex conjugate of $\Delta_0(k)$ corresponds to exchanging
$\Delta^>_F$ and $\Delta^<_F$.
 
 
\subsubsection{Perturbation theory in fermion-scalar theories}

Combining all results obtained so far separately for scalars and fermions, we can perform perturbation theory in a generic fermion-scalar theory, such as a theory featuring generic Yukawa and quartic scalar interactions. Let us illustrate this point in the real-time formalism.

Also for theories involving fermions, like  for pure scalar theories (see the discussion around Eq.~(\ref{ZZ02})),  in the Feynman diagram representation of perturbation theory only vertices of type 1 or type 2 exist (no mixed vertices arise) and the type 2 vertices have sign opposite to the type 1 ones. However, there are propagator components $\{a, b\}=\{1, 2\}$
  and $\{a, b\}= \{2, 1\}$, which mix type 1 and type 2 particles (see Eqs.~(\ref{Propf1221})).

 We can, therefore, extend the circling notation introduced in Sec.~\ref{Cutting rules at finite temperature} to fermion-scalar theories: type 1 vertices are uncircled and type 2 ones are circled. This allows us to perform perturbation theory directly in the real-time formalism in the presence of both scalars and fermions. In Sec.~\ref{Gauge theory} we will include gauge fields to obtain a completely realistic setup.

\subsection{Gauge theory}\label{Gauge theory}

We now consider a generic Green's function of the form~(\ref{GreenO}), where now the $\hat O_i$ are generic local operators built out of gauge fields and/or matter fields. The matter fields are either scalars or fermions. The latter are assumed to be described by Dirac's theory.

We consider a general Yang-Mills theory, where the gauge group $G$ is a Lie group  with generators $T^a$, with $a=1, ... , \dim G$. We choose the $T^a$ such that the structure constants $f^{abc}$, defined by
\be [T^a,T^b] = i f^{abc} T^c, \ee
are totally antisymmetric.  The field strength $F^a_{\mu\nu}$ is given in terms of the gauge fields $A^a_\mu$ by
\be F^a_{\mu\nu} = \partial_\mu A^a_\nu-\partial_\nu A^a_\mu -C^{bca} A^b_\mu A^c_\nu,\ee
where $C^{bca}\equiv g f^{bca}$ and $g$ is the gauge coupling.

We start from the gauge-invariant  classical Lagrangian given by
\be \mathscr{L}= -\frac14 F^a_{\mu\nu}F^{a\mu\nu} + \mathscr{L}_M(\Phi,D\Phi), \ee
where $\mathscr{L}_M$ represents the matter Lagrangian, which we assume here to be an ordinary function of only the matter fields $\Phi$ and their covariant derivatives $D_\mu\Phi$, given by
\be D_\mu\Phi=(\partial_\mu+i t^a A^a_\mu)\Phi,\ee 
where $t^a\equiv g T^a$.

The momenta conjugate to $A^a_\mu$ are
\be \Pi_a^{\mu} \equiv \frac{\partial\mathscr{L}}{\partial \dot  A_\mu^a}  = F^{a}_{0\mu} \ee
and we note that the momentum conjugate to $A_0^a$ vanishes:
\be \Pi_a^0  =F^{a}_{00}= 0.  \ee
In the axial gauge, i.e.
\be A_3^a = 0 \ee
the canonical coordinates are only $A_i^a$ with $i=1,2$. Let us now choose this gauge. Therefore, the classical Hamiltonian density due to the gauge fields only, $\mathcal{H}_{cG}$, is given by
\bea \mathcal{H}_{cG} &=& \Pi_a^{i} \dot A^a_i +\frac14 F^a_{\mu\nu}F^{a\mu\nu}\nonumber \\ &=&\Pi_a^{i} \left(F^a_{0i}+\partial_i A_0^a+C^{bca} A^b_0 A^c_i\right) -\frac{1}{2}F^a_{0i}F^a_{0i}+\frac14 F^a_{ij}F^{a}_{ij} +\frac12F^a_{i3}F^a_{i3}-\frac{1}{2}F^a_{03}F^a_{03}\nonumber \\
&=&\Pi_a^{i} \left(\partial_i A_0^a+C^{bca} A^b_0 A^c_i\right)+\frac{1}{2}\Pi_a^{i} \Pi_a^{i}  +\frac14 F^a_{ij}F^{a}_{ij}+\frac12 \partial_3 A^a_i\partial_3 A^a_i-\frac12 \partial_3 A^a_0\partial_3 A^a_0. \label{HamG}\eea
The full classical Hamiltonian density is
\be \mathcal{H}_{c} = \mathcal{H}_{cG} + \mathcal{H}_{cM}, \ee
where $\mathcal{H}_{cM}$ is the classical Hamiltonian density of the matter fields.

As long as the matter Hamiltonian is at most quadratic in the momenta, readapting the derivation of the path integral  given in textbooks on quantum field theory at zero temperature\footnote{See e.g.~\cite{Weinberg2}. One can use the fact that  the Hamiltonian in~(\ref{HamG})
 has a term linear and one quadratic 
 in the momenta and  the quadratic one has a field-independent coefficient; so in the general path-integral formula one can integrate 
 over 
 all momenta (one has a Gaussian integral) and the result leads to the expression in the exponential appearing in the path integral evaluated at the stationary point, which precisely leads to the Lagrangian. }
 one finds
\be  \Tr (\rho \mathcal{T}\hat O_1(x_1) ... \hat O_n(x_n)) =\frac1{``O_i\to 1"}\int \delta\Phi \delta A \, \delta(A_3) \exp\left(i\int_C d^4x\mathscr{L}\right)O_1(x_1) ... O_n(x_n), \label{GreenOt} \ee
where $\delta \Phi$ includes the scalar measure $\delta\varphi$ as well as the Dirac fermion measure $\delta\bar\psi\delta\psi$, also  
\be \delta A \equiv \prod_{x\mu a} dA_\mu^a(x), \qquad \delta(A_3) \equiv \prod_{x a} \delta(A_3^a(x))\ee 
and we have periodic and antiperiodic boundary conditions for bosons and fermions, respectively (see~(\ref{PerCon}) and~(\ref{AntiPpsi})). 
Like in Secs.~\ref{Path integral} and~\ref{Fermionic path integral}, the contour $C$ connects the times $t_0$ and $t_0-i\beta$ and includes the times $x_1^0, ... , x_n^0$; one can choose $C$ as in Fig.~\ref{contour} to work in the real-time formalism or set $t_0=0$ and go from $0$ to $-i\beta$ along the imaginary axis to work in the imaginary-time formalism. 

At this point one can show that the right-hand side of~(\ref{GreenOt}) is a particular case of the Faddeev-Popov path integral
\be \frac1{``O_i\to 1"}\int \delta\Phi \delta A \, \Delta_f(A) \delta(f(A)) \exp\left(i\int_C d^4x\mathscr{L}\right)O_1(x_1) ... O_n(x_n), \label{fullGreen}\ee
where the function $f$ identifies the generic gauge condition through $f_a(A,x)=0$ (for example, in the axial gauge $f_a(A,x) = A^a_3(x)$), 
\be \Delta_f(A) \equiv \left(\int \mathcal{D} U \delta(f(A^U))\right)^{-1}, \quad \delta(f(A))\equiv \prod_{xa}\delta(f_a(A,x)), \ee
$\mathcal{D} U$ is the invariant measure on the group\footnote{For a generic function $F$ on the group $G$ the integral of $F$ on $G$ with the invariant measure is 
\be \int \mathcal{D}U F(U) \equiv \int_G \mu(U) d^nU F(U),\ee
where $d^nU$ should be thought of as the differential of all parameters needed to identify the generic element of the group ($n=\dim G$) and $\mu(U)$ has the property
 \be \int_G \mu(U) d^nU F(U) =\int_G \mu(U) d^nU F(\bar U \cdot U) ,\ee
where $\bar U$ is any fixed element of $G$ and a dot here represents the product between group elements.} and $A^U$ is the gauge field configuration  that is obtained from the gauge field configuration  $A$ acting with an element $U=\exp(i\alpha^aT^a)$ of the gauge group.  Indeed, for $f(A) = A_3$ 
the quantity $\Delta_f(A)$ is independent of $A$ for $A_3=0$ and, therefore, cancels with the same quantity in the denominator, which is indicated in~(\ref{fullGreen}) with $``O_i\to 1"$. So we can work in the axial gauge as well as in other gauges. A choice of gauge is a choice of $f$.

Just like at zero temperature, barring the Gribov ambiguity, we can generically substitute $\Delta_f(A)$ with an integral over Grassmann variables, $c^a$ and $\bar c^a$, known as Faddeev-Popov ghosts:
\be \Delta_f(A) \to \int \delta\bar c\,  \delta c \, \exp\left(i\int_C d^4x d^4y \, \bar c^a(x) M_{ab}(x,y) c^b(y)\right),  \label{GhostInt}\ee
where 
\be M_{ab}(x,y) = \frac{\delta f_a(A^U,x)}{\delta \alpha^b(y)}. \ee 
Since the operator $M_{ab}(x,y)$ acts on a space of periodic functions, the Faddeev-Popov ghosts obey periodic boundary conditions although they are Grassmann variables.

Just like at zero temperature, the propagators of the gauge fields and the Faddeev-Popov ghosts depend on the gauge (on the choice of $f$). One novelty of the finite temperature case is that, like for fermions and scalars, the contour $C$ in Fig.~\ref{contour} leads to a doubling of the degrees of freedom and so a four-component propagator for each field. The form of the four-component  propagators of the gauge fields and the Faddeev-Popov ghosts can be easily obtained (for a given gauge-fixing function $f$) from the zero-temperature expression and the results of Secs.~\ref{Scalar theory} and~\ref{Fermions}. 

Also in a generic gauge theory, like  for pure scalar theories (see the discussion around Eq.~(\ref{ZZ02})),  in the Feynman diagram representation of perturbation theory only vertices of type 1 or type 2 exist (no mixed vertices arise) and the type 2 vertices have sign opposite to the type 1 ones. However, there are propagator components $\{a, b\}=\{1, 2\}$
  and $\{a, b\}= \{2, 1\}$, which mix type 1 and type 2 particles.
  
  We are now able to perform perturbation theory both in the real and imaginary-time formalism. We will apply the results obtained in this section to particle production and phase transitions in Secs.~\ref{Weakly-coupled particle production} and~\ref{Phase transitions in field theory}, respectively.

\newpage
\section{Weakly-coupled particle production}
\label{Weakly-coupled particle production}

Let us  now provide some physical application of the TFT formalism developed so far. Here we discuss the production of particles that are weakly-coupled to a thermal bath, i.e.~a system of other particles in thermal equilibrium. The produced particles are not necessarily in thermal equilibrium as their couplings are small.  
We work as usual in the rest frame of the thermal bath, such that its density matrix is simply $\rho=\exp(-\beta H)/Z$. 

Here we will make use of the real-time formalism, which we have developed in full generality in Sec.~\ref{Thermal Green's functions}. This formalism is indeed the most convenient one to compute particle interaction rates as one does not have to perform an analytic continuation on time.

\subsection{Production of a spin-0 particle}\label{Production of a spin-0 particle}

The first example we illustrate is the production of a weakly-coupled spin-0 particle, described by a real scalar field $\Phi$. In the interaction picture this field can be decomposed like in~(\ref{varphiDec}), using its creation and annihilation operators. The interaction between $\Phi$ and the thermal bath is given by a term $\lambda \Phi O$ in the Lagrangian, where $O$ is a local scalar operator made of the fields in thermal equilibrium and $\lambda$ is a small coupling constant.

The $S$ matrix element for such particle  production is (at leading order in $\lambda$)
\be S_{if}(q)  \simeq i\lambda \int d^4x \, \langle f, q|\Phi(x) O(x) |i\rangle = \frac{i\lambda}{\sqrt{2Vq_0}} \int d^4x \,e^{iqx} \langle f|O(x) |i\rangle,\ee
where $|i\rangle$ and $|f,q\rangle$ are the initial and final states, respectively, we work in the orthonormal basis of eigenstates of $P^\mu$ and $q$ is the four-momentum of the produced particle. The production probability averaged over the initial state and summed over $f$ is
\be  \frac{1}{Z(\beta)}\sum_{if} e^{-\beta E_i} |S_{if}(q)|^2 \simeq \frac{\lambda^2}{2Vq_0Z(\beta)}\sum_{if} e^{-\beta E_i} \int d^4xd^4y \, e^{iq(x-y)} \langle i|O(y)|f\rangle\langle f|O(x)|i\rangle,  \ee
where $E_i$ is the energy of $|i\rangle$, namely $H|i\rangle = E_i|i\rangle$. Since $\lambda$ is small, i.e.~$\Phi$ is weakly-coupled, we can neglect $\lambda$ in everything that multiplies $\lambda^2$ in the expression above and write 
\be\sum_f\langle i|O(y)|f\rangle\langle f|O(x)|i\rangle= \langle i|O(y)O(x)|i\rangle. \ee
Therefore, at leading order in $\lambda$ 
\be  \frac{1}{Z(\beta)}\sum_{if} e^{-\beta E_i} |S_{if}(q)|^2 \simeq  \frac{\lambda^2}{2Vq_0} \int d^4xd^4y \, e^{iq(x-y)}  \langle O(y)O(x)\rangle, \ee
where
\be \langle O(y)O(x)\rangle = \frac{1}{Z(\beta)}\sum_i e^{-\beta E_i} \langle i|O(y)O(x)|i\rangle.\ee
Using now translation invariance $\langle O(y)O(x)\rangle=\langle O(0)O(x-y)\rangle$ one obtains
\be  \frac{1}{Z(\beta)}\sum_{if} e^{-\beta E_i} |S_{if}(q)|^2 \simeq  \frac{\Omega}{2Vq_0} \Pi^<(q), \ee
where $\Omega$ is the spacetime volume and 
\be \Pi^<(q) = \lambda^2 \int d^4 x \, e^{iqx}\langle O(0)O(x)\rangle \ee
is nothing but the non time-ordered self energy of $\Phi$ in momentum space. Therefore, the differential production rate per unit of time and volume is (in the infinite $V$ and $\Omega/V$ limit)
\be \boxed{\frac{d\Gamma}{d^3q} = \frac{\Pi^<(q)}{2(2\pi)^3 q_0}, \qquad \mbox{(at leading order in $\lambda$)}}\ee
while the total production rate per unit of time and volume is 
\be \Gamma = \int d^3 q \, \frac{\Pi^<(q)}{2(2\pi)^3 q_0}.\qquad\mbox{(at leading order in $\lambda$)}\ee
So the production rate can be extracted from the non time-ordered self energy of $\Phi$.

\begin{figure}[t]
\begin{center}
  \vspace{-1.3cm} \includegraphics[scale=0.6]{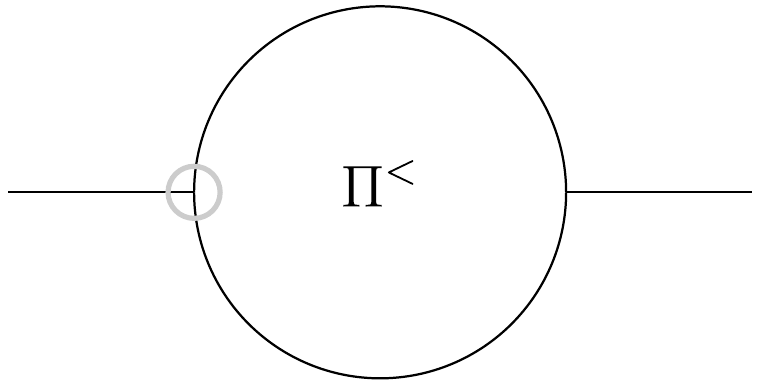}  
  \vspace{-.3cm}
    \caption{\em The diagram corresponding to the non time-ordered self energy of the weakly-coupled particle. The  external vertex on the left is circled, the external vertex on the right is uncircled and one sums over all possible ways of circling the internal vertices.}
\label{SelfBubble}
  \end{center}
\end{figure}

On the other hand, as we have seen in Sec.~\ref{Path integral}, in a generic scalar theory with fields  $\hat\varphi$ the non time-ordered 2-point function can be written as 
\be \langle \hat\varphi(0)\hat\varphi(x)\rangle = \lim_{\sigma \to 0} \langle \mathcal{T} \hat\varphi(x_0,\vec x)\hat\varphi(-i\sigma,0) \rangle, \ee
where $\mathcal{T}$ is the time-ordered product on the contour $C$ in Fig.~\ref{contour}. This is because the point $-i\sigma$ along $C$ is after all points, like $x_0$, on the real axis in the complex time plane. The Green's function $\langle \mathcal{T} \hat\varphi(x_0,\vec x)\hat\varphi(-i\sigma,0) \rangle$ can be obtained  by taking the functional derivatives of the generating functional $\mathcal{Z} (J)$ with respect to $J(x) = J_1(x)$ and $J(-i\sigma, 0) = J_2(0)$, where the notation in~(\ref{J12Def}) has been used. Therefore, $\langle\mathcal{T} \hat\varphi(x_0,\vec x)\hat\varphi(-i\sigma,0) \rangle$ coincides with the component 12 of the full propagator in the theory with degrees of freedom doubled, discussed at the end of Sec.~\ref{Path integral}.
As a result, denoting the vertices of type 1 as uncircled and those of type 2 as circled,  $\Pi^<(q)$ is given by the diagram in Fig.~\ref{SelfBubble}. The bubble in Fig.~\ref{SelfBubble} can contain an arbitrary number of internal vertices and one should sum over all possible ways of circling the internal vertices keeping the  external vertex on the left circled and the external vertex on the right uncircled. One then should use the momentum-space propagators in~(\ref{Prop1122}) and~(\ref{Prop1221}) for $\sigma\to 0$ to connect the various vertices of type 1 and 2 (Kobes-Semenoff rules). 

It is important to note that this result, although only at leading order in $\lambda$, is  valid to all orders (and even non-perturbatively) in the couplings of the thermalized sector other than $\lambda$.

Given the general relation in~(\ref{Gtfb}), which links $\Pi^<$ and $\Pi^>$, and the rule in~(\ref{KobesSemenoff}), it is clear that we can equivalently extract the production rate from the imaginary part of the self energy.

The method illustrated here has been applied to compute the thermal production of some interesting very weakly-coupled spin-0 particles, such as axions\footnote{For a recent application to other beyond-the-SM scalars see~\cite{Hardy:2024gwy}.}~\cite{Graf:2010tv,Salvio:2013iaa,DEramo:2021psx,DEramo:2021lgb,Bouzoud:2024bom}, hypothetical particles~\cite{Weinberg:1977ma} that are the low-energy manifestation of a symmetry (known as the Peccei-Quinn symmetry~\cite{Peccei:1977hh}) able to explain why strong interactions do not violate CP.

\subsection{Production of a massless spin-1 particle}\label{Production of a massless spin-1 particle}

Let us now illustrate the production of a weakly-coupled massless spin-1 particle, e.g.~a photon, described by a gauge field $A_\mu$. In the interaction picture this field can be decomposed like in~(\ref{AmuDec}), using its creation and annihilation operators. The interaction between $A_\mu$ and the thermal bath is given by a term $\lambda A_\mu J^\mu$ in the Lagrangian, where $J^\mu$ is a vector local operator made of the fields in thermal equilibrium and $\lambda$ is a small coupling constant. Note that gauge invariance implies the conservation law $\partial_\mu J^\mu = 0$ (current conservation).

The $S$ matrix element for such spin-1 particle  production with a given polarization, $r$, is (at leading order in $\lambda$)
\be S^{(r)}_{if}(q)  \simeq i\lambda \int d^4x \, \langle f, q|A_\mu(x) J^\mu(x) |i\rangle = \frac{i\lambda}{\sqrt{2Vq_0}} \int d^4x \,e^{iqx} \epsilon_{r\mu}(q)\langle f|J^\mu(x) |i\rangle,\ee
where $|i\rangle$ and $|f,q\rangle$ are the initial and final states, respectively, we work in the orthonormal basis of eigenstates of $P^\mu$ and $q$ is the four-momentum of the produced particle. The production probability averaged over the initial state and summed over $f$ and the polarization, $r$, is
\be  \frac{1}{Z(\beta)}\sum_{if r} e^{-\beta E_i} |S^{(r)}_{if}(q)|^2 \simeq \frac{\lambda^2}{2Vq_0Z(\beta)}\sum_{ir} \epsilon_{r\mu}(q)\epsilon_{r\nu}(q)e^{-\beta E_i} \int d^4xd^4y \, e^{iq(x-y)} \langle i|J^\mu(y)J^\nu(x)|i\rangle.\nonumber  \ee
Like in zero-temperature quantum field theory, using $\partial_\mu J^\mu=0$ we can substitute
\be \sum_r\epsilon_{r\mu}(q)\epsilon_{r\nu}(q) \to - \eta_{\mu\nu} \ee
Therefore, at leading order in $\lambda$ 
\be  \frac{1}{Z(\beta)}\sum_{ifr} e^{-\beta E_i} |S^{(r)}_{if}(q)|^2 \simeq -\eta_{\mu\nu} \frac{\lambda^2}{2Vq_0} \int d^4xd^4y \, e^{iq(x-y)}  \langle J^\mu(y)J^\nu(x)\rangle, \ee
where
\be \langle J^\mu(y)J^\nu(x)\rangle = \frac{1}{Z(\beta)}\sum_i e^{-\beta E_i} \langle i|J^\mu(y)J^\nu(x)|i\rangle.\ee
Therefore, we can write
\be  \frac{1}{Z(\beta)}\sum_{ifr} e^{-\beta E_i} |S_{if}^{(r)}(q)|^2 \simeq   -\eta^{\mu\nu}\frac{\Omega}{2Vq_0} \Pi_{\mu\nu}^<(q), \ee
where $\Omega$ is the spacetime volume and  
\be \Pi_{\mu\nu}^<(q) = \lambda^2 \int d^4 x \, e^{iqx}\langle J_\mu(0)J_\nu(x)\rangle \label{Pimunub}\ee
(translational invariance has been used).
$\Pi_{\mu\nu}^<(q)$ is nothing but the non time-ordered self energy of $A_\mu$ in momentum space. Therefore, the differential production rate per unit of time and volume is (in the infinite $V$ and $\Omega/V$ limit)
\be \boxed{\frac{d\Gamma}{d^3q} = \frac{-\eta^{\mu\nu}\Pi_{\mu\nu}^<(q)}{2(2\pi)^3 q_0}, \qquad \mbox{(at leading order in $\lambda$)}}\ee
while the total production rate per unit of time and volume is 
\be \Gamma = \int d^3 q \, \frac{-\eta^{\mu\nu}\Pi_{\mu\nu}^<(q)}{2(2\pi)^3 q_0}.\qquad\mbox{(at leading order in $\lambda$)}\ee
So the production rate can be extracted from the non time-ordered self energy of $A_\mu$. 

On the other hand, using an argument analogous to the one presented at the end of Sec.~\ref{Production of a spin-0 particle}, and denoting again the  vertices of type 1 as uncircled and those of type 2 as circled,
one also finds  that $\Pi_{\mu\nu}^<(q)$ is given  by the diagram in Fig.~\ref{SelfBubble}, but with the external scalar legs substituted with vector legs. 

Like in Sec.~\ref{Production of a spin-0 particle}, this result, although only at leading order in $\lambda$, is  valid at all orders (and even non-perturbatively)  in the couplings of the thermalized sector other than $\lambda$.

The method illustrated here has been applied to compute the thermal production of some spin-1 particles. These include, of course, photons (see~\cite{Ghiglieri:2020dpq} for a review), but also, more recently, hypothetical particles associated with a somewhat hidden U(1) gauge symmetry\footnote{See also~\cite{Redondo:2008ec,Vogel:2013raa} for related calculations.}~\cite{Salvio:2022hfa}, known as dark photons~\cite{Holdom:1985ag} (see~\cite{ Fabbrichesi:2020wbt} for  a recent review).

\subsection{Production of a spin-1/2 particle}\label{Production of a spin-1/2 particle}

We now turn to the production of a weakly-coupled spin-1/2 particle, such as an electron, described by a fermion field $\psi$. In the interaction picture $\psi$ can be decomposed using the creation and annihilation operators, like in~(\ref{DiracFieldDec}). As an interaction between $\psi$ and the thermal bath we take a term $\lambda \bar\psi O + \lambda^* \bar O \psi$ where  $O$ is a local fermion operator made of the fields in thermal equilibrium and $\lambda$ is a small coupling constant.

Now, at leading order in $\lambda$, the $S$ matrix element for such spin-1/2 particle  production with a given spin state, $r$, is
\be S^{(r)}_{if}(q)  \simeq i\lambda \int d^4x \, \langle f, q|\bar\psi(x) O(x) |i\rangle = \frac{i\lambda}{\sqrt{Vq_0}} \bar u_{qr} \int d^4x \,e^{iqx} \langle f|O(x)|i\rangle\ee
where $|i\rangle$ and $|f,q\rangle$ are the initial and final states, respectively, we work in the orthonormal basis of eigenstates of $P^\mu$ and $q$ is the four-momentum of the produced particle. 
The corresponding production probability averaged over the initial state and summed over $f$ and $r$ is
\bea  \frac{1}{Z(\beta)}\sum_{if r} e^{-\beta E_i} |S^{(r)}_{if}(q)|^2 &\simeq& \frac{|\lambda|^2}{Vq_0Z(\beta)}\sum_{ir}  \bar u_{qr\alpha} u_{qr\beta}e^{-\beta E_i} \int d^4xd^4y \, e^{iq(x-y)} \langle i|\bar O_\beta(y)O_\alpha(x)|i\rangle\nonumber   \\
&=& \frac{|\lambda|^2 \Omega}{Vq_0} \int d^4x \, e^{iqx} \left(\frac{\slashed{q}+m}{2} \right)_{\beta\alpha} \langle \bar O_\beta(0)O_\alpha(x)\rangle,\eea
where Eq.~(\ref{sumuv}) has been used. Thus,  the differential production rate per unit of time and volume this time can be written as (in the infinite $V$ and $\Omega/V$ limit)
\be \boxed{\frac{d\Gamma}{d^3q} = \frac{\Tr\left(\frac{\slashed{q}+m}{2} \Pi^<(q)\right)}{(2\pi)^3 q_0}, \qquad \mbox{(at leading order in $\lambda$)}},\ee
with 
\be  \Pi_{\alpha\beta}^<(q) = |\lambda|^2 \int d^4x \, e^{iqx} \langle \bar O_\beta(0)O_\alpha(x)\rangle. \ee 
The total production rate per unit of time and volume is 
\be \Gamma = \int d^3 q \, \frac{\Tr\left(\frac{\slashed{q}+m}{2} \Pi^<(q)\right)}{(2\pi)^3 q_0}.\qquad\mbox{(at leading order in $\lambda$)}\ee
Like for the spin-0 and spin-1 particle production, we can then make use of the circling notation to compute the differential and total production rates. These rates, although only at leading order in $\lambda$, are  valid at all orders (and even non-perturbatively)  in the couplings of the thermalized sector other than $\lambda$.

The classic applications of these methods include the thermal production of leptons or quarks. More recently, it has been used to compute the interaction rate of (hypothetical) right-handed counterparts of observed neutrinos in the early universe~\cite{Giudice:2003jh,Salvio:2011sf} (see also~\cite{Laine:2011pq} and,  for a review,~\cite{Laine:2022pgk}). This is a key ingredient of a mechanism, known as thermal leptogenesis~\cite{Fukugita}, to explain the observed matter/antimatter asymmetry.

\subsection{Fermion-pair production} 

The formalism of Sec.~\ref{Thermal Green's functions} can also be applied to compute the production of a weakly-coupled pair of fermions of equal mass $m$, e.g. a muon or electron fermion-antifermion pair. We assume that such pair is produced via a massless gauge field $A_\mu$, for example a photon field. We consider  again an interaction  $\lambda A_\mu J^\mu$ in the Lagrangian, where $\lambda$ is a small coupling constant (we will work at leading order in $\lambda$). The fermion pair in question is created by the conserved current $J^\mu$:
\be J^\mu = \bar\psi \gamma^\mu\psi + ...\ee
where $\psi$ is the Dirac field that contain the annihilation operators of the fermion and the creation operators of the antifermion in question, see Eq.~(\ref{DiracFieldDec}), and the dots are possible additional contributions to $J^\mu$ due to other particles.

From four-momentum conservation it follows that the vector boson associated with $A_\mu$ cannot be on shell. This vector boson only virtually mediates the interaction between the fermion pair and other particles of the thermal bath (contained in $J^\mu$). Therefore, the probability amplitude for this process is of order $\lambda^2$.

At the leading non-vanishing order in $\lambda$ the production probability amplitude is, using the Feynman gauge\footnote{Using other gauges, where the gauge-field momentum-space propagator $\tilde G_{\mu\nu}(k)$ differs from that in Feynman gauge by terms proportional to $k_\mu k_\nu$, leads to the same result thanks to current conservation, $\partial_\mu J^\mu=0$.},
\be S^{(r_1r_2)}_{if}(p_1,p_2)  \simeq i\frac{\lambda^2}{(p_1+p_2)^2} \frac{\bar u_{p_1r_1}\gamma^\mu v_{p_2r_2}}{\sqrt{VE_1}\sqrt{VE_2}} \int d^4x \,e^{i(p_1+p_2)x}  \langle f|J_\mu(x) |i\rangle,\ee
where $|i\rangle$ is the initial state, $| f\rangle$ is the result of applying the annihilation operators of the fermion and antifermion on the final state, $p_1, E_1, r_1$ and $p_2, E_2, r_2$ are the momenta, energies and spin states of the fermion and antifermion respectively. 

The production probability averaged over the initial state and summed over $f$ and the spin states is
\be  \frac{1}{Z(\beta)}\sum_{if r_1 r_2} e^{-\beta E_i} |S^{(r_1r_2)}_{if}(p_1,p_2)|^2 \simeq \left(\frac{\lambda}{(p_1+p_2)^2}\right)^2\frac{ \Omega  \, \sigma^{\mu\nu}}{2VE_12VE_2} \, \Pi_{\mu\nu}^<(p_1+p_2),\nonumber  \ee
where $\Pi_{\mu\nu}^<$ is again the non time-ordered self energy of $A_\mu$ in momentum space, Eq.~(\ref{Pimunub}),  
\be \sigma^{\mu\nu} \equiv 4\sum_{r_1 r_2}(\bar u_{p_1r_1}\gamma^\mu v_{p_2r_2})^*\bar u_{p_1r_1}\gamma^\nu v_{p_2r_2}. \label{SpinSum} \ee 
and $\Omega$ is the spacetime volume.
Therefore, the differential fermion-pair production rate per unit of time and volume is given by (in the infinite $V$ and $\Omega/V$ limit)
\be d\Gamma \simeq \left(\frac{\lambda}{(p_1+p_2)^2}\right)^2\frac{d^3p_1d^3p_2}{2(2\pi)^3E_12(2\pi)^3E_2} \, \sigma^{\mu\nu}\Pi_{\mu\nu}^<(p_1+p_2). \label{GammaDiff1}\ee
The spin sum in~(\ref{SpinSum}) can be performed with the help of Eqs.~(\ref{sumuv}), to obtain
\be \sigma^{\mu\nu}= \Tr\left(\gamma^\mu(\slashed{p_1}+m)\gamma^\nu(\slashed{p_2}-m)\right) = 4\left[p_1^\mu p_2^\nu+p_1^\nu p_2^\mu - (p_1 p_2+m^2)\eta^{\mu\nu}\right],\ee
where the following well-known formul\ae~for the traces of two and four gamma matrices  have been used:
\be \Tr(\gamma^\mu\gamma^\nu) = 4\eta^{\mu\nu}, \qquad \Tr(\gamma^\mu\gamma^\alpha\gamma^\nu\gamma^\beta) = 4(\eta^{\mu\alpha}\eta^{\nu\beta}-\eta^{\mu\nu}\eta^{\alpha\beta}+\eta^{\mu\beta}\eta^{\nu\alpha}).\ee

By integrating~(\ref{GammaDiff1}) and inserting $1=\int d^4q \, \delta(q-p_1-p_2)$ one obtains the integrated fermion-pair production rate per unit of time and volume:
\be \Gamma \simeq \lambda^2 \int\frac{d^4q}{(2\pi)^4}\frac{L^{\mu\nu}(q)\Pi_{\mu\nu}^<(q)}{(q^2)^2}, \label{GammaInt}\ee  
where $L^{\mu\nu}(q)$ is the function of the four-momentum $q$ defined by
\be L^{\mu\nu}(q) \equiv (2\pi)^4\int \frac{d^3p_1d^3p_2}{2(2\pi)^3E_12(2\pi)^3E_2} \sigma^{\mu\nu}  \delta(q-p_1-p_2). \label{LmunuDef} \ee 
Since $L^{\mu\nu}(q)$ is  Lorentz covariant, that is for any Lorentz transformation, $\Lambda^{\mu}_{~\nu}$, we have $L^{\mu\nu}(\Lambda q)= \Lambda^{\mu}_{~\rho}\Lambda^{\nu}_{~\sigma} L^{\rho\sigma}(q)$, we can decompose
\be L^{\mu\nu}(q) = a_1(q^2) q^\mu q^\nu + a_2(q^2) q^2 \eta^{\mu\nu},\ee 
where $a_1$ and $a_2$ are dimensionless quantities that can depend on $q$ only through $q^2$. Moreover, one finds $\sigma^{\mu\nu}q_\mu q_\nu = 0$ where $q=p_1+p_2$ so 
\be a_2(q^2) = - a_1(q^2). \ee
To compute the remaining function $a_1$ we note
\be \eta_{\mu\nu} L^{\mu\nu}(q) = -3q^2a_1(q^2), \qquad \eta_{\mu\nu} \sigma^{\mu\nu}= -8(p_1 p_2 +2m^2)\ee   
and so\footnote{The calculation  of $\eta_{\mu\nu} L^{\mu\nu}(q)$ (performing the integral in~(\ref{LmunuDef})) can be easily done by going to the rest frame, $\vec q = 0$, where the integral over $\vec p_2$ simply fixes $\vec p_2 = -\vec p_1$ through $\delta(\vec p_1+\vec p_2)$. The remaining Dirac delta $\delta(q_0-2E_1)$ can then be used to perform the integral over $\vec p_1$ as the integrand only depends on $|\vec p_1|$.} 
\be a_1(q^2) =  \frac{1}{6\pi}\left(1+\frac{2m^2}{q^2}\right)\sqrt{1-\frac{4m^2}{q^2}}.  \ee 

Finally, by inserting in~(\ref{GammaInt}) the expression of $L^{\mu\nu}$ we have just calculated and using the current conservation, $\partial_\mu J^\mu=0$ that implies $q^\mu \Pi_{\mu\nu}^<(q) = 0$, we can find a useful formula for a differential rate in units of $d^4q$:
\be \frac{d\Gamma}{d^4q} \simeq    \frac{\lambda^2}{96\pi^5}\left(1+\frac{2m^2}{q^2}\right)\sqrt{1-\frac{4m^2}{q^2}}\,\frac{-\eta^{\mu\nu}\Pi_{\mu\nu}^<(q)}{q^2}. \label{GammaDiff2}\ee

Therefore, the fermion-pair production can be computed, like the massless spin-1 production, by evaluating $\eta^{\mu\nu}\Pi_{\mu\nu}^<(q)$ along the lines described at the end of Sec.~\ref{Production of a massless spin-1 particle}.

\newpage
\section{Phase transitions in field theory}\label{Phase transitions in field theory}

In this last section we provide an introduction to phase transitions in the framework of thermal field theory, as an application of the formalism of Sec.~\ref{Thermal Green's functions}. We will work as usual in the plasma rest frame. In this case the angular momentum and the four-momentum are conserved.

\subsection{Effective scalar action} 

 In  field theory the role of the order parameter can be played by the expectation value of some fields. Since the angular momentum is conserved the expectation values of vector and fermion fields vanish and we can focus on the expectation values of scalar fields, $\langle\hat\varphi\rangle$. As we will see, these quantities can be computed by solving the field equations generated by a functional of the scalar fields known as the effective (scalar) action $\Gamma(\varphi)$,
 which we will construct shortly.
 
  Here we will make use of the imaginary-time formalism. As discussed in Sec.~\ref{Thermal Green's functions}, this corresponds to choosing a vanishing initial time $t_0=0$ and going  from $0$ to $-i\beta$ along the imaginary axis (as a possible choice for the contour $C$).
   The imaginary-time formalism is particularly convenient because $\langle\hat\varphi\rangle$ is spacetime independent at equilibrium as a consequence of the four-momentum conservation and, therefore, the derivative terms in $\Gamma(\varphi)$ are not relevant in determining $\langle\hat\varphi\rangle$. As a result, using the imaginary-time formalism does not require a final analytic continuation to real time and, as always, unlike the real-time formalism, does not require a doubling of the degrees of freedom either.

   Combining the results of Secs.~\ref{Scalar theory},~\ref{Fermions} and~\ref{Gauge theory},  the generating functional of the thermal Green's functions involving scalar fields only, Eq.~(\ref{ScalarTGF}), can be obtained through a path integral over fermion, vector as well as scalar fields\footnote{In this section we restore $\hbar$ because, at some point, we will perform an expansion in $\hbar$.}:
 \be \mathcal{Z} (J) =\frac1{``J\to 0"} \int \delta\varphi \, e^{-I(\varphi)/\hbar+J\varphi/\hbar}, \label{ZJgen}\ee
 where
 \be e^{-I(\varphi)/\hbar} \equiv e^{-S_s(\varphi)/\hbar} \int \delta\bar\psi\delta\psi \,\delta A \, \Delta_f(A) \delta(f(A)) e^{-S_{fg}(\varphi, \bar\psi, \psi, A)/\hbar}, \label{IntFerGauge}\ee
 $S_s$ is the part of the (classical) Euclidean action $S$ that depends on the scalar fields only and $S_{fg}$ is the remaining piece that depends on the fermions and gauge fields too.
Moreover, we have introduced the convenient notation
 \be J\varphi \equiv \int d^4x J(x) \varphi (x),\ee
 where the integration is performed on the Euclidean spacetime. Since we are dealing  with a generic gauge theory we have included the Faddeev-Popov insertion $\Delta_f(A) \delta(f(A))$ corresponding to a gauge-fixing function $f$ (see Sec.~\ref{Gauge theory}). In~(\ref{ZJgen}) we have turned on the scalar field source only, $J$. One can, of course, perform a more general analysis where the sources for the fermion and gauge fields are turned on too, but since we are here interested in determining $\langle\hat\varphi\rangle$ this setup will be sufficient.

 Moreover, here we have temporarily restored the reduced Planck constant $\hbar$. This is because we will not be able to compute $\Gamma(\varphi)$ exactly, but only when $\hbar$ is small compared to the typical value of $|S|$; in this case we will be able to perform an expansion of $\Gamma(\varphi)$ in powers of $\hbar$ and compute the first terms. We have also conveniently rescaled the source term $J\varphi\to J\varphi/\hbar$: the $n$th functional derivatives of $\mathcal{Z} (J)$ gives in this case the $n$-point thermal Green's functions divided by $\hbar^n$. When working at finite temperature there is, however, a subtlety. The path integral is computed with periodic and antiperiodic boundary conditions for bosons and fermions, respectively (see Eqs.~(\ref{PerCon}) and~(\ref{AntiPpsi})). This is because we are interested in the thermal version of Green's functions, where we compute the trace of the time-ordered product of fields multiplied by $\exp(-\beta H/\hbar)$ and we interpret this operator as the imaginary-time evolution operator corresponding to an imaginary-time translation $\beta$. Therefore, $\beta$ here is not precisely $1/T$ but rather $\hbar/T$.
 
  We will, thus, perform an expansion in powers of $\hbar$ by keeping $\hbar/T$ fixed. As we shall see, this expansion corresponds to taking into account quantum as well as thermal corrections to the action. We now show that such an expansion corresponds to a loop expansion\footnote{In the presence of massive particles $\hbar$ also appears in $S$ in the form $m_i/\hbar$, where $m_i$ is the mass of the $i$th particle. We  will perform an expansion in powers of $\hbar$ by keeping $m_i/\hbar$ as well as $\hbar/T$ fixed.}. 
  Note that the expansion parameter $\hbar$ only appears in $\exp(-S/\hbar+J\varphi/\hbar)$ and so the propagators are multiplied by $\hbar$, while the vertices are divided by $\hbar$. We  can go from a Feynman diagram with a given number of loops to a Feynman diagram with the same external legs but with one more loop in three distinct ways.
 \begin{itemize}
 \item We can add a propagator and no new vertices: this is the case when the loop is obtained by joining together two existing vertices through a new propagator. 
 \item We can add one new vertex and two propagators: this is the case when we join together one existing vertex with a new vertex through a new propagator
 \item We can add two new vertices and three new propagators: this is the case when we join together two new vertices with a new propagator.
 \end{itemize}
 In all these cases the new diagram with an extra loop has an extra factor of $\hbar$ with respect to the old one. Therefore, $\hbar$ plays the role of a loop-counting parameter.
 

 Let us now provide the definition of $\Gamma(\varphi)$. This can be given even non-perturbatively in $\hbar$. One first introduces the new functional $W$ of the source $J$ through
 \be e^{W(J)/\hbar}\equiv \mathcal{Z} (J). \label{Wdef}\ee
 Next, the classical scalar field is defined by
 \be \varphi_c(x) \equiv \frac{\delta W}{\delta J(x)}, \label{varphic} \ee
 which should be thought of as a functional of $J$. Finally, $\Gamma$ is defined by a Legendre transformation
 \be \Gamma(\varphi_c) \equiv J\varphi_c - W(J).\label{GammaDef} \ee 
 In the last term $W$ depends on $J$ and, therefore, on $\varphi_c$, because, as written above, $\varphi_c$ is a functional of $J$.
 
 Now note that
 \be \frac{\langle\hat\varphi\rangle}{\hbar} = \left.\frac{\delta \mathcal{Z}}{\delta J} \right|_{J=0} = e^{W(J)/\hbar}\frac1{\hbar}\left.\frac{\delta W}{\delta J} \right|_{J=0}\ee
 and so, using $W(0) =0$, 
 \be \langle\hat\varphi\rangle =  \varphi_c |_{J=0}. \label{VEVJ0}\ee
 On the other hand, by using the definition of $\varphi_c$ and $\Gamma$ given above, one finds
 \be \frac{\delta \Gamma(\varphi_c)}{\delta \varphi_c(x)} = J(x). \ee
 Therefore, looking at~(\ref{VEVJ0}), one sees that the expectation value we are interested in, $\langle\hat\varphi\rangle$, is a stationary point of the effective action. In other words, $\langle\hat\varphi\rangle$ is a solution of the field equations generated by $\Gamma$,
 \be \frac{\delta \Gamma(\varphi_c)}{\delta \varphi_c(x)}(\langle\hat\varphi\rangle) = 0 \ee  This is why $\Gamma$ is called the effective action.  
  
 It is interesting to find the relation between $\Gamma(\varphi_c)$ and the free energy $F$, which we define through 
 \be Z\equiv e^{-\beta F/\hbar}, \label{ZFlink}\ee 
  where $Z$ is the partition function and $\beta= \hbar/T$. More precisely, the quantity $F$ defined above  is  known as the Helmholtz free energy.
  Combining~(\ref{Wdef}) and~(\ref{GammaDef}) one finds  $$\mathcal{Z} (J) =\exp(J\varphi_c/\hbar - \Gamma(\varphi_c)/\hbar).$$ On the other hand, the partition function in the presence of the external source  $J$ is
 \be Z_J = \mathcal{C}\int \delta \varphi \, e^{-I(\varphi)/\hbar+J\varphi/\hbar}, \label{ZJpart}\ee
 where $\mathcal{C}$ is a $J$-independent constant that appears when performing the integration on the conjugate momentum fields in the bosonic sector. So, we find that $\mathcal{Z} (J)$ and $Z_J$ are equal  up to $J$-independent factors and, therefore,
 \be F(\varphi_c)= \frac1{\beta} (\Gamma(\varphi_c)- J\varphi_c) +... \label{FGammaLink} \ee
 where the dots are $\varphi_c$-independent terms
 (recall that $J$ is related to $\varphi_c$ through Eq.~(\ref{varphic})). 
Since the system  minimizes $F$, 
according to~(\ref{FGammaLink}), 
we should look for a configuration $\varphi_c$ that minimizes $\Gamma(\varphi_c)- J\varphi_c$. Ultimately we are interested in the partition function without  the external source; therefore, setting  $J=0$, we should minimize $\Gamma$ to find the most favourable $\langle\hat\varphi\rangle$ at a given temperature. 
  This result will be important in determining $\langle\hat\varphi\rangle$ for various types of phase transitions in Secs.~\ref{first-order phase transitions} and~\ref{Second-order phase transitions} and in understanding the phenomenon of thermal vacuum decay in first-order phase transitions in Sec.~\ref{Thermal vacuum decay in first-order phase transitions}.

  
 Let us now turn to the perturbative expansion. When $\hbar$ is treated as a small parameter, Eq.~(\ref{ZJgen}) shows that the path integral is dominated by the scalar field configurations that correspond to the minimum of the functional $I(\varphi)-J\varphi$. These configurations satisfy
 \be \frac{\delta I}{\delta\varphi} = J  \label{IvarphiJ}\ee
 and, therefore, depend on $J$. The solution of this equation corresponding to the minimum of $I(\varphi)-J\varphi$ will be denoted $\varphi_J$.
 We can, therefore, expand the exponent in~(\ref{ZJgen}) in powers of the new integration variable $\eta\equiv \varphi-\varphi_J$. The Jacobian of this change of integration variable is 1 so
 \be \mathcal{Z} (J) = \frac{1}{``J\to 0"}\exp\left(-I(\varphi_J)/\hbar+J\varphi_J/\hbar\right) \int \delta\eta \, \exp\left( -\frac1{2\hbar} \eta\, \left.\frac{\delta^2 I}{\delta \varphi^2}\right|_{\varphi_J} \eta + ...\right), \label{ZJsemi} \ee
 where 
 \be \eta\, \left.\frac{\delta^2 I}{\delta \varphi^2}\right|_{\varphi_J} \eta \equiv \int d^4x d^4y \left.\frac{\delta^2 I}{\delta\varphi(x)\delta\varphi(y)}\right|_{\varphi=\varphi_J}\eta(x) \eta(y) \ee
 and the dots represent terms with more than two $\eta$s, which are subleading in the expansion we are performing. Note that we are interested in the functional derivatives of $\mathcal{Z} (J)$ with respect to $J$ evaluated at $J=0$ (see Eq.~(\ref{ZJ0})). Therefore, we are interested in $\varphi_J$ for infinitesimal values of $J$, which is very close to the solution of $\frac{\delta I}{\delta\varphi} =0$. By using the definition of $W(J)$ and performing the Gaussian integral in~(\ref{ZJsemi}) one obtains
 \be W(J) = -I(\varphi_J)+I(\varphi_0) +J\varphi_J -\frac{\hbar}{2}\log\left(\det\left(\left.\frac{\delta^2 I}{\delta \varphi^2}\right|_{\varphi_J}\right)/``J\to 0"\right) + ...\, . \label{WJexp}\ee
 Here $\varphi_0$ is $\varphi_J$ at $J=0$, namely the solution of $\frac{\delta I}{\delta\varphi}=0$ (see Eq.~(\ref{IvarphiJ})). From the definition of $\varphi_c$ in~(\ref{varphic}) then it follows
 \be \varphi_c(x) \equiv \frac{\delta W}{\delta J(x)} = - \frac{\delta I(\varphi_J)}{\delta J(x)} +\varphi_J(x) + J \frac{\delta\varphi_J}{\delta J(x)}-\frac{\hbar}{2}\frac{\delta}{\delta J(x)}\log\left(\det\left(\left.\frac{\delta^2 I}{\delta \varphi^2}\right|_{\varphi_J}\right)/``J\to 0"\right) + ... \, . \label{varphicJstep}\ee
 On the other hand, using~(\ref{IvarphiJ})
 \be \frac{\delta I(\varphi_J)}{\delta J(x)} = \int d^4 y\,   \frac{\delta I(\varphi_J)}{\delta\varphi_J(y)}\frac{\delta\varphi_J(y)}{\delta J(x)}  = \int d^4 y\,   J(y)\frac{\delta\varphi_J(y)}{\delta J(x)} \equiv J  \frac{\delta\varphi_J}{\delta J(x)}\ee
so, inserting in~(\ref{varphicJstep}),
\be \varphi_c(x) = \varphi_J(x) -\frac{\hbar}{2}\frac{\delta}{\delta J(x)}\log\left(\det\left(\left.\frac{\delta^2 I}{\delta \varphi^2}\right|_{\varphi_J}\right)/``J\to 0"\right) + ... \, .  \ee 
 This result tells us, among other things, that the difference between $\varphi_c$ and $\varphi_J$ is of order $\hbar$.  So, using~(\ref{GammaDef}) and~(\ref{WJexp}),
 \be \Gamma(\varphi_c) =J\varphi_c+ I(\varphi_c +(...))-I(\varphi_0) -J(\varphi_c+(...)) +\frac{\hbar}{2}\log\left(\det\left(\left.\frac{\delta^2 I}{\delta \varphi^2}\right|_{\varphi_c}\right)/``\varphi_c\to \varphi_0"\right) + ...\,  \ee 
 and noting 
 \be  I(\varphi_c +(...)) =  I(\varphi_c)+\frac{\delta I}{\delta\varphi_c}(...) = I(\varphi_c)+ \frac{\delta I}{\delta\varphi_J}(...)+... = I(\varphi_c)+ J(...)+...\, ,\ee
 where the dots in the brackets are $\mathcal{O}(\hbar)$, while the other dots are $\mathcal{O}(\hbar^2)$, one obtains (renaming for simplicity $\varphi_c\to \varphi$) 
 \be \Gamma(\varphi) = I(\varphi)+\frac{\hbar}{2}\log\left(\det\left(\frac{\delta^2 I}{\delta \varphi^2}\right)\right) -``\varphi\to \varphi_0"+ ... \, .  \label{GammaI1} \ee

 Note that $I(\varphi)= S_s(\varphi) + I_{fg}(\varphi)$, where  $I_{fg}$ is the contribution of the integration over fermion and gauge fields, see Eq.~(\ref{IntFerGauge})).  
We will compute this contribution at one-loop level for a generic renormalizable gauge theory (but, for simplicity, we only include Dirac masses for fermions here\footnote{In Sec.~\ref{Fermion one-loop effective potential} we will include both Dirac and Majorana masses for completeness.}), 
 \be S = \int d^4x\left(\frac12 D_\mu\varphi D_\mu\varphi+\mathcal{V}(\varphi)-\bar\psi(i\slashed{D}-\mu_F(\varphi))\psi +\frac14 F^a_{\mu\nu}F_{\mu\nu}^a\right),\ee
 where here $\mathcal{V}$ represents the tree-level (zero-loop) scalar potential, 
 \be\mu_F(\varphi) \equiv (m+y\varphi)P_L+(m^\dagger+y^\dagger\varphi)P_R,\ee
  $m$ is the fermion mass matrix, $y$ represents the Yukawa matrices and $P_L$ and $P_R$ are the projectors on the left and right-handed components, respectively. Moreover, $D_\mu\varphi$ and $D_\mu\psi$ are the covariant derivatives of scalars and fermions, respectively, and $\slashed{D}=\gamma_\mu D_\mu$ where here $\gamma_\mu$ are the Euclidean gamma matrices that satisfy $\{\gamma_\mu,\gamma_\nu\}=-2\delta_{\mu\nu}$. Note that for these theories
 \be S_s(\varphi)=   \int d^4x\left(\frac12 \partial_\mu\varphi \partial_\mu\varphi+\mathcal{V}(\varphi) \right)\label{SsRen}\ee
 and, as always, $S_{fg} = S-S_s$.  The action $S$ appears in the path integral divided by $\hbar$ so it is convenient to rescale both the fermion and gauge fields, $\psi\to\sqrt{\hbar}\psi$, $\bar\psi\to\sqrt{\hbar}\bar\psi$, $A^a_\mu\to \sqrt{\hbar}A^a_\mu$, as well as the ghost fields, $c\to\sqrt{\hbar}c$, $\bar c\to\sqrt{\hbar}\bar c$ (see Eq.~(\ref{GhostInt})). These rescalings produce in the integration measures the same rescalings, which, however, cancel with the denominator $``J\to0"$  in~(\ref{ZJgen}). Forthermore, from~(\ref{IntFerGauge}) it is clear that at one-loop level (up to $\mathcal{O}(\hbar)$ in $\Gamma(\varphi)$) one can consider only the pieces quadratic in $\psi$, $\bar\psi$ and $A^a_\mu$ in $S$ and, as long as the gauge-fixing function $f$ is independent of $\varphi$, one can neglect the ghost contribution.

 Let us consider now the integration over $\psi$ and $\bar\psi$ in~(\ref{IntFerGauge}). By recalling that the integration over the Grassmann variables $\psi$ and $\bar\psi$ produces a determinant of the operator between $\bar\psi$ and $\psi$ in $S$, this leads to the following  fermion one-loop contribution to $I(\varphi)-I(\varphi_0)$:
 \be \Delta I_F(\varphi) = -\hbar \log\left(\frac{\det(i\slashed{\partial}-\mu_F(\varphi))}{\det(i\slashed{\partial}-\mu_F(\varphi_0))}\right). \label{FerContI} \ee 
Note that $\mu_F(\varphi)$ plays the role of a background-dependent fermion mass matrix. Also, at this order, $\varphi_0$ in the expression above can be computed as the solution of the classical scalar field equations, $\frac{\delta S_s}{\delta\varphi} = 0$.
 
 It is generically hard to compute the determinant in~(\ref{FerContI}) for a generic spacetime-dependent scalar background. However, we are interested here in spacetime-independent scalar backgrounds because spacetime translation invariance is preserved. We will compute $\Delta I_F(\varphi)$ for spacetime-independent $\varphi$ and $\varphi_0$ in Sec.~\ref{Effective potential}. However, it should be kept in mind that when one includes spacetime-dependent backgrounds, which is, for example, needed to study thermal vacuum decay (the subject of Sec.~\ref{Thermal vacuum decay in first-order phase transitions}) a more general calculation is generically required.
 
 \subsection{Effective potential}\label{Effective potential}
 
 The effective potential is defined as 
 \be V_{\rm eff}(\varphi) \equiv \frac1{\beta V}\Gamma(\varphi), \quad \mbox{for a spacetime-independent $\varphi$.} \label{GammaToV}\ee
  The denominator $\beta V$ (the four-dimensional spacetime volume) has been introduced to cancel the integration over $d^4x$ in $\Gamma(\varphi)$.
  
  \subsubsection{Fermion one-loop effective potential}\label{Fermion one-loop effective potential}
  
  Let us start by computing the one-loop fermion contribution to the effective potential. Taking $\varphi$ and $\varphi_0$ to be spacetime independent, Eq.~(\ref{FerContI}) can be written
\be\hspace{-0.1cm} \Delta I_F(\varphi) = -\frac{\hbar}{2} \log\left(\frac{\det(i\slashed{\partial}-\mu_F(\varphi))}{\det(i\slashed{\partial}-\mu_F(\varphi_0))}\frac{\det(i\slashed{\partial}+\mu^\dagger_F(\varphi))}{\det(i\slashed{\partial}+\mu^\dagger_F(\varphi_0))}\right) = -\frac{\hbar}{2} \log\left(\frac{\det(-\partial^2+\mu_F(\varphi)\mu_F^\dagger(\varphi))}{\det(-\partial^2+\mu_F(\varphi_0)\mu^\dagger_F(\varphi_0))}\right), \nonumber \ee
where we have used the fact that exchanging left and right-handed spinors corresponds to replacing $\mu_F$ with $\mu_F^\dagger$, the sign of fermion masses can be changed by an appropriate chiral transformation on the fermion fields 
 and we defined $\partial^2\equiv  \partial_\mu\partial_\mu$. Using now the formula $\log\det(...) = \Tr\log(...)$, 
\be \Delta I_F(\varphi) = -\frac{\hbar}{2} \Tr\log\left(-\partial^2+M_F^2(\varphi)\right) - ``\varphi\to\varphi_0",\ee
where
\be M_F^2(\varphi) \equiv \mu_F(\varphi)\mu_F^\dagger(\varphi). \ee 
Recalling  that the fermion fields satisfy antiperiodic boundary conditions and working in a finite space volume $V$, we obtain
\be  \Delta I_F(\varphi) =-\frac{\hbar}{2} V\sum_f n_f \sum_{n=-\infty}^{+\infty} \int \frac{d^3p}{(2\pi)^3} \log\left(p_n^2+m_f^2(\varphi)\right)- ``\varphi\to\varphi_0", \label{DeltaIfS} \ee
where $n_f$ is the spinor dimension ($n_f=2,4$ for a  Weyl\footnote{The Weyl case will be discussed at the end of this subsection.} and Dirac spinor, respectively), the $m_f^2(\varphi)$ are the non-negative eigenvalues of the matrix $M_F^2(\varphi)$, known as the background-dependent (fermion) squared masses,
  and $p_n^2\equiv\omega_n^2+\vec p^{\, 2}$, with 
\be \omega_n=\frac{(2n+1)\pi}{\beta}.\ee
The $\omega_n$ are known as the Matsubara frequencies (for fermions). 
  Using now~(\ref{GammaToV}), the fermion contribution 
to the effective potential reads
\be V^F_{\rm eff}(\varphi) = -\frac{\hbar}{2\beta}\sum_f n_f \sum_{n=-\infty}^{+\infty} \int \frac{d^3p}{(2\pi)^3} \log\left(p_n^2+m_f^2(\varphi)\right)- ``\varphi\to\varphi_0",  \label{VeffF}\ee

The sum over $n$ in~(\ref{VeffF}) converges thanks to the subtraction of the $``\varphi\to\varphi_0"$ term. To compute that sum  
we introduce a variable $y\equiv \beta \omega/(2\pi)$, where $\omega$  is defined  by \be\omega\equiv \sqrt{\vec p^{\, 2}+m_f^2(\varphi)}.\ee
Then, calling $y_0$ the value of $y$ at $\varphi=\varphi_0$, the sum over $n$ in~(\ref{VeffF}) reads\footnote{The third step in~(\ref{ProofSerF}) can be understood as follows. The function of the real variable $x$ 
 \be C_m(x)= \prod_{n=0}^m \left[1+\left(\frac{x}{\pi (n+1/2)}\right)^2\right], \ee
 where $m\in \mathbb{N}$, is a polynomial having all the first $2m+2$ zeros of $\cosh x$
 (that is $x =i (n+1/2)\pi$ with $n=-m-1,-m, ..., 0, 1, 2, ... ,  m$) and $C_m(x)= \cosh x +{\cal O}(x^2)$. 
 On the other hand, the Taylor polynomial of $\cosh x$ of degree $2m+2$ has approximately the same roots for large $m$ (with an error that goes to zero as $m\to \infty$) and converges to $\cosh x$ for any $x$ so
 \be\lim_{m\to \infty}C_m(x) = \cosh x.\ee
 }
\bea &&\sum_{n=-\infty}^{+\infty} \log\left((n+1/2)^2+y^2\right)- ``y\to y_0"= 
  2\sum_{n=0}^{+\infty} \log\left((n+1/2)^2+y^2\right) - ``y\to y_0" \nonumber \\
&&= 2\log\prod_{n=0}^{+\infty}\left(1+\frac{y^2}{(n+1/2)^2}\right)- ``y\to y_0" =  2\log\left(\cosh(\pi y)\right)- ``y\to y_0"\nonumber \\
&& = 2\pi y +2 \log (1+e^{-2\pi y})- ``y\to y_0". \label{ProofSerF}\eea
By inserting now in~(\ref{VeffF}), one obtains
\be V^F_{\rm eff}(\varphi) = -\hbar\sum_f n_f   \int \frac{d^3p}{(2\pi)^3} \left(\frac{\omega}{2}+\frac1{\beta}\log\left(1+e^{-\beta\sqrt{\vec p^{\, 2}+m_f^2(\varphi)}}\right)\right)- ``\varphi\to\varphi_0".  \label{VeffF2}\ee
The first term in the expression above is the known zero-temperature quantum correction due to fermion fields 
\be -\frac{\hbar}{2}\sum_f n_f   \int \frac{d^3p} {(2\pi)^3} \sqrt{\vec p^{\, 2}+m_f^2(\varphi)}- ``\varphi\to\varphi_0", \label{VFq} \ee
which, as well-known, needs the renormalization procedure: one can render it finite by appropriately redefining the coefficients in the tree-level potential and rescaling the fields as in standard zero-temperature QFT. The second term proportional to $1/\beta$ in~(\ref{VeffF2}) is the thermal correction, which is instead finite and  can be written as follows
\be -\frac{\hbar}{2\pi^2\beta^4}\sum_f  n_f  J_F(m_f^2(\varphi)\beta^2)  - ``\varphi\to\varphi_0", \label{VFT}\ee
where
\be J_F(x)\equiv \int_0^\infty dp\, p^2 \log\left(1+e^{-\sqrt{p^2+x}}\right). \label{JFdef} \ee
 Recalling that $\beta=\hbar/T$, which is kept fixed in our expansion, we see that the one-loop thermal correction does not vanish in the classical limit. Therefore, our loop expansion also contains classical thermal corrections. It must be noted, however, that at finite temperature we cannot generically obtain the leading classical correction only from the one-loop part of the effective potential because the higher-loop contributions can also contain classical corrections.

 Let us conclude the section on the fermion contribution with a discussion of the Weyl (two-component) fermion formalism as an alternative to the Dirac (four-component) fermion formalism, which we have used so far. The Weyl formalism is more convenient to describe Majorana masses, which can appear in a given model in addition to Dirac masses. An example is the SM extension obtained by adding three right-handed neutrinos~\cite{Asaka:2005pn,Salvio:2015cja}; given that these neutrinos do not transform under any gauge symmetry, they can feature Majorana masses.
 
 With the Weyl-fermion formalism the relevant fermion Euclidean action is 
 \be S_f(\varphi,\bar\psi,\psi) = -\int d^4x \left(\bar
 \psi i \slashed{\partial}\psi+\frac12 \left(\psi \mu_F(\varphi)\psi +\mbox{h.c.}\right)\right), \ee
 where we have neglected the gauge fields in the  fermion action because they are irrelevant at the one-loop level, as discussed above. 
In the Weyl formalism we adopt the following notation.
\begin{itemize}
\item $\psi$ and $\bar\psi$ are two-component spinors. Spinors on the left have upper spinor indices, $\psi^\alpha$, $\bar\psi^\alpha$, while spinors on the right have lower spinor indices, $\psi_\alpha$, $\bar\psi_\alpha$ and $\psi^\alpha\equiv \psi_\beta \epsilon^{\beta\alpha}$, $\bar\psi_\alpha\equiv \epsilon_{\alpha\beta}\bar\psi^\beta$, where $\epsilon^{\alpha\beta}$ and $\epsilon_{\alpha\beta}$ are the antisymmetric symbols with $\epsilon^{12}=1$ and $\epsilon_{12}=-1$, such that $\epsilon^{\alpha\beta}\epsilon_{\beta\gamma} = \delta^\alpha_{~\gamma}$.
\item The kinetic term $\bar
 \psi i \slashed{\partial}\psi$ is now constructed with the $2\times 2$ matrices $\bar\sigma^\mu$ (where in Euclidean spacetime $\{\bar{\sigma}^{\mu}\}\equiv (i, -\vec{\sigma})$ and $\vec{\sigma}$ represents the three Pauli matrices) as 
 \be\bar
 \psi i \slashed{\partial}\psi = \bar
 \psi i \bar\sigma^\mu \partial_\mu\psi =\bar\psi^\alpha i\bar\sigma^{\mu~\beta}_{~\alpha} \partial_\mu\psi_\beta. \ee
  Neglecting boundary terms
 \be \int d^4x   \, \bar
 \psi i \slashed{\partial}\psi = \frac12  \int d^4x   \, \bar
 \psi i \slashed{\partial}\psi +
 \psi i \slashed{\partial}\bar \psi ,  \ee 
 where 
 \be \psi i \slashed{\partial}\bar \psi = \psi i \sigma^\mu \partial_\mu \bar\psi=  \psi^\alpha i \sigma^{\mu~\beta}_{~\alpha}\partial_\mu \bar\psi_\beta \ee
 and $\{\sigma^{\mu}\}\equiv (i, \vec{\sigma})$. Here we have used  $\sigma_2 \left(\bar \sigma^\mu\right)^T \sigma_2 = \sigma^\mu$. 
 \item Finally,
 \be \mu_F(\varphi) \equiv m + y \varphi,\label{muFWeyl}\ee
 where $m$ is the fermion mass matrix and $y$ represents the Yukawa matrices in this notation and 
 \be \psi \mu_F(\varphi)\psi \equiv  \psi_{\beta i}\epsilon^{\beta\alpha} \mu_F^{ij} \psi_{\alpha j}, \qquad (\psi \mu_F(\varphi)\psi)^\dagger = \bar\psi_{\alpha j} (\mu_F^{ij})^* \epsilon^{\beta\alpha}\bar\psi_{\beta i}\equiv \bar\psi \mu_F^\dagger \bar\psi. \ee
 \end{itemize}
 The matrix $\mu_F(\varphi)$ generically has complex elements, but can be taken to be symmetric without loss of generality. Note that we can put  $\mu_F(\varphi)$ in diagonal form through a unitary transformation\footnote{This is known as the complex Autonne-Takagi factorization, see e.g.~\cite{Youla}.} acting on $\psi$ (this transformation can also have unit determinant  if one does not require any reality condition on its elements and leave the fermion measure invariant). We then  work with a field basis where $\mu_F$ is diagonal.

  With this notation 
  \be S_f(\varphi,\bar\psi,\psi) = -\frac12 \int d^4x\ba {cc} \left(\psi\right. & \left.\bar\psi \right) \\ \phantom{d} & \phantom{f} \ea  \left(\ba {cc} \mu_F(\varphi) & i\sigma^\mu\partial_\mu \\ i\bar\sigma^\mu\partial_\mu & \mu^\dagger_F(\varphi) \ea \right)  \left( \bac \psi \\
\bar\psi\ea \right) \label{SfOp}\ee 
and   the fermion one-loop contribution to $I(\varphi)-I(\varphi_0)$ is given by :
 \be \Delta I_F(\varphi) = - \hbar \left(\log \int  \delta\bar\psi\delta\psi \, e^{-S_f(\varphi,\bar\psi,\psi)} - ``\varphi\to\varphi_0"\right).\ee
 The Grassmann integral above has the form 
 \be \int d^{2m}\eta \, e^{-\frac12 \eta_i A_{ij} \eta_j}, \ee 
 where the $\eta_i$ are $2m$ Grassmann variables ($m$ is a positive integer) and $A_{ij}=-A_{ji}$. According to a well-known theorem (see again e.g.~\cite{Youla}) there exists a unitary matrix $U$ such that the  matrix $A$ can be expressed as $A= U^T \mathcal{A} U$ where $\mathcal{A}$ is a block diagonal matrix $\mathcal{A} = \mbox{diag}\left(A_1, A_2, ... , A_m\right)$ and the single block is a $2\times 2$ antisymmetric matrix
 \be A_i = \left(\ba {cc} 0 & a_i \\ -a_i & 0 \ea\right). \ee 
 If one allows the $a_i$ to be complex one can impose $\det U=1$.  The Jacobian of the transformation $U\eta = \eta'$ is then 1 and we can easily perform the integration on the Grassmann  variables to obtain
  \be \int d^{2m}\eta \, e^{-\frac12 \eta_i A_{ij} \eta_j}  = \prod_i a_i  = (\det \mathcal{A})^{1/2}=(\det A)^{1/2}.\ee
In our case this leads to
 \be \Delta I_F(\varphi) = -\frac{\hbar}2 \log\left(\frac{\det(-\partial^2+M_F^2(\varphi))}{\det(-\partial^2+M_F^2(\varphi_0))}\right) \label{DeltaFW} \ee 
 where
\be M_F^2(\varphi) \equiv \mu_F(\varphi)\mu_F^\dagger(\varphi), \ee 
with $\mu_F(\varphi)$ defined in~(\ref{muFWeyl}),
and we used  the fact that  $\sigma^\mu\partial_\mu$ and $\bar\sigma^\mu\partial_\mu$ commute
 \be [\sigma^\mu\partial_\mu,\bar \sigma^\nu\partial_\nu ] = [\sigma^\mu,\bar \sigma^\nu]\partial_\mu\partial_\nu =[\sigma^i,\bar \sigma^j]\partial_i\partial_j = 2\epsilon_{jik} \sigma^k\partial_i\partial_j= 0.\ee 
 and
 \be \sigma^\mu\bar\sigma^\nu+ \sigma^\nu\bar\sigma^\mu  =-2\delta^{\mu\nu}\ee 
 (recall that we work in Euclidean spacetime).
 Note that the expression in~(\ref{DeltaFW}) is formally equal to the corresponding expression found at the beginning of this section in the four-component formalism, although the definition of $\mu_F(\varphi)$ is different here. Then, from now on, we can perform the same steps done in the four-component formalism. Therefore, one finds the fermion contribution to the effective potential given in Eqs.~(\ref{VFq}) and~(\ref{VFT}), where now $n_f=2$ and the background-dependent squared masses $m_f^2(\varphi)$ are the non-negative eigenvalues of $\mu_F(\varphi)\mu_F^\dagger(\varphi)$, with $\mu_F(\varphi)$ defined in~(\ref{muFWeyl}). 

\subsubsection{Gauge one-loop effective potential}

We now turn to the one-loop gauge field contribution to the effective potential. To compute this contribution  it is convenient to adopt a $\xi$-dependent gauge: one first chooses the gauge-fixing function $f(A,x)=\partial_\mu A^a_\mu(x)+\tau(x)$, where $\tau$ is some fixed function on the spacetime and then one performs a path integration over $\tau$ with weight $\exp(-\int d^4x \, \tau^2(x)/(2\xi))$. For $\xi\to0$ one recovers the Landau gauge, $\partial_\mu A^a_\mu=0$. The gauge contribution $\Delta I_G(\varphi)$ to $I(\varphi)-I(\varphi_0)$ can then be obtained as
\be e^{-\Delta I_G(\varphi)/\hbar} =\frac1{``\varphi\to\varphi_0"} \int \delta A \, \exp\left(\int d^4x\left[ -\frac{\partial_\mu A^a_\nu\partial_\mu A^a_\nu}{2} +\frac{\xi-1}{2\xi} (\partial_\mu A^a_\mu)^2 - \frac{M^2_{ab}(\varphi)}{2} A_\mu^aA_\mu^b\right]\right), \nonumber  \ee
where $M^2_{ab}(\varphi)$ are the elements of the background dependent gauge-field squared-mass matrix, $M_G^2(\varphi)$, which is positively defined. 
In the Landau gauge, where $\partial_\mu A^a_\mu=0$, one obtains
\be \Delta I_G(\varphi) = \frac{3\hbar}{2}\log\det(-\partial^2+M^2_G(\varphi))- ``\varphi\to\varphi_0", \ee
where the factor $3$ comes from the fact that a massive vector field has $3$ degrees of freedom.

Recalling now that the gauge fields satisfy periodic boundary conditions and working in a finite space volume $V$, we obtain
\be  \Delta I_G(\varphi) =\frac{3\hbar}{2} V\sum_g  \sum_{n=-\infty}^{+\infty} \int \frac{d^3p}{(2\pi)^3} \log\left(p_n^2+m_g^2(\varphi)\right)- ``\varphi\to\varphi_0", \label{DeltaIGS} \ee
where $p_n^2=\omega_n^2+\vec p^{\,2}$ with, for gauge fields, 
\be \omega_n=\frac{2n\pi}{\beta},\ee
and the $m_g^2(\varphi)$ are the eigenvalues of the matrix $M_G^2(\varphi)$, known as the background-dependent (vector) squared masses. The $\omega_n$ are called Matsubara frequencies (for bosons). Using now~(\ref{GammaToV}), the gauge-field contribution 
to the effective potential reads
\be V^G_{\rm eff} = \frac{3\hbar}{2\beta}\sum_g \sum_{n=-\infty}^{+\infty} \int \frac{d^3p}{(2\pi)^3} \log\left(p_n^2+m_g^2(\varphi)\right)- ``\varphi\to\varphi_0".  \label{VeffG}\ee
The sum over $n$ in~(\ref{VeffG}) again converges thanks to the subtraction of the $``\varphi\to\varphi_0"$ term. To compute that sum  
we introduce a variable $y\equiv \beta \omega/(2\pi)$, where, for gauge fields, $\omega$  is defined  by \be\omega\equiv \sqrt{\vec p^{\, 2}+m_g^2(\varphi)}.\ee
Then, calling $y_0$ the value of $y$ at $\varphi=\varphi_0$, the sum over $n$ in~(\ref{VeffG}) reads\footnote{The third step in~(\ref{ProofSerG}) can be understood as follows. The function of the real variable $x$ 
 \be S_m(x)= x\prod_{n=1}^m \left[1+\left(\frac{x}{\pi n}\right)^2\right], \ee
 where $m\in \mathbb{N}$, is a polynomial having all the first $2m+1$ zeros of $\sinh x$
 (that is $x =i n\pi$ with $n=0, \pm1, \pm 2, ... \pm m$) and $S_m(x)= \sinh x +{\cal O}(x^3)$. 
 On the other hand, the Taylor polynomial of $\sinh x$ of degree $2m+1$ has approximately the same roots for large $m$ (with an error that goes to zero as $m\to \infty$) and converges to $\sinh x$ for any $x$ so
 \be\lim_{m\to \infty}S_m(x) = \sinh x.\ee
 }
\bea &&\sum_{n=-\infty}^{+\infty} \log\left(n^2+y^2\right)- ``y\to y_0"=2\log y+2\sum_{n=1}^{+\infty} \log\left(n^2+y^2\right) - ``y\to y_0" \nonumber \\
&&= 2\log y+2\log\prod_{n=1}^{+\infty}\left(1+\frac{y^2}{n^2}\right)- ``y\to y_0" = 2\log y+2\log\left(\frac{\sinh(\pi y)}{\pi y}\right)- ``y\to y_0"\nonumber \\
&& = 2\log\left(\sinh(\pi y)\right)- ``y\to y_0" = 2\pi y +2 \log (1-e^{-2\pi y})- ``y\to y_0". \label{ProofSerG}\eea
%
So, by inserting in~(\ref{VeffG}) one obtains
\be V^G_{\rm eff}(\varphi) = 3\hbar\sum_g   \int \frac{d^3p}{(2\pi)^3} \left(\frac{\omega}{2}+\frac1{\beta}\log\left(1-e^{-\beta\sqrt{\vec p^{\, 2}+m_g^2(\varphi)}}\right)\right)- ``\varphi\to\varphi_0".  \label{VeffF2}\ee
The first term in the expression above is the known zero-temperature quantum correction due to gauge fields 
\be \frac{3\hbar}{2}\sum_g   \int \frac{d^3p} {(2\pi)^3} \sqrt{\vec p^{\, 2}+m_g^2(\varphi)}- ``\varphi\to\varphi_0", \ee
which, just like the fermion contribution, needs the renormalization procedure. The second term proportional to $1/\beta$ is the thermal correction, which is instead finite and can be written as follows
\be \frac{3\hbar}{2\pi^2\beta^4}\sum_g J_B(m_g^2(\varphi)\beta^2)  - ``\varphi\to\varphi_0", \ee
where
\be J_B(x)\equiv \int_0^\infty dp\, p^2 \log\left(1-e^{-\sqrt{p^2+x}}\right). \label{JBdef}\ee
Again one can note that our loop expansion also contains classical thermal corrections.

\subsubsection{Scalar one-loop effective potential}\label{Scalar one-loop effective potential}
We conclude this section on the effective potential by computing the scalar one-loop contribution. As we have seen, the fermion and gauge fields give a correction to $I$ of order $\hbar$. So to compute the scalar one-loop contribution we can substitute $I$ with the tree-level scalar action $S_s$ in the term proportional to $\hbar$ in~(\ref{GammaI1}). Using~(\ref{SsRen}) one then obtains the following scalar one-loop contribution to the effective action:
\be \Gamma_S(\varphi) = \frac{\hbar}{2}\log\det\left(-\partial^2+M_S^2(\varphi)\right)-``\varphi\to\varphi_0",\ee 
where $M_S^2$ is the background-dependent squared-mass matrix for scalars, which is by definition the Hessian matrix of $\mathcal{V}$. By repeating the same steps performed for the gauge fields we obtain the following scalar contribution to the effective potential at one-loop
\be V^S_{\rm eff}(\varphi) = \frac{\hbar}{2}\sum_s   \int \frac{d^3p} {(2\pi)^3} \sqrt{\vec p^{\, 2}+m_s^2(\varphi)}+\frac{\hbar}{2\pi^2\beta^4}\sum_s J_B(m_s^2(\varphi)\beta^2)   - ``\varphi\to\varphi_0", \ee
where the first term is the zero-temperature contribution, the second term represents the thermal correction and  the $m_s^2(\varphi)$ are the eigenvalues of $M_S^2(\varphi)$, they are the  background-dependent (scalar) squared masses. 

It must be noted that $M_S^2$, unlike $\mu_F\mu_F^\dagger$ and $M_G^2$,  is not positively defined. This is because the tree-level potential $\mathcal{V}$ is not always convex. When some of the scalar eigenvalues $m_s^2$ are negative $V^S_{\rm eff}$ has a non-vanishing imaginary part.  
This pathology is a manifestation of the breaking of the perturbative (loop) expansion. This occurs because the regions where $\mathcal{V}$ is not convex are too far from the minima of the scalar action $S_s$ (recall that our expansion is around the minima of $I(\varphi)-J\varphi$ for infinitesimal $J$ and $I$ reduces to $S_s$ in the scalar case). When some of the sizable scalar eigenvalues  are negative one must use a non-perturbative approach to address this problem, such as the lattice\footnote{Lattice methods have been used in~\cite{KRS,KLRS1,KLRS2,Gould,Gould2} to study the electroweak phase transition, which, because of some issues, including the one mentioned here (see also~\cite{Linde:1980ts,Linde:1978px}), cannot be studied with purely perturbative methods.}.

\subsubsection{Full one-loop effective potential}\label{Full one-loop effective potential}

To summarize the full one-loop effective potential reads (in units where $\hbar=1$)
\be V_{\rm eff}(\varphi) = \mathcal{V}(\varphi) +\mathcal{V}_1(\varphi) +\frac{T^4}{2\pi^2}\left(\sum_b n_b J_B(m_b^2(\varphi)/T^2)-\sum_f n_f J_F(m_f^2(\varphi)/T^2)\right),  \label{VeffSumm}  \ee
where $\mathcal{V}_1(\varphi)$ is the zero-temperature one-loop correction to the effective potential, $n_b=1,3$ for scalar and vector fields, respectively, and $n_f=2,4$ for Weyl and Dirac spinors, respectively.

As we have discussed in Sec.~\ref{Scalar one-loop effective potential}, $V_{\rm eff}$ has an imaginary part for values of $\varphi$ such that the tree-level potential $\mathcal{V}$ is not convex due to the scalar contributions. These values always exist when symmetries are only broken by the Higgs mechanism. To trust perturbation theory and avoid non-perturbative methods (which would be too difficult to discuss here) we now illustrate an alternative symmetry breaking mechanism first introduced by Coleman and E.~J.~Weinberg~\cite{CW}. Coleman and E.~J.~Weinberg considered only a single scalar field, but later Gildener and S.~Weinberg~\cite{GW} generalized the analysis to an arbitrary number of scalars. We will refer to this mechanism as the radiative symmetry breaking (RSB) one, for reasons that will be clear soon.

  In the RSB mechanism one assumes that in the tree-level (zero-loop) potential there are no dimensionful parameters,
  \be \mathcal{V}(\varphi) = \frac{\lambda_{\alpha\beta\gamma\delta}}{4!} \varphi_\alpha\varphi_\beta\varphi_\gamma\varphi_\delta, \label{Vns} \ee  
  where the indices $\alpha, \beta, ...$ run over all scalars in the theory. The $\lambda_{\alpha\beta\gamma\delta}$ are the full set of (dimensionless) quartic couplings.
  The mass scales then emerge radiatively from loops in a way we discuss now.
  
  The basic idea is that, since at quantum level the couplings depend on the energy $\mu$ 
as dictated by the renormalization group, there can be some specific energy 
at which the potential  in Eq.~(\ref{Vns}) develops a flat direction. Such flat direction can be written as $\varphi_\alpha = \nu_\alpha \chi$, where $\nu_\alpha$ are the components of a unit vector $\nu$, i.e.~$\nu_\alpha \nu_\alpha =1$, and $\chi$ is a single scalar field, 
which parameterizes this direction. 
After renormalization, the potential $\mathcal{V}$ along the flat direction, therefore, reads
\be \mathcal{V}(\nu \chi) = \frac{\lambda_\chi (\mu)}{4}\chi^4,  \label{Vvarphi}\ee 
where 
\be \lambda_\chi(\mu) \equiv\frac1{3!} \lambda_{\alpha\beta\gamma\delta}(\mu)\nu_\alpha \nu_\beta \nu_\gamma \nu_\delta. \label{lambdaphi}\ee
Having a flat direction along $\nu$ for $\mu$ equal to some specific value $\tilde\mu$ means 
 \be \lambda_\chi(\tilde\mu)\equiv\lambda_{\alpha\beta\gamma\delta}(\tilde\mu)\nu_\alpha \nu_\beta \nu_\gamma \nu_\delta=0. \ee
 
 Besides the potential in~(\ref{Vvarphi}) loop corrections also generate other terms $\mathcal{V}_1+\mathcal{V}_2+...$, where $\mathcal{V}_i$ represents the $i$-loop correction. 
 The explicit expression of $\mathcal{V}_1$ is well known. We can recover it here by recalling that the effective potential does not depend on $\mu$. Indeed, the renormalization changes the couplings, the masses and  the fields, but leaves the action invariant. So we can write
 \be \mu \frac{d}{d\mu}(\mathcal{V}+\mathcal{V}_1+...) =0.\ee
 Using~(\ref{Vvarphi}), the solution of this equation at the one-loop level reads
 \be \mathcal{V}+\mathcal{V}_1=\frac{\lambda_\chi (\mu)}{4}\chi^4+ \frac{\beta_{\lambda_\chi}}4\left(\log\frac{\chi}{\mu}+a_s\right)\chi^4,  \ee
 where 
 \be \beta_{\lambda_\chi} \equiv  \mu\frac{d\lambda_{\chi}}{d\mu} \ee
 and  $a_s$ is a renormalization-scheme-dependent quantity. Setting now $\mu=\tilde\mu$, where $\lambda_\chi=0$, one obtains
 \be \mathcal{V}_{\rm CW} \equiv \mathcal{V}+\mathcal{V}_1 = \frac{\bar \beta_{\lambda_\chi}}4\left(\log\frac{\chi}{\chi_0}-\frac14\right)\chi^4,\label{CWpot}\ee
where
 \be \bar\beta_{\lambda_\chi} \equiv \left[\beta_{\lambda_\chi} \right]_{\mu=\tilde\mu}, \qquad \chi_0\equiv \frac{\tilde\mu}{e^{1/4+a_s}}.\ee
 We see that the flat direction gets some steepness at loop level.  The new field value $\chi_0$ is a stationary point of $\mathcal{V}_{\rm CW}$.
Moreover, $\chi_0$ is a point of minimum when $\bar\beta_{\lambda_\chi}>0$. Therefore, when the conditions 
  \beq\left\{
\begin{array}{rcll}
\lambda_\chi(\tilde\mu)  &=& 0 & \hbox{(flat direction),}\\
 & & \\ 
\beta_{\lambda_\chi}(\tilde\mu)  &>& 0 & \hbox{(minimum condition),}
\end{array}\right.
\label{eq:CWgen}
\eeq
 are satisfied quantum corrections generate a minimum of the potential at a non-vanishing value of $\chi$, i.~e.~$\chi_0$. In that case $\chi_0$ is the (radiatively induced) zero-temperature vacuum expectation value of $\chi$. 
 
  This non-trivial minimum can generically break global and/or local symmetries and thus generate the  particle masses. Consider for example a term in the (real-time) Lagrangian density $\Lag$ of the form 
 \be \Lag_{\chi h}\equiv \frac12 \lambda_{\alpha\beta} \varphi_\alpha\varphi_\beta |H|^2,\label{LvarphiH}\ee 
 where $H$ is the SM Higgs doublet and the $\lambda_{\alpha\beta}$ are some of the quartic couplings. Substituting the coefficients with the corresponding running quantities and setting $\mu=\tilde\mu$ and $\varphi$ along $\nu$, \be  \Lag_{\chi H} = \frac12 \lambda_{\chi h}(\tilde\mu) \chi^2 |H|^2, \ee
 where 
 \be \lambda_{\chi h}(\mu) \equiv  \lambda_{\alpha\beta}(\mu) \nu_\alpha\nu_\beta.\ee
 Thus, by evaluating this term at the minimum $\chi=\chi_0$ we obtain the Higgs squared mass parameter
  \be \mu_h^2 = \frac12\lambda _{\chi h}(\tilde\mu) \chi_0^2. \ee
 In order to generate the particle masses through the usual Higgs mechanism, we need $\mu_h^2>0$, namely we have the additional condition
 \be \lambda _{\chi h}(\tilde\mu) >0. \ee

Since the parameter $\mu_h^2$ is directly related to the Higgs squared mass we cannot use this mechanism to generate $\mu_h^2$ when $\chi_0$ is much below the EW scale. This is mainly due to collider limits on extra light scalars with strong coupling to the Higgs as well as the requirement of validity of perturbation theory. Of course, it is still possible that the SM with an explicit scale-symmetry breaking parameter is coupled to an approximately scale-invariant sector that features RSB (see e.g.~\cite{Salvio:2023blb,Ghoshal:2020vud}). In this case perturbation theory can be compatible with a $\chi_0$ much smaller than the EW scale and such scale-invariant sector must be ``dark", i.e.~must have only tiny couplings with the SM particles.

Let us see now that the imaginary part of the effective potential in~(\ref{VeffSumm}) vanishes in the CW case. The Hessian matrix  of the classical potential in~(\ref{Vns}) is given by
\be M_{\alpha\beta}^2  \equiv \frac{\partial^2\mathcal{V}}{\partial\varphi_\alpha\partial\varphi_\beta} =\frac12 \lambda_{\alpha\beta\gamma\delta} \varphi_\gamma\varphi_\delta. \ee 
By evaluating this Hessian matrix at the CW flat direction we obtain that $M_{\alpha\beta}^2$ is proportional to $\chi^2$ via some quartic couplings, namely
\be M_{\alpha\beta}^2(\chi)=\frac12\lambda_{\alpha\beta\gamma\delta}\nu_\gamma\nu_\delta\chi^2. \ee
All the  $m_s^2$ must be non negative to have a bounded-from-below classical potential. To prove this first note that 
the requirement that the classical potential in~(\ref{Vns}) is bounded from below also implies that potential has its absolute minimum at $\varphi=0$ because no scales are present in~(\ref{Vns}) and this minimum vanishes. Also the classical potential vanishes at the flat direction, otherwise that direction would not be flat. So, for a classical potential bounded from below and with a flat direction $\varphi=\nu\chi$
\be \mathcal{V}(\varphi) = \mathcal{V}(\nu \chi)+\frac{\partial \mathcal{V}}{\partial\varphi_\alpha}(\nu \chi) \delta\varphi_\alpha+\frac12\frac{\partial^2\mathcal{V}}{\partial\varphi_\alpha\partial\varphi_\beta}(\nu \chi) \delta\varphi_\alpha\delta\varphi_\beta=\frac12\frac{\partial^2\mathcal{V}}{\partial\varphi_\alpha\partial\phi_\beta}(\nu \chi) \delta\varphi_\alpha\delta\varphi_\beta,  \ee
where $\delta\varphi\equiv \varphi-\nu\chi$ is taken here infinitesimal. As a result, if there were negative eigenvalues of $M_{S\alpha\beta}^2(\chi)$   the classical potential would become smaller than its value at the flat direction, but we have seen that this is not possible for  a bounded-from-below potential. We can, therefore, conclude that the CW  symmetry breaking supports the validity of perturbation theory.

\begin{figure}[t]
\begin{center}
  \includegraphics[scale=0.5]{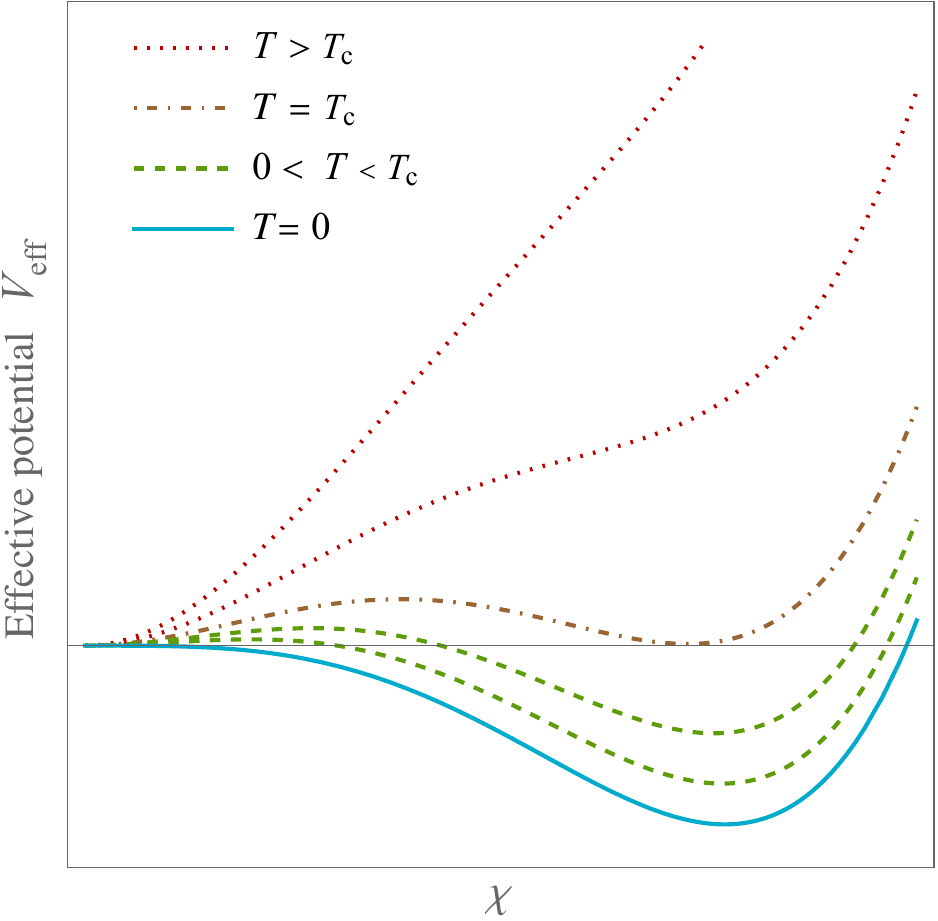} \qquad \includegraphics[scale=0.5]{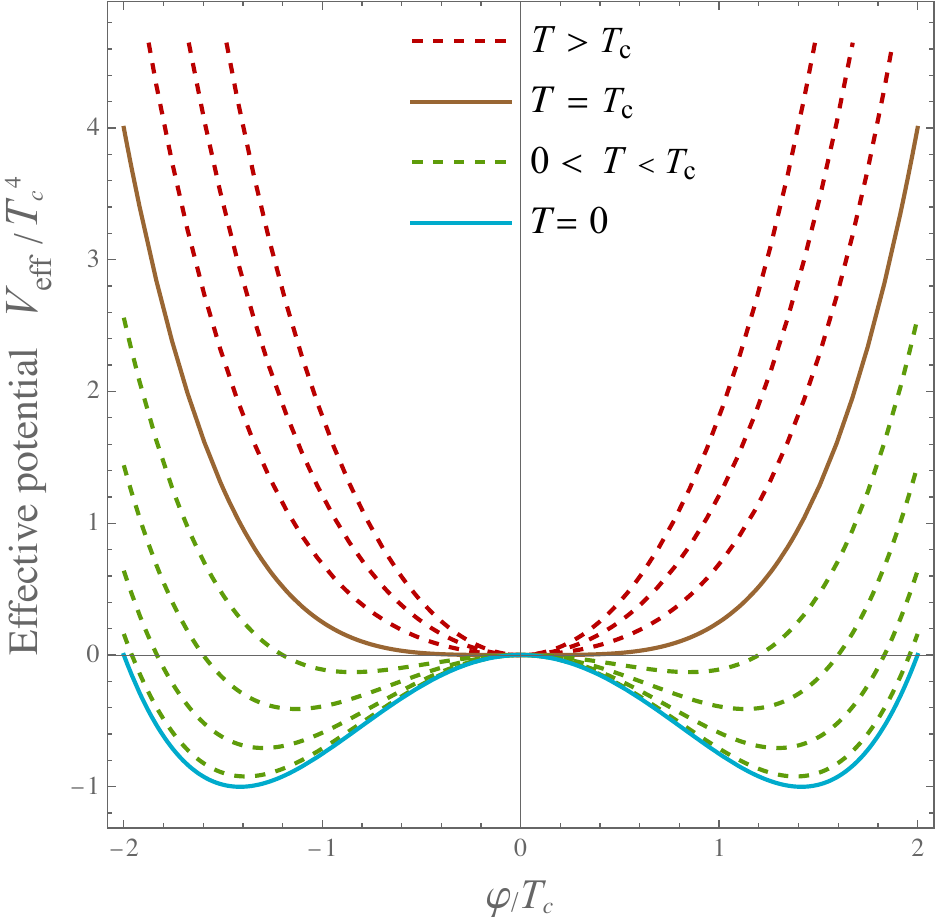}  
    \caption{\em {\bf Left:} The temperature-dependent effective potential built around the Coleman-Weinberg potential (azure solid line). This effective potential features a first-order phase transition. {\bf Right:} The temperature-dependent effective potential of Eq.~(\ref{Ex2PT}), which features a second-order phase transition. Here $B=\lambda=1$.}\label{CWpotf}
  \end{center}
\end{figure}

 \subsection{First-order phase transitions}\label{first-order phase transitions}

 As we pointed out at the beginning of Sec.~\ref{Phase transitions in field theory}, the role of the order parameter can be played by $\langle\hat\varphi\rangle$ in field theory. 
Here we discuss one class of phase transitions, known as first-order phase transitions, where the  order parameter changes discontinuously between two equilibrium states as the temperature crosses a critical value $T_c$.

First-order phase transitions always occur in the presence of the RSB mechanism discussed in Sec.~\ref{Full one-loop effective potential} as we now show~\cite{Salvio:2023qgb,Salvio:2023ynn} (see also~\cite{Salvio:2020axm}).
Let us start by noting that the first three derivatives of the CW potential in~(\ref{CWpot}) vanishes at the origin, $\chi=0$. On the other hand, it can be shown that $J_B(x)$ and $J_F(x)$ defined in~(\ref{JBdef}) and~(\ref{JFdef}), respectively, feature in their small-$x$ expansion a term linear in $x$ with a coefficient that is positive in $J_B$ and negative in $J_F$
\bea J_B(x) &=& J_B(0)+J_B^{(1)} x + ...,  \\
J_F(x) &=&J_F(0)+J_F^{(1)} x + ... , \eea
where the dots are the $\mathcal{O}(x^2)$ terms and
\bea J_B^{(1)}&=&\frac12 \int_0^{\infty} dp\frac{p}{e^p-1}=\frac{\pi^2}{12},  \\
J_F^{(1)} &=&-\frac12 \int_0^{\infty} dp\frac{p}{e^p+1} =-\frac{\pi^2}{24}. 
\eea
Since $J_B$ and $J_F$ appear in the effective potential in the way described by Eq.~(\ref{VeffSumm}), this implies that the effective potential has a minimum at the origin, $\chi=0$, at any temperature.  
Going to small enough temperature the absolute minimum should be approximately the $T=0$ one given by the CW potential, but since there is always a positive quadratic term thanks to the finite-temperature contributions the full effective potential  always features a barrier between $\chi=0$ and $\chi=\chi_0$ (see the left plot in Fig.~\ref{CWpotf}). The order parameter then changes discontinuously between two equilibrium states after the temperature has crossed the critical value $T_c$ where the two minima are degenerate.

In the left plot of Fig.~\ref{CWpotf} we illustrate how the effective potential typically changes varying the temperature. As $T$ approaches $T_c$ from above a barrier starts to form. The two minima become degenerate at $T=T_c$ and for $T<T_c$ the absolute minimum is the one with $\chi\neq0$, which tends to $\chi_0$ as $T\to 0$.

  \subsection{Second-order phase transitions}\label{Second-order phase transitions}

Here we  discuss another class of phase transitions, known as second-order phase transitions, where the order parameter changes continuously as the temperature crosses a critical value $T_c$.

For example, this can happen if the effective potential, computed with some non-perturbative method that avoids any sizable imaginary part, has the form 
\be V_{\rm eff}(\varphi) = B(T^2-T_c^2)\varphi^2+\frac{\lambda}{4} \varphi^4, \label{Ex2PT}\ee
where, for simplicity, we have assumed that there is a single relevant scalar, $\varphi$, and $B$ and $\lambda$ are real constants. The temperature dependence of this effective potential is illustrated in the right plot of Fig.~\ref{CWpotf}. No barrier emerges at any temperature. So when the non-vanishing point of minimum appears,  $\varphi=0$ becomes a maximum and the order parameter continuously rolls towards the minimum. 

  \subsection{Thermal vacuum decay in first-order phase transitions}\label{Thermal vacuum decay in first-order phase transitions}

In a first-order phase transition, when $T<T_c$, the state of the system changes due to  quantum and thermal tunneling through the barrier, which separates two equilibrium states. The example of the Coleman-Weinberg potential is  shown in the left plot of Fig.~\ref{CWpotf}, but we provide here a general analysis of this tunneling phenomenon that applies to any first-order phase transition.
Let us now describe this process in the thermal field theory we have constructed\footnote{For further details see~\cite{Coleman,CallanColeman,Linde1,Linde2}.
For a textbook treatment see e.g. Sec.~23.8 of~\cite{Weinberg2}.}. 

Let us start by conventionally setting $V_{\rm eff}(0)=0$, where $\chi=0$ is identified with the metastable (a.k.a.~``false") vacuum, which is only a local minimum. The transition between $\chi=0$ and the ``true" vacuum $\chi=\chi_0$ (the absolute minimum) cannot be homogeneous in spacetime: only when 
\be \chi\to 0 \qquad \mbox{as}  \quad r\equiv\sqrt{t_E^2+\vec{x}^2}\to \infty, \label{BCbounce}
\ee
 the effective action $\Gamma(\chi)$ can be finite. Since the effective action corresponds to the free energy via Eq.~(\ref{FGammaLink}) and we are ultimately interested in setting the external source to zero, see Eq.~(\ref{VEVJ0}), we also note that the  true and false vacua are separated by a positive free energy barrier. So one would need an infinite free energy if the transition were homogeneous.

This implies a departure from thermal equilibrium: recall that the average $\langle\hat\chi\rangle$ should be spacetime independent at equilibrium as a consequence of spacetime translation invariance. However,
note that interactions between the system and the environment can lead to a time-dependence of the temperature and/or the Hamiltonian.
For example, a cosmological spacetime, which depends on time,  is not time-translation invariant and induce a departure from thermal equilibrium. For example, suppose that effects of this type introduce a dependence of the temperature $T$ on time $t$, such that one finds  the out-of-equilibrium density matrix $\rho(t) = \exp(-H/T(t))/Z$, where $Z =  \Tr\exp(-H/T(t))$. In this case the average reads 
\be \langle\hat\chi(t,\vec{x})\rangle = \Tr(\rho(t)\hat\chi(t,\vec{x})). \ee
 Since $[P_\mu,H]=0$, the unitary operator corresponding to any spacetime translation commutes with $\rho(t)$ and so $\langle\hat\chi(t,\vec{x})\rangle = \Tr(\rho(t) \hat\chi(0))$, which is not the same as  $\langle\hat\chi(0,0)\rangle = \Tr(\rho(0)\hat\chi(0))$. The expectation value of the  field is no longer spacetime independent. Nevertheless, one can think of this departure from equilibrium as a sequence of quasi-equilibrium states with some time-dependent temperature.

The non-homogeneous transition is described by a {\it regular} field configuration $\chi_b$ (known as the ``bounce"), which can interpolate between the true and the false vacua. It thus represents a bubble of the true vacuum inside the false vacuum. First-order phase transitions occur through this nucleation process, similarly to boiling water. The bounce is a minimum of $\Gamma(\chi)$, as we now show, and it thus has to solve the associated field equations
\be \left.\frac{\delta \Gamma(\chi)}{\delta \chi(x)}\right|_{\chi=\chi_b} =0.\ee
 Note that \be \exp(-F/T) \equiv Z = \sum_i \exp(-  E_i/T),\ee 
where $E_i$ is the energy of the microstate $i$ and the sum over $i$ ranges over all possible microstates. If there are three macrostates, the false vacuum, the true vacuum and the transition state described by the bounce, the microstates can be roughly divided in three sets corresponding to these three macrostates. So, the probability of set $a$ with free energy  $F_a$ is approximately given by $\exp(-F_a/T)$. Since the bounce describes the transition, $\exp(- F_b/T)$, where $F_b$ is the free energy of the bounce, approximately gives us the transition probability. By recalling the relation between the free energy and the effective action we can rewrite  this  probability in terms of $\exp(-\Gamma(\chi_b))$. Now we see that the field configuration we are interested in has to solve the field equations associated with $\Gamma(\chi_b)$ (together with the mentioned boundary conditions, namely regularity and also Condition~(\ref{BCbounce})): other configurations would have a larger action and give a negligible contribution to the transition probability.

 Since $V_{\rm eff}$ in phase transitions lead to non-linear field equations, finding the bounce is generically a complicated non-linear problem, which typically requires numerical methods. 

\begin{figure}[t]
\begin{center}
  \includegraphics[width=.9\linewidth]{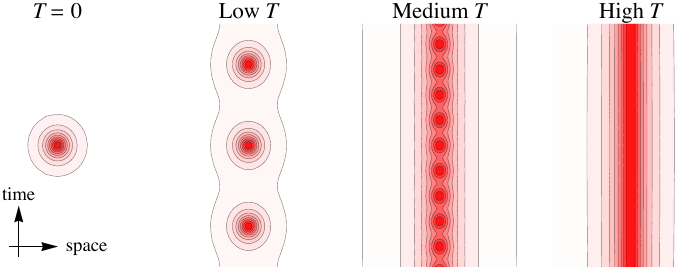}
  \end{center}
\caption{\em 
Bounces at various temperatures. The vertical and horizontal axis represent the Euclidean time direction and the  spatial radius, respectively.
At $T=0$ (left-most panel) the bounce solution represents a bubble.
At finite temperature, the bounce solution becomes a series of bubbles
placed at distance $\beta$ in the time direction.
At large temperature (right-most panel) the bounce no longer depends on time. Figure reproduced from~\cite{Salvio:2016mvj}.
 \label{fig:ThermalBounce} }
\end{figure}

Because of the time periodicity with period $\beta$, in raising the temperature from very small to very large values one interpolates between a four-dimensional bubble and a three-dimensional (time-independent) bubble~\cite{Linde1}, as illustrated in Fig.~\ref{fig:ThermalBounce}.

 Note that this formalism can readily be applied to the zero-temperature limit. In this case the role of the critical temperature $T_c$ is played by a critical surface in the space of  parameters of the zero-temperature effective action. Recently, it has been shown that even the   parameters of the SM (or some of its realistic extensions) could lie in such a critical surface, with interesting cosmological consequences~\cite{Buttazzo:2013uya,Bezrukov:2014bra,Hamada:2014wna,Salvio:2017oyf,Salvio:2018rv,Salvio:2022mld}.


First-order phase transitions are very interesting for various reasons. One is that they represent a violent process because of the bubble nucleation we have described. Such process leads to the production of interesting phenomena such as gravitational waves and black holes. These in turn can produce observable effects that can be measured by some  experiments. A second reason is that in the Standard Model of particle physics (SM) there are no first-order phase transitions. Therefore, the observation of such a phase transition would be a clear signal of beyond-the-SM physics and, therefore, a discovery of great importance.

\newpage

\section{Summary and discussion}

Here the fundamentals and some important applications of TFT have been reviewed. The relevant formul\ae~have been derived from first principles assuming only knowledge of non-statistical quantum field theory. 

We began in Sec.~\ref{Density matrix and ensemble averages} with a detailed discussion of the density matrix for relativistic quantum theories where the conserved quantities are the four-momentum, the angular momentum and a set of conserved charges corresponding to internal symmetries. In the same section it was shown that the thermal Green's functions play a crucial role in TFT, as they allow us to compute, among other things, the rates of inclusive processes.

Thus, a detailed discussion of the thermal Green's functions was provided, first without interactions in Sec.~\ref{Thermal free fields} and then in the general case in Sec.~\ref{Thermal Green's functions}. However, in those sections, we assumed that the system is at equilibrium and that the average angular momentum vanishes. All relevant types of fields were considered:  scalars, fermions and gauge fields. For all cases, including fermions, path-integral representations for the thermal Green's functions have been derived from first principles (for negligible chemical potentials), with a formalism that includes the real-time and imaginary-time formalisms as particular cases.   

These results already allowed us to understand some of the most important applications of TFT, the weakly-coupled particle production in a thermal bath (Sec.~\ref{Weakly-coupled particle production}) and the phase transitions in field theory (Sec.~\ref{Phase transitions in field theory}), which have attracted, and continue to attract, significant interest in the fundamental-physics community. 
The rates of particle thermal production are conveniently re-expressed in terms of exact thermal 2-point functions.  The thermal phase transitions in field theory have been explicitly and extensively described in the case where the phase transition is associated with an RSB, as in this case one can use perturbative methods. However, in Sec.~\ref{Thermal vacuum decay in first-order phase transitions} the phenomenon of thermal vacuum decay in first-order phase transitions has been explained in a general theory, illustrating how the bounce solution emerges in this out-of-equilibrium process and allows us to compute the corresponding decay probability. 

Many other (more advanced) developments and applications of TFT are present in the literature, such as the thermal Green's functions at finite density~\cite{Landsman:1986uw} (see also~\cite{Podo:2023ute}), the thermal production of gravitons~\cite{Grasso:2003cq,Ghiglieri:2015nfa,Ghiglieri:2020mhm,Ghiglieri:2024ghm} and some non-perturbative numerical description of phase transitions in field theory~\cite{KRS,KLRS1,KLRS2,Gould,Gould2}. The introductory treatment of TFT provided here, appropriately extended, may allow for an understanding of these more advanced topics too.

\subsection*{Acknowledgments}
I thank R.~Frezzotti, J.~Ghiglieri and all the students, particularly A.~Cipriani and F.~Rescigno, who provided useful feedback on the manuscript.

\vspace{1cm}

 \footnotesize
\begin{multicols}{2}

\end{multicols}

  \end{document}